\documentclass[format=acmsmall, review=false, screen=true]{acmart}

\usepackage{booktabs} % For formal tables

\usepackage[ruled]{algorithm2e} % For algorithms

\SetAlFnt{\small}
\SetAlCapFnt{\small}
\SetAlCapNameFnt{\small}
\SetAlCapHSkip{0pt}
\IncMargin{-\parindent}
\usepackage{algpseudocode}
\usepackage{subfig}
\usepackage{enumitem}
\usepackage{graphicx}
\usepackage{multicol}
\usepackage{amssymb}
\usepackage{hhline}
\usepackage[labelfont=bf]{caption}
\usepackage{subfig}
\usepackage{float}
\usepackage{color}
\usepackage{graphicx}
\usepackage{multirow}
\usepackage{tabularx}
\usepackage{array,graphicx}
\usepackage{booktabs}
\usepackage{pifont}
\usepackage[utf8]{inputenc}
\usepackage[english]{babel}
\usepackage{amsmath}
\newtheorem{defn}{Definition}[section]



\acmJournal{TWEB}
\acmVolume{9}
\acmNumber{4}
\acmArticle{39}
\acmYear{2018}
\acmMonth{7}
\copyrightyear{2009}
\setcopyright{acmlicensed}
\acmDOI{0000001.0000001}

\received{July 2018}
\received[revised]{2018}
\received[accepted]{2018}

\begin{document}
% Title portion. Note the short title for running heads
\title[Privacy in Social Media]{Privacy in Social Media: Identification, Mitigation and Applications}%:Leakage, Atatcks and Applicatins

\author{Ghazaleh Beigi}
\affiliation{%
  \institution{Arizona State Univesity}
  \city{Tempe}
  \state{AZ}
  \country{USA}}
\email{gbeigi@asu.edu}
\author{Huan Liu}
\affiliation{%
	\institution{Arizona State Univesity}
	\city{Tempe}
	\state{AZ}
	\country{USA}}
\email{huan.liu@asu.edu}

\begin{abstract}
The increasing popularity of social media has attracted a huge number of people to participate in numerous activities on a daily basis. This results in tremendous amounts of rich user-generated data. This data provides opportunities for researchers and service providers to study and better understand users' behaviors and further improve the quality of the personalized services. Publishing user-generated data risks exposing individuals' privacy. Users privacy in social media is an emerging task and has attracted increasing attention in recent years. These works study privacy issues in social media from the two different points of views: identification of vulnerabilities, and mitigation of privacy risks. Recent research has shown the vulnerability of user-generated data against the two general types of attacks, identity disclosure and attribute disclosure. These privacy issues mandate social media data publishers to protect users' privacy by sanitizing user-generated data before publishing it. Consequently, various protection techniques have been proposed to anonymize user-generated social media data. There is a vast literature on privacy of users in social media from many perspectives. In this survey, we review the key achievements of user privacy in social media. In particular, we review and compare the state-of-the-art algorithms in terms of the privacy leakage attacks and anonymization algorithms. We overview the privacy risks from different aspects of social media and categorize the relevant works into five groups 1) graph data anonymization and de-anonymization, 2) author identification, 3) profile attribute disclosure, 4) user location and privacy, and 5) recommender systems and privacy issues. We also discuss open problems and future research directions for user privacy issues in social media.
\end{abstract}




\maketitle

\renewcommand{\shortauthors}{G. Beigi et al.}

\section{Introduction}
%privacy concern has been an issue since the emergence of social media.
\footnote{This paper is currently under review.}Explosive growth of the Web in the last decade has drastically changed the way billions of people all around the globe conduct numerous activities such as surfing the web, creating online profiles in social media platforms, interacting with other people, and sharing posts and various personal information in a rich environment. This results in tremendous amounts of user-generated data. The centralization of massive amounts of user information and the availability of up-to-date data which is consistently tagged and formatted, makes social media platforms an attractive target for organizations seeking to collect and aggregate this information either for legitimate purposes or malicious goals~\cite{bonneau2009prying}. For example, the user-generated data provides opportunities for researchers and business partners to study and understand individuals at unprecedented scales~\cite{backstrom2007wherefore,beigi2018securing}. This information is also crucial for online vendors to provide personalized services and a lack of it would result in a deteriorating quality of online personalization service.

On the other hand, this tremendous amount of user-generated data risks exposing individuals' privacy as it is rich in content including a user's relationships and other sensitive and private information~\cite{ji2016general,narayanan2009anonymizing,beigi2018social}. This data also makes online users traceable and accordingly, users become severely vulnerable to potential risks ranging from persecution by governments to targeted frauds. For example, users may share their vacation plans publicly on Twitter without knowing that this information could be used by adversaries for break-ins and thefts in the future~\cite{TextAnonymization,mao2011loose}. Moreover, sensitive information that users do not explicitly disclose such as location~\cite{li2012towards,mahmud2014home}, age~\cite{wang2016your} and trust/distrust relationships~\cite{beigi2016exploiting,beigi2016signed}, can be easily inferred from their activities on social media.

Privacy issues could be raised when the data get published by a data publisher or service provider. In general, two types of information disclosures have been identified in the literature: identity disclosure and attribute disclosure attacks~\cite{duncan1986disclosure,lambert1993measures,li2007t}. Identity disclosure occurs when an individual is mapped to an instance in a released dataset. Attribute disclosure happens when the adversary could infer some new information regarding an individual based on the released data. Attribute disclosure becomes more probable when there is accurate disclosure of people's identities.
Similarly, privacy leakage attacks in social media could be also categorized into either identity disclosure or attribute disclosure. These user privacy issues mandate social media data publishers to protect users' privacy by sanitizing user-generated data before it is published publicly.% in order to address privacy concerns. 

Data anonymization is a complex problem and its goal is to remove or perturb data to prevent adversaries from inferring sensitive information while ensuring the utility of the published data. One straightforward anonymization technique is to remove ``Personally Identifiable Information" (a.k.a. PII) such as names, user ID, age and location information. This solution has been shown to be far from sufficient in preserving privacy~\cite{backstrom2007wherefore,narayanan2008robust}. An example of this insufficient approach is the anonymized dataset published for the Netflix prize challenge. As a part of the Netflix prize contest, Netflix publicly released a dataset containing movie ratings of 500,000 subscribers. The data was supposed to be anonymized and all PII are removed from it. Narayanan et al.~\cite{narayanan2008robust} propose a de-anonymization attack which map users' records in the anonymized dataset to corresponding profiles on IMDB. In particular, the results of this work show that the structure of the data carry enough information for a potential breach of privacy to re-identiy anonymized users.

Consequently, various protection techniques have been proposed to anonymize user-generated social media data. In general, the ultimate goal of an anonymization approach is to preserve social media user privacy while ensuring the utility of published data. As a counterpart to this research direction, another group of works investigate the potential privacy breaches from social media user data by introducing new attacks. These group of works find the gaps in anonymizing user-generated data and further improve anonymization techniques. %In particular, the aim of de-anonymization attacks is to reduce breach of privacy by probing the potential drawbacks of anonymization techniques. 

There is vast literature on privacy of users in social media from many perspectives. The goal of this article is to provide a comprehensive review of existing works on user privacy issues and solutions in social media and give a guidance on future research directions. The contributions of this paper are summarized as follows:
\begin{itemize} [leftmargin=*]
	\item We overview the traditional privacy models for structured data and discuss how these models are adopted for privacy issues in social media. We formally define two types of privacy leakage disclosures that covers most of the existing definitions in the literature.
	\item We categorize privacy issues and solutions on social media into different groups including 1) graphs data anonymization and de-anonymization, 2) author identification, 3) user profile attributes disclosure, 4) location and privacy and 5) recommendation systems and privacy. We then give an overview of existing works in each group with a principled way to group representative methods into different categories.
	\item We discuss several open issues and provide future directions for privacy in social media.
\end{itemize}

The remainder of this survey is organized as follows. In Section 2, we present an overview of traditional methods and formally define two types of privacy disclosures. In Section 3, we review the state-of-the-art methods for privacy of social media graphs. More specifically, Section 3.1. covers de-anonymization attacks on social media graphs and Section 3.2. covers anonymization techniques which are proposed for preserving privacy of graph data against de-anonymization attacks. We review author identification works in Section 4. In Sections 5 and 6, we overview state-of-the-art de-anonymization techniques for inferring users profile attributes and location information. In Section 7, privacy issues and solutions in recommendation systems are reviewed. Finally, we conclude this article in Section 8 by discussing the open issues and future directions .

%Privacy of individuals, algorithms and to-be-released datasets have been studied a lot. There is vast literature on privacy from many perspectives. There is a thorough survey~\cite{verykios2004state} on privacy preserving data mining which studies different approaches to modify data to preserve privacy. Another work from Agrawal et al~\cite{agrawal2000privacy} proposes perturbing data values by adding random noise to them in order to preserve privacy of users while retain the statistical properties of the original data. Another set of works focus on developing privacy preserving association mining rules to minimize privacy loss~\cite{rizvi2002maintaining,evfimievski2004privacy}.

\section{Traditional Privacy Models}
Privacy preserving techniques were first introduced for tabular and micro data. With the emergence of social media, the issue of online user privacy was raised. Researchers then focus on studying privacy leakage issues as well as anonymization and privacy preserving techniques specialized for social media data. There are two types of information disclosure in the literature: identity disclosure and attribute disclosure attacks~\cite{duncan1986disclosure,lambert1993measures,li2007t}. We can formally define identity disclosure attack as:
\begin{defn}{\textbf{ Identity Disclosure Attack}}. Given $T = (\mathbf{G}, \mathbf{A}, \mathbf{B})$, which is a snapshot of a social media platform with a social graph $\mathbf{G} =(V,E)$ where $V$ is the set of users and $E$ demonstrates the social relations between them, a user behavior $\mathbf{A}$ and an attribute information $\mathbf{B}$, the identity disclosure attack is to map all users in the list of target users $V_t$ to their known identities. For each $v \in V_t$, we have the information of her social friends and behavior.
\end{defn}

Attribute disclosure attack for social media data could be also formally defined as:
\begin{defn}{\textbf{ Attribute Disclosure Attack}}. Given $T = ( \mathbf{G}, \mathbf{A}, \mathbf{B})$, which is a snapshot of a social media platform with a social graph $\mathbf{G} =(V,E)$ where $V$ is the set of users and $E$ demonstrates the social relations between them, a user behavior $\mathbf{A}$ and an attribute information $\mathbf{B}$, the attribute disclosure attack is used to infer the attributes $a_v$ for all $v \in V_t$ where $V_t$ is a list of targeted users. For each $v \in V_t$, we have the information of her social friends and behavior.
\end{defn}
%In general, two types of information disclosures have been identified in the literature, identity disclosure and attribute disclosure attacks~\cite{duncan1986disclosure,lambert1993measures,li2007t}. Identity disclosure happens when an individual is mapped to an instance in a released dataset. Attribute disclosure happens when the adversary could infer some new information regarding an individual based on the released data. Identity disclosure could also help to have more accurate disclosure of people's attributes.
%Similarly, privacy leakage attacks in social media could be also categorized into either identity disclosure or attribute disclosure. 
Network graph de-anonymization and author identification are examples of identity disclosure attacks that exists in social media. Examples of attribute disclosure attack include the disclosure of users' profile attributes, location, and preferences information in recommendation systems. 

Before we discuss privacy leakage in social media, we first overview the traditional privacy models for structured data. Traditional privacy models such as $k$-anonymity~\cite{sweeney2002k}, $l$-diversity~\cite{machanavajjhala2006diversity}, $t$-closeness~\cite{li2007t} and differential privacy~\cite{dwork2008differential} are defined over structured databases and cannot be applied to unstructured user generated data in social media platforms. The reason is that quasi-identifiers and sensitive attributes are not clear in the context of social media data. These techniques are further adopted for social media data which we will discuss more in the next sections. Last but not least, we discuss the related work and highlight the differences between this work and other surveys in existing literature. 
\subsection{k-anonymity, l-diversity and t-closeness}
$k$-anonymity was one of the first techniques introduced for protecting data privacy~\cite{sweeney2002k}. The aim of $k$-anonymity is to anonymize each instance in the dataset so that it is indistinguishable from at least $k-1$ other instances with respect to certain identifying attributes. $k$-anonymity could be achieved through suppression or generalization of the data instances. The goal here is to anonymize the data such that $k$-anonymity is preserved for all instances in the dataset with a minimum number of generalizations and suppressions while maximizing the utility of the resultant data. It has been shown that this problem is NP-hard~\cite{aggarwal2005approximation}. $k$-anonymity was initially defined for tabular data, but then researchers start to adopt it for solving privacy issues in social media data. In social media related problems, $k$-anonymity ensures that users cannot be identified and there are $k-1$ other users with the same set of features which makes these $k$ users indistinguishable. These features may include users' attributes and structural properties. 

Although $k$-anonymity is among the first techniques proposed for protecting the privacy of datasets, it is still vulnerable against specific types of privacy leakage. Machanacajjhala et al.~\cite{machanavajjhala2006diversity} introduces two simple attacks which defeats $k$-anonymity. The first attack is homogeneity attack in which the adversary can infer an instance's (in this case, a users in social media) sensitive attributes when sensitive values in an equivalence class lack diversity. In the second attack the adversary can infer an instance's sensitive attributes when he has access to background knowledge even in the case that the data is $k$-anonymized. The second attack is known as background knowledge attack. Variations of background knowledge attacks are proposed and used for inferring social media users' attributes. The background knowledge could be users' friends' or behavioral information. We will discuss more about different types of the attribute inference attacks problem in Sections 6 and 7.

To protect data against homogeneity and background knowledge attacks, Machanacajjhala et al.~\cite{machanavajjhala2006diversity} introduce the concept of $l$-diversity. It ensures that the sensitive attribute values in each equivalence class are diverse. More formally, a set of records in an equivalence is $l$-diverse if the class contains at least $l$ \textit{well represented} values for the sensitive attributes. The dataset is then $l$-diverse if every class is $l$-diverse. Two instantiations of the $l$-diversity concept are then introduced, entropy $l$-diversity and recursive $(c,l)$-diversity. With entropy $l$-diversity, each equivalence must not only have enough different sensitive values, but also each sensitive value must be distributed evenly enough. More formally, the entropy of the distribution of sensitive values in each equivalence class is at least $log(l)$. For recursive $(c,l)$-diversity, the most frequent value should appear frequent enough in the dataset. Interested readers could refer to the work of~\cite{machanavajjhala2006diversity} for more details. 

After $l$-diversity, Li et al.~\cite{li2007t} studies the vulnerabilities of $l$-diversity and introduce a new privacy concept, $t$-closeness. They show that $l$-diversity cannot protect the privacy of data when the distribution of sensitive attributes in the equivalence class is different from the distribution in the whole dataset. If the distribution of sensitive attributes is skewed, then $l$-diversity presents a serious privacy risk. This attack is known as the skewness attack. $l$-diversity is also vulnerable against similarity attacks. This attack can happen when the sensitive attributes in an equivalence class are distinct but semantically similar~\cite{li2007t}. Li et al.~\cite{li2007t} thus introduce a new privacy concept $t$-closeness which ensures that the distribution of a sensitive attribute in any equivalence class is close to the distribution of the sensitive attribute in the overall table. More formally speaking, an equivalence class satisfies $t$-closeness if the distance between the distribution of a sensitive attribute in this class and the distribution of the attribute in the whole dataset is no more than a certain threshold. The whole dataset is said to have $t$-closeness if all equivalence classes have $t$-closeness. It's valuable to mention that $t$-closeness protects the data against attribute disclosure but not identity disclosure. 

$k$-anonymity, $l$-diversity and $t$-closeness are further adopted for unstructured social media data. Table~\ref{kanonymity} summarizes different approaches that leverage adopted versions of these techniques for privacy problems in social media. These works are discussed more in the following sections.

\begin{table}[h]
	\center
	\small
	\caption{$k$-anonymity, $l$-diversity and $t$-closeness applications in user privacy in social media.}
	\label{kanonymity}
	\begin{tabular}{|l|p{4cm}|p{3cm}|}
		\hline
		Technique & Type of Information & Paper\\ \hline
		$k$-degree anonymity & graph structure & \cite{liu2008towards}\\\hline
		$k$-neighborhood anonymity & graph structure & \cite{zhou2008preserving}\\\hline
		$k$-automorphism & graph structure &\cite{zou2009k}\\\hline
		$k$-isomorphic & graph structure &\cite{cheng2010k} \\\hline
		$k$-anonymity & graph structure and attribute information & \cite{yuan2010personalized}\\\hline
		$(\theta,k)$-matching anonymity & graph structure and attribute information & \cite{andreou2017identity}\\ \hline
		$(k,d)$-anonimity & graph structure and attribute information & \cite{backes2015closed,backes2016profile} \\ \hline
		$l$-diversity & attribute information & \cite{machanavajjhala2006diversity}\\\hline
		$t$-closeness & attribute information & \cite{li2007t} \\ \hline
	\end{tabular}
	
\end{table}

\subsection{Differential Privacy}
Differential privacy is a powerful technique which protects a user's privacy during statistical query over a database by minimizing the chance of privacy leakage while maximizing the accuracy of queries. It is introduced by Dwork et al.~\cite{dwork2006calibrating,dwork2008differential} and provides a strong privacy guarantee. The intuition behind differential privacy is that the risk of user's privacy leakage should not be increased as a result of participating in a database~\cite{dwork2008differential}. In particular,  it imposes a guarantee on the data release mechanism rather than the dataset itself. The privacy risk is also evaluated according to the existence or absence of an instance in the database. Differential privacy assumes that data instances are independent from each other and guarantees that existence of an instance in the database does not pose a threat to its privacy as the statistical information of data would not change significantly in comparison to the case that the instance is absent~\cite{dwork2006calibrating,dwork2008differential}. This way, the adversary cannot infer whether an instance is in the database or not or which record is associated with it~\cite{kifer2011no}. Differential privacy can be more formally defined as:

\begin{defn}{\textbf{Differential Privacy}}. Given a query function $f(.)$, a mechanism $K(.)$ with an output range $\mathcal{R}$ satisfies $\epsilon$-differential privacy for all datasets $\mathcal{D}_1$ and $\mathcal{D}_2$ differing in at most one element \textit{iff}:
	\begin{equation}\label{dp}
	\frac{Pr[K(f(\mathcal{D}_1))=R \in \mathcal{R}]}{Pr[K(f(\mathcal{D}_2))=R \in \mathcal{R}]} \leq e^{\epsilon}
	\end{equation}
\end{defn}

Here $\epsilon$ is called privacy budget and large values of $\epsilon$ (e.g., 10) results in large $e^{\epsilon}$ and indicates that large output difference could be tolerated and hence we have large privacy loss. This is because the adversary can infer the change in the database according to the large change of the query function $f(.)$. On the other hand, small values of $\epsilon$ (e.g., 0.1) indicate that small privacy loss could be tolerated. Query function $f(.)$ can be thought of as a request about value of a random variable and mechanism $K(.)$ is also a randomized function which can be considered as an algorithm that returns the results for the query function, possibly with some noise. To make it more clear, let's assume that we have a dataset containing every patient information. An example of the query function $f(.)$ could be the question: \textit{How many people have the disease $x$?}. The mechanism $K(.)$ could be any algorithm that finds the answer to this question. The output range $\mathcal{R}$ for mechanism $K(.)$ in this example is, $\mathcal{R} = \{1,2,...,n\}$ where $n$ is the total number of patients in the dataset.

Differential privacy models could be either interactive or non-interactive. Assume that the data consumer executes a number of statistical queries on the same dataset. In the interactive models, the data publisher responds to the customer with $K(f(\mathcal{D}))$, where $K(.)$ perturbs the query results to achieve the privacy guarantees. In non-interactive models, the data publisher designs a mechanism $K(.)$ which transforms the original data $\mathcal{D}$ into a new anonymized dataset $\mathcal{D}' = K(f(\mathcal{D}))$. The perturbed data $\mathcal{D}'$ is then returned to the consumer which is ready for arbitrary statistical queries.

A common way of achieving differential privacy is through adding random noises, i.e. Laplacian or Exponential to the query answers~\cite{dwork2008differential}. The Laplacian mechanism is a popular technique for providing $\epsilon$-differential privacy which adds Laplace noise drawn from Laplace distribution. Since $\epsilon$-differential privacy is defined over the query function and holds for all datasets according to Eq.~\ref{dp}, the amount of added noise only depends on the sensitivity of the query function. Sensitivity of the query function is further defined as:

\begin{equation}
	\Delta(f) = \max \| f(\mathcal{D}_1) - f(\mathcal{D}_2) \|_1
\end{equation}
for any $\mathcal{D}_1$ and $\mathcal{D}_2$ which differ in at most one element. $\|.\|_1$ denotes the $l_1$ norm. 

The added Laplacian noise is then drawn from $Lap(\Delta(f)/\epsilon) \propto e^{-\epsilon /\Delta(f)} $ and the output result considering differential privacy constraint will be $K(f(\mathcal{D})) = f(\mathcal{D}) + Y$, where $Y\sim Lap(\Delta(f)/\epsilon)$. The mechanism $K(.)$ works best when $\Delta(f)$ is small as it introduces the least noise. The larger the sensitivity of a query, the less privacy risks can be tolerated as removing any instance from the dataset would change the output of the query more. Note that the sensitivity basically captures how a great difference (between the value of $f(.)$ on two datasets differing in a single element) must be hidden by the additive noise generated by the data publisher.

Note that recent studies show that the dependency between instances in the dataset will hurt the privacy guarantees provided by the differential privacy~\cite{kifer2011no,liu2016dependence}.

There also exists a relaxed version of $\epsilon$-differential privacy, known as $(\epsilon, \delta)$-differential privacy which was developed to deal with very unlikely outputs of $K(.)$~\cite{dwork2006calibrating,dwork2008differential}. It could be defined as:

\begin{defn}{\textbf{$(\epsilon, \delta)$-differential privacy}}. Given a query function $f(.)$, a mechanism $K(.)$ with an output range $\mathcal{R}$ satisfies $(\epsilon, \delta)$-differential privacy for all datasets $\mathcal{D}_1$ and $\mathcal{D}_2$ differing in at most one element \textit{iff}:
	
	\begin{equation}\label{dp2}
	Pr[K(f(\mathcal{D}_1))=R \in \mathcal{R}] \leq e^{\epsilon} \times Pr[K(f(\mathcal{D}_2))=R \in \mathcal{R}]
	\end{equation}
	where $\epsilon$ and $\delta$ are two model parameters related to the level of privacy guarantees and are considered to be very small numbers. 
\end{defn}

Table~\ref{diffprivacy} summarizes different works that utilize differential privacy in social media data. All these works are discussed more later.

\begin{table}[t]
	\center\small
	\caption{Differential privacy applications in user privacy in social media.}
	\label{diffprivacy}
	\begin{tabular}{|l|p{8cm}|}
		\hline
		Type of Information & Paper\\ \hline
		graph structure & \cite{sala2011sharing,proserpio2014calibrating,xiao2014differentially,wang2013preserving,liu2016dependence} \\ \hline
		recommender systems & \cite{mcsherry2009differentially,machanavajjhala2011personalized,zhu2013differential,jorgensen2014privacy,shen2014privacy,hua2015differentially,guerraoui2015d,zhu2016differential,meng2018personalized}\\ \hline
		textual data & \cite{TextAnonymization}\\ \hline
	\end{tabular}
	
\end{table}

\subsection{Related Work}
There are multiple relevant surveys related to the privacy of data and privacy preserving approaches~\cite{fungprivacy,secgraph,ji2016graph,abawajy2016privacy,sharma2012anonymisation,zheleva2012privacy,verykios2004state,agrawal2000privacy,rizvi2002maintaining,evfimievski2004privacy}. Fung et al.~\cite{fungprivacy} reviews privacy preserving data publishing methods for relational data such as $k$-anonymity, $l$-diversity, $t$-closeness and their other variations. These methods are compared in terms of privacy models, anonymization algorithms and information metrics. Zhelva et al.~\cite{zheleva2012privacy} review the concepts of privacy issues in tabular data and introduce new privacy risks in graph data. Multiple surveys focus on reviewing graph data privacy risks~\cite{secgraph,ji2016graph,abawajy2016privacy,sharma2012anonymisation}. Sharma et al.~\cite{sharma2012anonymisation} are among the first works which reviews $k$-anonymity and randomization based techniques for anonymizing graph data. Another overview by Abawajy et al.~\cite{abawajy2016privacy} presents the threat model for graph data and classified the background knowledge that is used by adversaries to breach the privacy of users. They also review and classify state-of-the-art approaches for anonymizing graph data. Ji et al.~\cite{secgraph,ji2016graph} conducted a survey on graph data anonymization, de-anonymization attacks and de-anonymizability quantification.

Another way of sanitizing data is by providing algorithms which are provably privacy-preserving and ensure no sensitive information leak from the data~\cite{zheleva2012privacy}. There is a thorough survey~\cite{verykios2004state} on privacy preserving data mining which studies different privacy preserving data mining approaches. Another work from Agrawal et al~\cite{agrawal2000privacy} proposes algorithms to perturb data values by adding random noise to them in order to preserve the privacy of users while retaining the statistical properties of the original data. Another set of works focus on developing privacy preserving association mining rules to minimize privacy loss~\cite{rizvi2002maintaining,evfimievski2004privacy}.

In this work, we go one step further and reviews all aspects of social media data which could lead to privacy leakage. Social media data is highly unstructured and noisy and inherently different from relational and tabular data. Therefore, other approaches are designed specifically to study privacy risks in the context of user-generated data in social media platforms. Different from previous works, we not only reviews state-of-the-art and recent approaches on social graph anonymization and de-anonymization, but we also survey other attribute and identity disclosure attacks which could be performed on the other aspects of user-generated social media data. In particular, we overview and summarize approaches that leverage users' activities on social media to infer their profile and location information. In addition to identity disclosure risks raised from social graphs, we survey author identification and user linkage across social media approaches that incorporate various pieces of user-generated information such as user profiles and textual posts. We introduce more risks and cover more recent works related to privacy leakage in social media which are not covered in the work of Zhelva et al.~\cite{zheleva2012privacy}. Furthermore, we include many new techniques related to the privacy of social graphs which are not included in previous surveys~\cite{secgraph,ji2016graph,abawajy2016privacy,sharma2012anonymisation}.

In summary, to the best of our knowledge, this is the first and most comprehensive work that systematically surveys, and analyzes the advances of research on privacy issues in social media.
\section{Social Graphs and Privacy}
A large amount of data generated by users in social media platforms has graph structure. Friendship and following/followee relations, mobility traces (e.g. WiFi contacts, Instant Message contacts) and spatio-temporal data (latitude, longitude, timestamps) all could be modeled as graphs. This mandates paying attention to privacy issues of graph data. We will first overview graph de-anonymization works and then survey the proposed solutions for anonymizing graph data.
\subsection{Graph De-anonymization}\label{deanony}
The work of Backstrom et al.~\cite{backstrom2007wherefore} were among the first works which studied the privacy breach problem according to the social network's graph structure. These attacks could be categorized as either a seed-based or seed-free approach according to whether pre-annotated seed users existed or not. Seed users are those whom their identity are clear for the attacker. Backstrom et al.~\cite{backstrom2007wherefore} is among the first seed-based approaches. This work introduces both active and passive attacks on anonymized social networks. In active attacks, the adversary creates $k$ new user accounts (a.k.a Sybils) and links them to the set of predefined target nodes before the anonymized graph is produced. Then it links these new accounts together to create a subgraph $H$. After publishing the anonymized graph, the attacher looks for the subgraph $H$ and then locates and re-identifies targeted nodes in the published graph. The main challenge in this approach is that the subgraph $H$ should be unique enough to be found efficiently regardless of $G$ with several million users. In passive accounts, the attacker is an internal user of the system and no new account is created. The attacker then de-anonymizes the users connected to him after the anonymized graph data is released. This attack is susceptible to Sybil defense approaches~\cite{al2017sybil} and wrongly assumes that attackers can always change the network before its release. 

Another work from Narayanan et al.~\cite{narayanan2009anonymizing} introduces an improved attack which does not need compromised accounts or Sybil users to perform the attack. This work assumes that the attacker has access to a different network whose membership has overlap with the original anonymized network. This auxiliary graph is also known as background or auxilary graph knowledge. It also assumes that the attacker has the information of a small set of users, i.e. seed users, who are present in both networks. Narayanan et al.~\cite{narayanan2009anonymizing} discuss different ways of collecting background knowledge. For example, if the attacker is a friend of a portion of the targeted users, he knows all the details about them~\cite{korolova2008link,stone2008autotagging}. Another approach is paying a set of users to reveal information about themselves and their friends~\cite{lewis2008tastes}. Crawling data via social media API or using compromised accounts as discussed in active attack are other approaches for gathering background knowledge. Social graph de-anonymization attack in social media could be then formally defined as:
\begin{defn}{\textbf{ Social Graph De-anonymization Attack}~\cite{narayanan2009anonymizing,fu2015effective}}. Given an auxiliary/ background graph $G_1 = (V_1, E_1)$ and a target anonymized graph $G_2 = (V_2, E_2)$, the goal of de-anonymization is to find identity disclosures in the form of $1-1$ mappings as many and accurately as possible. An identity disclosure indicates that the two nodes $i \in V_1$ and $j \in V_2$ actually correspond to the same user.
\end{defn}

\subsubsection{Seed-based De-anonymization}
Seed-based de-anonymziation approaches have two main steps. In the first step, a set of seed users are mapped from the anonymized graph to the background/auxiliary graph knowledge and thus are re-identified. In the second step, the mapping and de-anonymization is propagated from the seed users to the other remaining unidentified users. Similarly, the work of Narayanan et al.~\cite{narayanan2009anonymizing} starts from re-identifying seed users in an anonymized and auxiliary graph. Then, other users are re-identified by propagating mappings based on seed users pairs. Structural information such as user's degree, user's eccentricity, and edge directionality are used to heuristically measures the strength of match between users. A straightforward application of this de-anonymization attack with less heuristics is predicting links between users~\cite{narayanan2011link}.

Yartseva et al.~\cite{yartseva2013performance} propose a percolation-based de-anonymization approach which maps every pair of users in both graphs (background knowledge and anonymized graphs) that have more than $k$ neighboring mapped pairs. The only parameter of this work is $k$ which is a predefined mapping threshold and does not require a minimum number of users in the seed set. Another similar work from Korula et al.~\cite{korula2014efficient} propose a parallelizable percolation-based attack with provable guarantees. It again starts with a set of seed users who are previously mapped and then propagates the mapping to the remaining network. Two users will be mapped if they have a specific number of mapped neighbors. Their approach is robust to malicious users and fake social relationships in the network.

In another work, Nilizadeh et al.~\cite{nilizadeh2014community} propose a community based de-anonymization attack using the idea of divide-and-conquer. Community detection has been extensively studied in the literature of social network analysis~\cite{yang2013community,alvari2016identifying,yang2013overlapping} and has been used in variety of tasks such as trust prediction~\cite{gbeigi_trust} and guild membership prediction~\cite{alvari2014predicting,hajibagheri2018using}. In this work, the attacker first leverage community detection techniques to partition both graphs (i.e., anonymized and knowledge graphs) into multiple communities. It then maps communities by creating a network of communities in both graphs. Then users within mapped communities are re-identified and matched together. Mappings are then propagated to re-identify the remaining users. This attack uses similar heuristics as~\cite{narayanan2009anonymizing} to measure the mapping strength between users. 

Ji et al.~\cite{ji2015your,ji2016your} study de-anonymizability of social media graph data based on seed-based approaches under both the Erdos-Renyi and a statistical model. Similar to~\cite{ji2014structural}, they specified the structure conditions for both perfect and partial de-anonymization (i.e. partial de-anonymization can only re-identify a set of users). Chiasserini et al.~\cite{chiasserini2016social,fabiana2015anonymizing} also study the problem of user de-anonymization according to their structural information under the scale-free user relation model. This assumption is more realistic since users degree-distribution in social media follows power-law distribution, a.k.a scale-free. The results of their analysis show that the information of a large portion of users in the seed set is useless in re-identifying users in anonymized graph. This because of the large inhomogeneities in the users degree. This results suggests that given a network with $n$ users, the order of $n^{\frac{1}{2}+\epsilon}$ (for any arbitrarily small $\epsilon$) seeds are needed to successfully de-anonymize all users when seeds are uniformly distributed among the vertices. It has been also shown that as few as $n^\epsilon$ seeds are needed if the attacker has the option to select seeds according to their degree and scale-free property of social network. Chiasserini et al.~\cite{chiasserini2016social,chiasserini2018anonymizing} also propose a two-phased percolation graph matching based de-anonymization attack similar to~\cite{yartseva2013performance}. 

Bringmann et al.~\cite{bringmann2014anonymization} also propose an approach which uses $n^\epsilon$ seed nodes (for an arbitrarily small $\epsilon$) for a graph with $n$ nodes. This is an improvement over the state-of-the-art structure based de-anonymization techniques which need $\Theta (n)$ seeds~\cite{korula2014efficient}. This approach then finds a signature set for each node as the intersection of its neighbors and previously re-identified nodes. It then defines criterion that further is used to decide if two signatures originate from same nodes with high probability or not, i.e. if the similarity of two nodes signature is more than $n^c$ ($c>0$ is a constant), then the two nodes are mapped together. Local sensitivity hashing technique~\cite{indyk1998approximate} is also used to reduce the number of comparisons needed for the de-anonymization attack. Theoretical and empirical analysis of their work show that the attack is performed in quasilinear time.

Manasa et al.~\cite{peng2014two} propose another seed-based attack against anonymized social graphs which has two steps. In the first step, it identifies a seed sub-graph of users with known identities. As discussed earlier in~\cite{backstrom2007wherefore}, this sub-graph could be injected by an attacker or it could even be a small group of users which the attack is able to re-identify. In the second step, it extends the seed set based on the users' social relations and re-identifies the remaining users. For each mapping iteration, the algorithm re-examines previous mapping decisions, given new evidences regarding re-identified nodes. This attack does not have any limitation on the size of the initial seed and the number of links between seeds. Another recent work by Chiasserini et al.~\cite{chiasserini2018anonymizing} incorporates clustering for de-anonymization attacks. Their attack uses various levels of clustering and their theoretical results highlight that clustering can potentially reduce the number of seeds in percolation based de-anonymization attacks due to its wave-like propagation effect. This attack is a modified version of~\cite{yartseva2013performance} which starts from a small set of seed users and then expands seed set to the closest neighbors of the users in the seed set and repeat the re-identification procedure. In this version, two users are mapped if they have a sufficiently large number of neighbors among the mapped pairs. 

\subsubsection{Seed-free De-anonymizatoin}
The efficiency of most of seed-based approaches depends on the size of seed set. However, seed-free approaches do not have this problem since they do not need the information of users in the form of a seed set to de-anonymize other users. Recently, some powerful seed-free de-anonymization attacks have been developed for social media graph data~\cite{ji2014structural,ji2016structural,pedarsani2011privacy}. Pedarsani et al.~\cite{pedarsani2011privacy} present a Bayesian model which starts from the users with the highest degree and iteratively solves a maximum weighted bipartite graph matching problem. This algorithm iteratively updates fingerprints of all users. A bipartite graph $G=(A,B,E)$ is a graph whose vertices can be divided into two disjoint sets $A$ and $B$. The goal in the maximum bipartite graph matching problem is to find a maximum matching between two partites so that each vertex is the endpoint of exactly one of the chosen edges. 

Moreover, Ji et al.~\cite{ji2014structural,ji2016structural} propose
to use optimization based methods to minimize an error function iteratively. More specifically, in each iteration of this attack, two candidate sets of users are selected from the anonymized and background graphs. Then users in the set from the anonymized graph are mapped (de-anonymized) to users in background graph by minimizing an error function defined by the edge difference caused by a mapping scheme. In particular, Ji et al.~\cite{ji2014structural} quantify the structure-based de-anonymization under the Configuration model~\cite{CMMOdel} and drive structural conditions for perfect and partial de-anonymization. The configuration Model generates a random graph given a degree sequence by randomly assigning edges to match the given degree sequence~\cite{CMMOdel}.

Another recently developed group of techniques leverages additional sources of information besides structural network to re-identify social media users in anonymized data. This information includes user interactions (e.g., commenting, tweeting) or non-personal identifiable information which is associated with users and is shared publicly such as gender, education, country and interests~\cite{gong2018attribute}. This combination of structural and exogenous sources of information could increase the of risk user privacy. Zhang et al.~\cite{zhang2014privacy} study the privacy breach problem in anonymized heterogeneous networks. They first introduce a privacy risk measure based on the potential loss of the user and the number of users who have same value. They then propose a de-anonymization algorithm which incorporates the defined privacy risk measure. For each target user, this framework first finds a set of candidates based on entity attribute matches in the heterogeneous network and then narrows down this candidate set by comparing the neighbors (which are found via heterogeneous links) of the target user and each candidate.

Fu et al.~\cite{fu2014anonymizing,fu2015effective} propose to use structural and descriptive information. Descriptive information is defined as attribute information such as name, gender, birth year. This work first proposes a new definition of user similarity, i.e., two users are similar if their neighbors match to each other as well. However, similarity of neighbors also depends on the similarity of users. Therefore, Fu et al. model similarity as a recursive problem and solves it iteratively. Then, they reduce the de-anonymization problem to a complete weighted bipartite graph matching which is solved with Hungarian algorithm~\cite{kuhn2010hungarian}. These weights here are calculated based on the users similarities. 

In another work, the effect of user attribute information as an exogenous source of information on de-anonymizing social networks is studied~\cite{qian2016anonymizing}. In particular, this work incorporates semantic background knowledge of adversary in the de-anonymization process and models it using knowledge graphs~\cite{knowledgegraphs}.% Knowledge graph is a graph with entities as nodes and links demonstrate the relations between entities. This graph is represented in the form of RDF triples indicating links, i.e., $\langle subject, relation, object\rangle$.
 This approach simultaneously de-anonymizes and infers users attributes (we will discuss user profile attribute inference attack later in section~\ref{AttrInf}). The adversary first models both the de-anonymized dataset and the background knowledge as two knowledge graphs. Then, she makes a complete adversary weighted bipartite graph. Each weight indicates the structural and attribute similarity between corresponding nodes in the anonymized and knowledge graphs. The de-anonymization problem will be then reduced to a maximum weighted bipartite matching problem which can be furthered reduced to a minimum cost maximum flow problem.% After each of de-anonymization iterations, adversary's knowledge graph will be updated with the information of mapped node in anonymized graph to make it richer and more accurate.
 Attacker prior semantic knowledge could be obtained via different ways such as common sense, statistical information, personal information and network structural information.

Ji et al.~\cite{ji2017sag} also study the same problem and show theoretically and empirically that using attribute information alongside structural information could result in a great privacy loss even in an anonymized dataset in comparison to the cases where the data only consists of structural information. They further propose the De-SAG de-anonymization framework  which incorporates both attribute and structural information. It first augments both types of information into a structure-attribute graph. De-SAG has two variants, i.e. user-based and set-based. In user-based De-SAG, the proposed de-anonymization approach first selects the $k$ most similar candidates to the target user from background/auxiliary knowledge graph based on similarity of their attributes. $k$ is a pre-defined parameter which controls the efficiency-accuracy trade-off in de-anonymization. Next, the target user will be mapped to one of the selected candidates based on their structural similarity. In set-based De-SAG, for each iteration, two sets of users are selected from anonymized graph and knowledge graph, respectively. Then, the de-anonymization problem reduces to a Maximum Weighted Bipartite graph Matching problem and users in these two sets are mapped to each other using Hungarian algorithm~\cite{kuhn2010hungarian}. These steps are repeated till no users remains unidentified. Note that the similarity of users are again calculated according to their attribute and structural information. The results of De-SAG show that users are re-identified 10 times more accurately than state-of-the-art structure based de-anonymization techniques~\cite{ji2014structural,korula2014efficient}.

In another work by Lee et al.~\cite{lee2017blind}, a blind de-anonymization technique is proposed in which the adversary does not need to have any background information. Inspired by the idea of $dK$-series for chatacterizing structural characteristics of a graph, they propose $nK$-series to describe structural features of each user by exploiting his multi-hop neighbors information. In particular, $nKi$ captures the degree histogram of the user's $i$-hop neighbors. Then, a structure score is calculated for each user (in both the anonymized graph and the background knowledge graph) based on his diversity score (calculated according to $nK$-series scores) and his relationships with all other non-reidentified users in the network. It then uses this information to re-identify all users in the anonymized social graph by leveraging pseudo relevance feedback support vector machines. Backes et al.~\cite{backes2017walk2friends} develop an attack which infers relations between users (i.e., edges between nodes in graph data) based on the users mobility profiles without using any additional information about existing relations between users. Their approach first constructs mobility profile for each user and then infer the social links between users based on the similarity of their mobility profile. The intuition behind this attack is that friends have more similar profiles in comparison to strangers. To infer users' mobility profiles, it first obtains random walk traces from the user-location bipartite graph and then uses skip-gram~\cite{mikolov2013distributed} to obtain features in a continuous vector space.

Beigi et al.~\cite{beigi2018securing} also introduce a new adversarial attack for social media data that does not need to have any background information before initiating the attack. This attack is designed for heterogeneous social media data which consists of different aspects (e.g., textual, structural, location, etc.) and shows that anonymizing all aspects of data is not sufficient when it is done without considering the hidden relationships between different data aspects. This attack first extracts the most revealing information for each user in the anonymized dataset, and then finds a set of candidate users based on the extracted information. Each user is finally mapped to the most probable candidate user. Sharad et al.~\cite{sharad2014automated} propose to formulate the problem of graph de-anonymizatin in social networks as a learning task. They use 1-hop and 2-hop neighborhood degree distributions to represent users in a graph. The intuition behind this selection is that two nodes refer to the same user if their neighborhoods also matches to each other. For each pair of users selected at random from background knowledge and anonymized graphs, their proposed approach first extracts structural features from user's 1-hop and 2-hop neighborhood. These features help the machine learning model to learn the degree deviation for identical and non-identical user pairs. It then trains a classifier on these features and predicts whether two pair of nodes are the same nodes in different ego-nets or not. They use decision tree and random forest as classifiers. In another work, Sharad et al.~\cite{sharad2016change} go even further and propose a new generation of de-anonymization attacks which is heuristic free, seedless and is considered as a learning problem. They use the same set of structural features as proposed in~\cite{sharad2014automated} and then de-anonymize the sanitized graph by re-identifying users with high degree first and then use them to attack low-degree nodes. They divide nodes into three categories based on their degrees and produce an initial set of mappings of all nodes with highest degrees. The mappings are used to filter out some of the nodes. Mappings are then frozen and propagated to the remaining nodes to discover new set of mappings.
\subsubsection{Theoretical Analysis and De-anonymization}
Another set of works studies de-anonymization attacks from the theoretical perspective of view. For example, Liu et al.~\cite{liu2016dependence} theoretically study the vulnerability of differential privacy mechanisms against de-anonymization attacks. Differential privacy provides protection against even the strongest attacks in which the adversary knows the entire dataset except one entry. However, differential privacy assumes the independence between dataset entities which is not correct in most real-world applications. This work introduces a new attack in which the probabilistic dependence between dataset entries are calculated and then leveraged to infer users' sensitive information from differentially private queries. The attack is also tested on graph data in which users' degree distributions is published differentially privately.

Lee et al.~\cite{lee2017quantify} also study the theoretical quantification for relating the anonymized graph data vulnerability against de-anonymization attacks. In particular, they study the relation between application specific anonymized data utility (i.e., quality of data) and capability of de-anonymization attacks. They define local neighborhood utility and global structure utility. They theoretically show that under certain conditions for each of defined utilities, the probability of successful de-anonymization approaches one with the increase of number of users in data. Their foundations could be used to evaluate the effectiveness of the de-anonymization/anonymization techniques.

Recent research by Fu et al.~\cite{fu2017anonymization2} studies the conditions under which the adversary can perfectly de-anonymize user identities in social graphs. In particular, they theoretically study the cost of quantifying the quality of the mappings. Community structures are also parameterized and leveraged as side information for de-anonymization. They study two different cases in which the community information is available for both background knowledge and anonymized graphs or only for one of them. They showed that perfectly de-anonymizing graph data with community information in polynomial time is NP-hard. They further propose two algorithms with approximation guarantees and lower time complexity by relaxing the original optimization problem. The main drawback of this study is the assumption of disjoint communities which fails to reflect the real-world situations. Wu et al.~\cite{wu2018social} extend Fu et al.'s study by considering overlapping communities. In contrast to Fu et al.'s work~\cite{fu2017anonymization2} which uses Maximum a Posteriori estimation to find the correct mappings, Wu et al. introduces a new cost function Minimum Mean Square Error which minimizes the expected number of mismatched users by incorporating all possible true mappings.

There are surveys by Ji et al.~\cite{ji2016graph,secgraph}, Lee et al.~\cite{lee2017quantify} and Abawaji et al.~\cite{abawajy2016privacy} on quantification and analysis of graph de-anomyziation techniques which studies a portion of covered works here in terms of scalability, robustness and practicability. Interested readers can refer to these surveys for further readings~\cite{ji2016graph,secgraph,lee2017quantify,abawajy2016privacy}.

\subsection{Graph Anonymization}
Another research direction in protecting privacy of users in graph data is studying graph anonymization techniques. Existing anonymization approaches use different techniques and mechanisms and could be categorized mainly into five categories: $k$-anonymity based approaches~\cite{liu2008towards,zhou2008preserving,zou2009k,yuan2010personalized,cheng2010k}, Edge manipulation techniques~\cite{ying2009graph}, cluster based techniques~\cite{hay2008resisting,bhagat2009class,khairnar2014anonymization,liu2016linkmirage,thompson2009union,mittal2012preserving}, random walk based techniques~\cite{liu2016smartwalk,mittal2012preserving}, and differential privacy based techniques~\cite{sala2011sharing,proserpio2014calibrating,xiao2014differentially,wang2013preserving}. We discuss each of these categories later.

\subsubsection{K-anonymity Based Approaches}
The aim of $k$-anonymity methods is to anonymize each user/node in the graph so that it is indistinguishable from at least $k-1$ other users~\cite{sweeney2002k}. Liu et al.~\cite{liu2008towards} proposed an anonymization framework for $k$-degree anonymization in which for each user, there are at least $k$ other users with the same degree. The goal of this approach to add/delete the minimum number of edges to preserve $k$-degree anonymity. This algorithm has two steps. In the first step, given the degree sequence of the original graph, a $k$-degree anonymized version of the degree sequence is constructed and then in the second step, the anonymized graph is built based on the anonymized degree sequence. In another work~\cite{zhou2008preserving}, Zhou et al. aim to achieve $k$-neighborhood anonymity. They consider the assumption that the adversary knows the subgraph constructed by the immediate neighbors of a target node. In the first step of the anonymization, 1-hop neighborhoods of all users are extracted and encoded in a way that isomorphic neighborhoods could be easily identified. In the second step, users with similar/isomorphic neighborhoods are grouped together until size of each group is at least $k$. Then, each group is anonymized satisfying $k$-neighborhood anonymity as each neighborhood has at least $k-1$ isomorphic neighborhoods in the same group. Eventually, this approach anonymizes the graph against neighborhood attacks.

Zou et al.~\cite{zou2009k} propose a $k$-automorphism based framework which protects the graph against multiple attacks including the neighborhood attack~\cite{zhou2008preserving}, degree based attack~\cite{liu2008towards}, hub-fingerprint attack~\cite{hay2008resisting} and subgraph attack~\cite{hay2008resisting}. A graph is $k$-authomorphic if there exists $k-1$ automorphic functions in the graph and for each user in the graph, the attacker cannot distinguish it from her $k-1$ symmetric vertices. The proposed approach first partitions the graph into $n$ blocks and then clusters blocks into $m$ groups (graph partitioning step). In the second step, alignments of blocks are obtained and original blocks are replaced with alignment blocks (block alignment step). In the last step, edge copy is performed to get the anonymized graph. Edge copy adds $k-1$ edges between $k-1$ pairs $(F_a(u), F_a(v)))$ $(a = 1,2,...k-1)$ where $F_a(.)$ is the automorphic function and $u$ and $v$ are users in the social graph. Authors also propose the use of generalized vertex ID's for handling dynamic data releases. Another similar work by Cheng et a.~\cite{cheng2010k} proposes a $k$-isomorphism anonymization approach. A graph is $k$-isomorphic if it is consisted of $k$ disjoint subgraphs and all subgraphs pairs are isomorphic. In the first step, the graph is partitioned into $k$ subgraphs with the same number of vertices. Then, edges are added or deleted so that these subgraphs are isomorphic. This approach protects the published graph against neighborhood attacks~\cite{zhou2008preserving}.

Yuan et al.~\cite{yuan2010personalized} incorporate semantic and graph information together to achieve personalized privacy anonymization. In particular they consider three different levels for attacker's knowledge regarding the target user, 1) only attribute information, 2) both attribute and degree information, and 3) combination of attribute, node degree and neighborhood's information. They accordingly propose three levels of protection to achieve $k$-anonymity. For level 1 protection, their approach considers label generalization. For the level 2 anonymization, it uses node/edge adding approach as well. For the level 3 protection, it uses edge label generalization.

\subsubsection{Edge Manipulation Based Approaches}
Edge manipulation and randomization algorithms for social graphs usually utilizes edge-based randomization strategies to anonymize data such as random edge adding/deleting and random edge switching~\cite{ying2009graph}. Ying et al.~\cite{ying2009graph} propose spectrum preserved edge editing which either adds $k$ random edges to the graph and remove another $k$ edges randomly or alternatively switches $k$ edges. In the switching technique, two random edges $(i_1, j_1)$ and $(i_2, j_2)$ are selected from the original graph edge set $E$ such that $\{(i_1,j_2)  \notin E \wedge  (i_2,j_1) \notin E \}$. Then edges $(i_1,j_1)$ and $(i_2,j_2)$ are removed and new edges $(i_1,j_2)$ and $(i_2,j_1)$ are added instead.  Backes et al.~\cite{backes2017walk2friends} also propose a randomization based approach to preserve the privacy of social links between users in graph data and counteract link inference attacks. In this specific type of attack, the adversary exploits users mobility traces to infer social links between users with the intuition that friends have more similar mobility profiles in comparison to the mobility profiles of two strangers~\cite{backes2017walk2friends}. They utilize three privacy preserving techniques: hiding, replacement, and generalization of user mobility information. Results show that data publishers need to hide 80\% of the location points or replace 50\% of them to prevent leakage of information of users social links.

\subsubsection{Clustering Based Techniques}
Clustering based approaches group users and edges and only reveal the density and size of the cluster so that individual attributes are protected. Hay et al.~\cite{hay2008resisting} propose an aggregation based method for graph data anonymization which is robust against three types of attacks: neighborhood, subgraph, and hub fingerprint. Hay et al.'s approach models the aggregate network structure by partitioning original graph and describing it at the level of partitions. Partitions are considered as nodes and edges between them makes the edges in the generalized graph. This generalized graph can be further used to randomly sample a graph from that can be published as the anonymized data.

Another cluster based work~\cite{bhagat2009class} proposes two approaches, label list and partitioning, which consider user attributes (i.e., labels) in addition to structural information. In the label list approach, a list of labels are allocated to each user which also includes her true label. This approach first clusters nodes into $m$ classes and then a set of symmetric lists is built deterministically for each class from the set of nodes in the corresponding class. In the partitioning approach, nodes are divided into classes and instead of releasing full edge information, only the number of edges between and within each class is released. This is similar to the generalization approach of Hay et al.~\cite{hay2008resisting}. Bhagat et al. also use a set of safety conditions to ensure that the released data does not leak information. The proposed partitioning approach is more robust than the label list technique when facing the attacks with richer background knowledge. However, the partitioning approach has lower utility than the label list as less information is revealed about the graph structure.

Thompson et al.'s approach~\cite{thompson2009union} protects the graph information against $i$-hop degree-based attack. They present two clustering algorithms, bounded $t$-means clustering and union-split clustering. These approaches group users with similar social roles into clusters with a minimum size constraint. Then they utilize the proposed inter-cluster matching anonymization method, which anonymizes the social graph by removing/adding edges according to the users' inter-cluster connectivity. The number of nodes and edges between and within clusters are then released similar to Hay et al.'s approach~\cite{hay2008resisting}. Another work~\cite{khairnar2014anonymization} proposes an incremental approach to partition graph data and release clusters centroids information as the anonymized data. Mittal et al.~\cite{liu2016linkmirage} also propose another clustering based aonymization technique which considers evolutionary dynamics of social graphs such as node/edge addition/deletion and consistently anonymizes the graph. It first dynamically clusters nodes and then perturbed the intra-cluster and inter-cluster links for changed clusters in a way that structural properties of social media graph is preserved. They leverage static perturbation method of~\cite{mittal2012preserving} to modify intra-cluster links and randomly connect marginal nodes to create fake inter-cluster links according to their degree. The obfuscated graph has higher indistinguishability which is defined from an information theoretic perspective. 

\subsubsection{Random Walk Based Approaches}
Another group of works utilizes random walk idea to anonymize graph data. The idea of random walk has been previously used in many security applications such as Sybil defense~\cite{al2017sybil}. Recent works also use this idea for anonymzing social graphs. The work of Mittal et al.~\cite{mittal2012preserving} introduces a random-walk based edge perturbation algorithm. According to this approach, for each node $u$, a random walk with the length $t$ will be performed starting from one of the $u$'s contacts, $v$ and an edge $(u,z)$ between destination node, $z$ and $u$ will be added with an assigned probability and the edge $(u,v)$ will be removed accordingly. This probability will decrease as more random walks are performed from $u$'s contacts. Later, Liu et al.~\cite{liu2016smartwalk} improve this approach such that instead of having a fixed length random walk with length $t$, they utilize a smart adaptive random which its length is learned based on the local structure characteristics. This method first predicts the local mixing timing for each node which is the minimum random walk length for a starting node to be within a given distance to stationary (distance) node. This mixing time is predicted based on the local structure and limited global knowledge of the graph and is further used to adjust the length of random walk for social graph anonymziation.

\subsubsection{Differential Privacy Based Approaches}
Differential privacy~\cite{dwork2008differential} was first proposed for providing a strong privacy guarantee for statistical database query. Recently many works extend differential privacy to the social graph data. Sala et al.~\cite{sala2011sharing} first use $dK$-series to capture sufficient graph structure at multiple granularities. $dK$-series is the degree distributions of connected components of size $K$ within a target graph~\cite{dimitropoulos2009graph,mahadevan2006systematic}. Then, they partition the statistical representation of the graph captured by $dK$-series into clusters and then use $\epsilon$-differential privacy mechanism to add noise to the representation in each cluster. Proserpio et al.~\cite{proserpio2014calibrating} propose another differentially private based approach which scales down the magnitude of added noise by reducing the contributions of challenging records.

In another work, Wang et al.~\cite{wang2013preserving} use $dK$-graph generation models to generate sanitized graphs. In particular, their approach first extracts various information form the original social graph such as degree correlations and then enforce differential privacy on the learned information and finally used perturbed pieces of information to generate an anonymized graph with $dK$-graph models. Different from the approach in Sala et al.~\cite{sala2011sharing}, in the specific case of $d=2$, noise is generated based on the smooth sensitivity rather than global sensitivity. The reason behind this specification is to reduce the magnitude of the added noise. Smooth sensitivity is a smooth upper bound on the local sensitivity when deciding the noise magnitude~\cite{nissim2007smooth}. Another work~\cite{xiao2014differentially} proposes an anonymization approach which satisfies edge $\epsilon$-differential privacy to hide each user's connections to other users. They propose to learn how to transform edges to connection probabilities via statistical Hierarchal Random Graphs (HRG) under differential privacy. In particular, their approach infers the HRG by learning the entire HRG model space and sampling an HRG by a Markov Chain Monte Carlo method and generating the sanitized graph according to the sampled HRG while satisfying differential privacy. Their results show that using edge probabilities can result in significant noise scale reduction in comparison to the case where the edges are used directly.

In another work from Liu et al.~\cite{liu2016dependence}, it has been shown that differential privacy is not robust to the de-anonymization attacks if there is dependence among dataset entries. Liu et al.~\cite{liu2016dependence} also propose a stronger privacy notion, dependent differential privacy in which it incorporates the probabilistic dependence between the tuples in a statistical database. They then propose an effective perturbation framework which provides privacy guarantees. Their result show that more noise should be added when there is dependency between tuples. The added noise is also dependent on the sensitivity of two tuples as well as the dependence relationship between them. They evaluate their proposed framework on graph data to sanitize the degree distribution of the given graph.

Ji et al.~\cite{ji2016graph,secgraph} and Abajaway et al.~\cite{abawajy2016privacy} study the defense and attacking performance of a portion of existing social graph anonymization and de-anonymization techniques. Ji et al.~\cite{ji2016graph,secgraph} have also performed a thorough theoretical and empirical analysis on a portion of existing related papers. Results demonstrate that anonymized social graphs are vulnerable to de-anonymization attacks.

\section{Authors in Social Media and Privacy}%Author Identification
People have the right to have anonymous free speech over different topics such as Politics~\cite{narayanan2012feasibility}. However, an author's identity can be unmasked by adversaries through providing her real name or IP address to a service provider. However, authors can use tools such as Tor to protect their identity at the network level~\cite{dingledine2004tor}. Manually generated content will always reflect some characteristics of the person who authored it. For example, some anonymous online author is prone to several specific spelling errors or has other recognizable idiosyncrasies~\cite{narayanan2012feasibility}. These characteristics could be enough to figure out whether authors of two pieces of content are same or not. Therefore, with material authored by the true identity of the author, the adversary can discover the identity of a content posted online by the same author anonymously. 

Identifying the author of a text according to her writing style, a.k.a stylometry, has been studied a long time ago~\cite{mendenhall1887characteristic,mosteller1964inference,stamatatos2009survey}. With the adverse of machine learning techniques, researches start to extract textual features and discriminate between 100--300 authors~\cite{abbasi2008writeprints}. The application of author identification includes identifying authors of terroristic threats and harassing messages~\cite{chaski2005s}, detecting fraud~\cite{afroz2012detecting}, and extracting author's demographic information~\cite{koppel2009computational}. 

Privacy implications of stylometry have been studied recently. For example, Rao et al.~\cite{rao2000can} investigate whether people who are posting under different pseudonyms to USENET newsgroup can be linked based on their writing style. They use a dataset of 117 people having 185 different pseudonyms and exploit function words and Principal Component Analysis (PCA) to perform matching between newsgroups posting and email domains. Another work from Koppel et al.~\cite{koppel2006authorship,koppel2011authorship}, studies author identification at the scale of over 10,000 blog authors. They use 4-grams of characters which is a context specific feature. The problem with this work is that it is not clear whether their approach is solving author recognition or context recognition. In another work, Koppel et al.~\cite{koppel2009computational} use both content-based and stylistic features to identify 10,000 authors in the blog corpus dataset. There are also several works on identifying authors of academic papers under blind review based on the citations of the paper~\cite{bradley2008author,hill2003myth} or other sources from unblind texts of potential authors~\cite{nanavati2011herbert}.

Narayanan et al.~\cite{narayanan2012feasibility} propose another author identification attack which exploits 1,188 real-valued features from each post, such as frequency of characters, capitalization of words (e.g., lowercase and uppercase words), syntactic structure (extracted by Stanford Parser~\cite{klein2003accurate}, e.g. noun phrases containing a personal pronoun, noun phrases containing a singular proper noun), distribution of word length, etc. These features capture the writing style of the author regardless of the topic at hand. This approach works for re-identifying large number of authors and has also been tested over a cross-context setting (i.e., two different blogs). However this approach will not work when authors anonymize their writing style.

Almishari et al.~\cite{almishari2012exploring} proposed a new linkage attack which investigates the linkability of prolific reviews that users post on social media platforms. More specifically, given a subset of information on reviews made by an anonymous user, this approach seeks to map it to a known identified record. This approach first extracts four types of tokens, unigrams, digrams, ratings and category of reviewed entity. Then, it uses Naive bayes and Kullback-Leibler divergence models to re-identify the anonymized information. This approach could be even used for identity disclosure attack across multiple platforms using people's posts and reviews.

Bowers et al.~\cite{Bower15} propose an anonymization approach which uses iterative language translation (ILT) to conceal one's writing style. This approach first translates English text into another foreign language (e.g., Spanish, Chinese, etc.) and then turns it back to English again for three iterations. Another work from Nathan et al.~\cite{mack2015best} evaluates Bowers's work by introducing a feature selection approach, namely Generative and Evolutionary Feature Selection (GEFES) over the set of predefined features which mask out non-salient previously extracted features. Both~\cite{Bower15} and~\cite{mack2015best} are tested  over a set of blog posts by users and the results show the efficiency of ILT-based anonymization. A recent work is also proposed by Zhang et al.~\cite{TextAnonymization} which anonymizes users' textual information before publishing user-generated data. This approach first introduces a verified version of differential privacy specified for textual data, namely, $\epsilon$-Text Indistinguishability to overcome the curse of dimensionality problem when original differential privacy is deployed on high-dimensional textual data. It then proposes a framework which perturbs user-keyword matrix by adding Laplacian noise to satisfy $\epsilon$-Text Indistinguishability. Results confirms both the utility and privacy of the data.
\section{Social Media Profile Attributes and Privacy}~\label{AttrInf}%Inference Attack}~\label{AttrInf}
A user's profile includes her self-disclosed demographic attributes such as age, gender, majors, cities she loved, etc. To address the privacy of users, social networks usually offer the option for users to limit the access to their attributes, i.e. they are only visible to friends or friends of friends. A user could also create a profile without explicitly disclosing any attribute information. A social network thus is a mixture of both private and public user information. However, there exists one privacy attack which focuses on inferring users' attributes. This attack is known as privacy inference attack and it leverages publicly available information of users in social networks to infer missing or incomplete attribute information~\cite{gong2016you}.

The attacker could be any party who is interested in this information such as social network service providers, cyber criminals, data brokers, advertisers. Data brokers benefit from selling individuals' information to other parties such as banks, advertisers, and insurance companies\footnote{\url{https://bit.ly/1AwePQE}}. Social network providers and advertisers leverage users' attribute information to provide more targeted services and advertisements. Cyber criminals exploit attribute information to perform targeted social engineering, spear phishing attacks\footnote{\url{http://www.microsoft.
		com/protect/yourself/phishing/spear.mspx}} and attacking personal information based backup authentication~\cite{gupta2013your}. This attribute information could be also used for linking users across multiple sites~\cite{goga2013exploiting,shu2017user} and records (e.g., vote registration records)~\cite{sweeney2002k,minkus2015city}. 
%This attack could be formally defined as:
%\begin{defn}{\textbf{ (Attribute Inference Attack)}~\cite{gong2016you}}. Given $T = (G, \mathbf{A}, \mathbf{B})$, which is a snapshot of social network $G=(V,E)$ with a user behavior and an attribute information, the attribute inference attack is to infer the attributes $a_v$ for all $v \in V_t$ where $V_t$ is a list of targeted user. For each $v \in V_t$, we have the information of her social friends and behavior.
%\end{defn}
%Binary representation is used for both user behaviors and attribute values and $m_b$ and $m_a$ denotes the number of distinct behaviors and attribute values, respectively. Element $i$-th in behavior vector $b_v$ for user $v$ indicates that whether $v$ has performed behavior $i$-th or not. The same setting holds for attribute value vector $a_v$. 
Existing attacks could be categorized into two groups, friend-based~\cite{he2006inferring,lindamood2009inferring,thomas2010unfriendly,mislove2010you,gong2014joint,zheleva2009join,dey2012estimating,backstrom2010find,mcgee2011geographic,jurgens2013s,rout2013s,compton2014geotagging,kong2014spot,jurgens2015geolocation} and behavior-based~\cite{weinsberg2012blurme,kosinski2013private,bhagat2014recommending,chaabane2012you,luo2014you}. We will discuss each of these  categories next.

\subsection{Friend-based Profile Attribute Inference}
Friend-based approaches use the homophily theory~\cite{mcpherson2001birds} which states that two friends are more probable to share similar attributes rather than two strangers. Following this intuition, if most of a user's friends study in Arizona State University, she is more likely studying in the same university. He et al.~\cite{he2006inferring}, first constructs a Bayesian network from a user's social neighbors and then uses it to model the causal relations among people in the network and thus obtains the probability that the user has a specific attribute. The main challenge in this approach is its scalability as Bayesian inference is not scalable to the millions of users in social networks. Another work by Lindamood et al.~\cite{lindamood2009inferring} uses Naive Bayes classification algorithm to infer a user's attributes by exploiting features from her node trait (i.e., other available attributes information) and link structures (i.e. friends). However, this approach is not usable for a user who does not share any attributes. In the other work~\cite{thomas2010unfriendly}, authors propose an approach which leverages friends' activities and information to infer a user's attributes. These features from friends and wall posts are then exploited into a multi-label classifier. The authors then propose a multi-party privacy approach which defends against attribute inference attacks. This approach enforces mutual privacy requirements for all users to prevent disclosure of users attributes and sensitive information.

Zhelva et al.~\cite{zheleva2009join} study how users sensitive attribute information could be leaked through their social relations and group memberships.
This friend-based attribute inference attack exploits social links and group information to infer sensitive attributes for each user. Authors propose various algorithms in which it was found LINK was the best among those that only use link information. This method models each user $u$ as a binary vector whose length is the size of the network (i.e., number of users in the network) and the value of each element $v$ is one if $u$ is connected to $v$. Then, different classifiers are trained over the users with a public profile and then attributes for users with private profiles could be inferred. The GROUP algorithm was the best among the methods which incorporates group information. This method first selects the groups that are relevant to the attribute inference problem using either feature selection approach (i.e., entropy ) or manually. Next, relevant groups are considered as features for each node and a classifier model is trained. In the last step, the attributes for targeted users are predicted using the classification model. Mislove et al. introduces a similar approach which leverages users' social links and communities information~\cite{mislove2010you}. Their approach takes some seed users with known attributes as the input and then finds the local communities around this seed set using available link information. Then it uses the fact that users in the same community share similar attributes. This approach then infers remaining users' attributes based on the communities they are a member of. The limitation is that this approach is not able to infer attributes for users who are not assigned to any local communities. 

Avello et al.~\cite{gayo2011all} propose a semi-supervised profiling approach named McC-Splat. They consider the attribute inference problem as a multiclass classifier. It then learns the attributes' weights according to the user's friends' attributes. Weights here indicate the users' likelihood in belonging to a given attribute value class. Finally, McC-Splat assigns the class with the highest percentile to the target user. The percentile is calculated according to the labeled individuals information. In the other work from Dey et al.~\cite{dey2012estimating}, the authors focus on predicting facebook users' ages considering their friendship network information. Although a user's friends list is not fully available for all users, this work uses reverse lookup approach to obtain a partial friend list for each user. Then, they designed an iterative algorithm which estimates users' ages based on friends' ages, friends of friends' ages and so on. They also incorporated other public information in each user's profile such as their high school graduation year to estimate their birth year. Another work~\cite{humbert2013nowhere}, seeks to find a targeted user based on her social network connections and the similarity of attributes between friends. It starts from a source user and continue crawling until it reaches the target user. The navigations are based on the set of target user's known attributes, friendship links between users and their attributes as well. Similarly, Labitzke et al.~\cite{labitzke2013online} also study whether profile information of Facebook users could be still leaked through their social relations.

Another set of works in this category focuses on predicting both network structure (i.e. links) and inferring missing users attribute information~\cite{yin2010linkrec,yin2010unified,gong2014joint}. The reason for simultaneously solving these two problems is that users with similar attributes tend to link to one another and individuals who are friends are likely to adopt similar attributes. The work of Yin et al.~\cite{yin2010linkrec,yin2010unified}, first creates a social-attribute network graph from an original social graph and user-attributes information, i.e. nodes in the graph are either users or attributes. Edges show the friendship between a pair of users or the relation between a user and attribute. Then, authors use random walk with restart algorithm~\cite{tong2006fast} to calculate link relevance and attribute relevance with regard to a given user. Similarly, Gong et al. transform the attribute inference attack problem to a link prediction problem in the social-attribute network graph. They generalized several supervised and unsupervised link prediction algorithms to predict the links between user-user and user-attributes.

\subsection{Behavior-based Profile Attribute Inference}
Unlike friend-based approaches, behavior-based inference attacks infer a user's attributes based on the publicly available information regarding her behaviors and public attributes of other users similar to her. For example, if a user is more engaged in liking and sharing posts that are mainly posted and liked by other female users, this user's gender is female with high probability. Weinsberg et al. ~\cite{weinsberg2012blurme} propose an approach which infers users’ attributes (i.e. gender) according to their behavior toward movies. In particular, each user is modeled with a vector with the size being the number of items. A non-zero value for each vector element demonstrates that the user has rated the item, and zero value means that user has not rated the item. Then, they use different classifiers such as logistic regression, SVM, and Naïve Bayes to infer users’ ages and their results revealed that logistic regression performed the best result. Accordingly, the authors propose a gender obfuscation method which adds movies and corresponding ratings to a given user’s profile such that it will be hard to infer the gender of the user while minimally impacting the quality of recommendations the user received. They use three different approaches for movie selection: random, sampled and greedy strategy. The sampled strategy picks a movie based on ratings distribution associated with the movies of the opposite gender. The greedy approach also selects a movie with the highest score in the list of movies for opposite gender. Ratings are also added for each movie based on either the average movie rating or the rating predicted using recommendation approaches such as matrix factorization. The greedy movie selection approach with predicted rating has the best results regarding user profile obfuscation. Kosinski et al.~\cite{kosinski2013private} follow a similar approach to~\cite{weinsberg2012blurme} and construct a feature vector for each user based on each users Facebook likes. Authors then use logistic regression classifier to train classifiers and infer various attributes for each user.

Another work from Bhagat et al.~\cite{bhagat2014recommending} proposes an active learning based attack which infers users’ attributes via interactive questions. In particular, their approach involves finding a set of movies and asking users to rate them. Each selection maximizes the confidence of the attacker in inferring users attributes. The work of~\cite{chaabane2012you} seeks to infer users attributes based on the different types of musics they like. This approach first extracts a user's interests and finds semantic similarity among them. It uses an ontologized version of Wikipeda related to each music and exploits topic modeling techniques (i.e. Latent Dirichlet Allocation, LDA~\cite{blei2003latent}) and learns semantic interest topics for each user. Then, a user is predicted to have similar attributes as those who like similar types of musics as the user. In another work from Luo et al.~\cite{luo2014you}, authors infer household structures of Internet Protocol Television (IPTV) based on the users' watching behavior (e.g., dynamics of watching time and TV programs). Their approach first extracts related features from IPTV log-data including TV programs topics and viewing behavior using LDA and low-rank model, respectively. Then, it combines graph-based semi-supervised learning with non-parametric regression and uses it to learn a classifier based on the extracted features for inferring the structure of household. A recent work published by Li et al.~\cite{li2017inferring} uses convolutionam neural network (CNN) to infer multi-valued attributes for a target user according to his ego network. A user's ego network is a subset of the original social network based on the user's friends and the social relations among them. CNN can capture the latent relationship between users' attributes and social links.

\subsection{Friend-based and Behavior-based Profile Attribute Inference}
Another category of approaches exploit both social link and user behavior information for inferring users attributes. Gong et al.~\cite{gong2016you,gong2018attribute} first make a social-behavior-attribute network (SBA) in which social structures, user behaviors and user attributes are integrated in a unified framework. Nodes of this graph are either users, behaviors or attributes and edges represents the relationship between these attributes. Then, they infer a target user's attributes through a vote distribution attack (VIAL) model. VIAL performs a customized random walk from a target user to all other users in the augmented SBA network and assigns probabilities to the users such that a user receives higher probability if it is structurally more similar to the target node in SBA network. The stationary probabilities of attribute nodes are then used to infer attributes of the target user, i.e., the attribute with maximum probability is assigned to the target user. Unlike most of the existing approaches which only use the information of users who have an attribute, a recent work from Ji et al.~\cite{jia2017attriinfer} incorporate information from users who do not have the attribute in the training process as well, i.e. negative training samples. This work associates a binary random variable with each user characterizing whether a user has an attribute or not. Then it learns the prior probability of each user having a specified attribute by incorporating the user's behavior information. Next, it models the joint probability of users as a pairwise Markov Random Field according to their social relationships and uses this model to infer posterior probability of attributes for each target user. Posterior probabilities are calculated using an optimized version of Loopy Belief Propagation.

\subsection{Exploiting Other Sources of Information for Profile Attribute Inference}
These approaches leverage sources of information other than social structures and behaviors, such as writing style~\cite{otterbacher2010inferring}, posted tweets~\cite{al2012homophily}, liked pages~\cite{gupta2013your}, purchasing behavior~\cite{wang2016your} and checked-in locations~\cite{zhong2015you}. A recent research combined identity and attribute disclosure across multiple social network platforms~\cite{andreou2017identity}. It defines the concept of $(\theta,k)$-matching anonymity as a measure of identity disclosure risk. Given a user and her identity in a source social network, a matching anonymity set is defined as the set of identities in the target social network with a matching probability of more than $\theta$. The user is $(\theta,k)$ anonymous if the size of the matching set is $k$. Another work by Backes et al.~\cite{backes2016profile} introduces a relative linkability measure that ranks identities within a social media site. In particular, it incorporates the idea of $k$-anonymity to define $(k,d)$-anonimity for each user $u$ in social media which captures the largest $k$ subset of identities (including $u$) who are within a similarity (or dissimilarity) threshold $d$ from $u$ considering their attributes. %Zhong et al.~\cite{zhong2015you} exploit semantics of user location check-ins from three points of view, spatiality, temporarily, and location knowledge. This information could be mined from online customer review sites and social networks. Then, Zhong et al.~\cite{zhong2015you} exploit these features in a tensor factorization model to extract low-rank representations of users' check-in preferences. This approach then trains a classifier over the extracted features to infer attributes such as age, blood type, marital status, educational background and sexual orientation.
A recent work from Liu et al.~\cite{liu2016dependence} also studies the vulnerability of differential privacy mechanism against the inference attack problem. As stated earlier, differential privacy provides protection against the adversary who knows the entire dataset except one entry. However, differential privacy considers the independence between dataset entities. Liu et al. introduce a new inference attack in which the probabilistic dependence between dataset entries are calculated and then leveraged to infer a user's location information from differentially private queries.

Different from all the works focusing on profile attribute inference, a recent work from~\cite{alufaisan2017hacking} brings evasion and poisoning attacks into this problem. As mentioned earlier, attribute inference could be interpreted as a classification problem (each attribute value is considered as a class) and leveraged information for this task could be also called as features. This work introduces five variants of evasion and poisoning attacks to interfere with the results of the profile attribute inference. It then uses Facebook likes data to show the effectiveness of the aforementioned attacks in inferring a user's sexual orientation and political view. Introduced attacks are as follows:
\begin{itemize}[leftmargin=*]
	\item \textbf{Good/Bad Feature Attack (Evasion)}: The adversary has knowledge of useful(good)/useless(bad) features for the inference task. She then adds good features from one attribute to another while removing bad features from each class for all users to introduce false signals for the predictor.
	\item \textbf{Mimicry Attack (Evasion)}: The goal is to make one class looks like the other class. Adversary first samples a subset of users from one class and then finds the set of the most similar users in the other class. Good (bad) features are added (removed) for users in the found subsets.
	\item \textbf{Class Altering Attack (Poisoning)}: In this attack, the adversary randomly chooses users from one class and then flips their class label. The number of contradictory profiles will then increase, which results in higher misclassification rate.
	\item \textbf{Feature Altering Attack (Poisoning)}: The goal is to increase the misclassification rate. She poisons the training data by randomly adding good feature values of one class to another class.
	\item \textbf{Fake Users Addition Attack (Poisoning)}: The attacker poisons the data by removing a set of real users and then injecting fake users into the training dataset. Feature values of fake users are selected randomly from the real users' feature values.
\end{itemize}

\section{Social Media Users Location and Privacy}
This location disclosure attack is a specific version of attribute inference attack in which the adversary focuses on inferring geo-location information for a given user. The location disclosure attack takes as input some geolocated data and produces some additional knowledge about target users. More precisely, the objective of this attack may be to: 1) predict the movement patterns of an individual, 2) learn the semantics of the target user mobility behavior, 3) link records of the same individual, and 4) identify points of interest~\cite{gambs2010show}. Existing works incorporates a given user's friends' known geo-location information~\cite{backstrom2010find,mcgee2011geographic,jurgens2013s,rout2013s,compton2014geotagging,kong2014spot,jurgens2015geolocation,mcgee2013location}. The work of~\cite{backstrom2010find} introduces a probabilistic model representing the likelihood of the target user's location based on her friends' location and geographic distance between them. \cite{kong2014spot} and~\cite{mcgee2011geographic} extend Backstrom et al.'s work~\cite{backstrom2010find} and find the target user's friends that are strong predictors of her location. 

In another work, Mcgee et al.~\cite{mcgee2013location} integrates social tie strength information to capture the uncertainty across multiple location granularities. The reason is that not all relationships in social media are the same and the location of friends with strong ties are more revealing of a user's location. Rout et al.~\cite{rout2013s} deploy a SVM classifier on a given set of features to predict the target user's location. These features include cities of the target user's friends, number of friends in the same city as the target user and number of reciprocal relationships the target user has per city. Jurgens et al.\cite{jurgens2013s} infer locations by proposing an iterative multi-pass label propagation approach. This approach calculates each target user's location as the geometric median of her friends' locations and it seeks to overcome the sparsity problem when the ground truth data is sparse. The work of~\cite{compton2014geotagging} extends~\cite{jurgens2013s} and limits the propagation of noisy locations by weighting different locations using information such as the number of times the users have interacted. 

Another work from Cheng et al.~\cite{cheng2010you} proposes a probabilistic framework which infers Twitter users' city level location based on the content of their tweets. The idea is that users' tweets include either implicit or explicit location-specific content, e.g.,  place names, or words or phrases more associated with certain locations (e.g., "howdy" for Texas). It uses lattice-based neighborhood smoothing technique to even out the word probabilities and overcome the tweet sparsity challenge. Hecht et al.~\cite{hecht2011tweets} also found that only $34\%$ of Twitter users do not provide their real location information or share fake locations or sarcastic comments to fool location inference approaches. They show that a user's location could be inferred using machine learning techniques through the implicit user behavior reflected in their tweets. In another work, Ryoo et al.~\cite{ryoo2014inferring} refine Cheng et al.'s city-level granularity location inference approach~\cite{cheng2010you} to 500 m distance bins. Having GPS-tagged tweets for a set of users, their approach builds geographic distributions of words and computes user location as a weighted center of mass from the user's words. It then uses a probabilistic model and computes the foci and dispersions by binning the distance between GPS coordinates and the word's center by 500m for computational scalability.

Li et al.~\cite{li2012towards} introduce a unified discriminative influence model which considers both users' social network and user-centric data (e.g., tweets) in order to solve the scarce and noisy data challenge for location inference. It first augments social network and user data in a probabilistic framework which is viewed as a heterogeneous graph with users and tweets as nodes and social and tweeting relations as edges. Every node in this graph is then associated with a location and the proposed probabilistic influence model measures how likely an edge is generated between two nodes considering their locations. This can further handle the noisy data challenge in location inference problem. It then predicts user's location either locally or globally. Another similar work from Li et al.~\cite{li2012multiple} exploits a user's tweets and social relations to build a complete location profile which infers a set of multiple long-term  geographic location scopes related to her which not only includes her home location, but also other related ones, e.g. work space. Their approach captures the locations related to social relations as well (e.g. Bob and Alice are friends as they both live in Texas). In particular, their approach is a probabilistic generative model which is consisted of three components, 1) location-based following model, 2) location-based tweeting model, and 3) partial information from users known locations.

Srivatsa et al.~\cite{srivatsa2012deanonymizing} propose a de-anonymization attack which exploits a user's friendship information in social media to de-anonymize users mobility traces. The idea behind this approach is that people meet those who have a relationship with them and thus they could be identified by their social relationships. This approach models mobility traces as contact graphs and identifies a set of seed users in both graphs, i.e. contacts graph and friendship in social network. In the second step, it propagates mapping from seed users to the remaining users in the graphs. This approach uses Distance Vector, Randomized Spanning Trees and Recursive Subgraph Matching heuristics to measure the mapping strength and propagate the measured strength through the network. 

Another work from Ji et al.~\cite{ji2014structure,ji2016general} improves the work of Srivasta et al.~\cite{srivatsa2012deanonymizing} in terms of accuracy and computational complexity. This work focuses on mapping anonymized users mobility traces to social media accounts. In addition to the users' local features, their approach incorporates users' global characteristics as well. Ji et al. define three similarity metrics: structural similarity, relative distance similarity and inheritance similarity. These similarities are then combined in a unified similarity. Structural similarity considers features such as degree centrality, closeness centrality, and betweenness centrality while relative distance similarity captures the distance between users and seed users. Inheritance similarity considers the number of common neighbors which have been mapped as well as the degree similarity between the users in mobility traces and social media network graph. Next, Ji et al.~\cite{ji2014structure,ji2016general} propose an adaptive de-anonymization framework which adaptively starts de-anonymizing from a core matching set which is consisted of a number of mapped users and $k$-hop mapping spanning set of them. 

In another work~\cite{mahmud2014home}, the location of Twitter users are inferred in different granularities (e.g., city, state, time zone, geographical region) based on their tweeting behavior (frequency of tweets per time unit) and the content of their tweets. This approach exploits external location knowledge (e.g., dictionary containing names of cities and states, and location based services such as Foursquare) and finds explicit references of locations in tweets. Then all features are fed into a dynamically weighted method which is an ensemble of the statistical and heuristic classifiers.

Another work from Wang et al.~\cite{wang2018you} links multiple users identities across multiple services/social media platforms (even with different types) according to the spatial-temporal locality of their activities, i.e. users mobility traces. This work also assumes that individuals can have multiple IDs/accounts. The motivation behind their algorithm is that IDs corresponding to the same person, are online at the same time in the same location and users' daily movement is predictable with repeated patterns. Wang et al. model users information as a contact graph where nodes are IDs (regardless of the service) and an edge represents connected IDs that have visited the same location. The weight of the edge demonstrates the number of co-location of two nodes. Then, a Bayesian matching algorithm is proposed to find the most probable matching candidates for a given target ID. A Bayesian inference method is then used to generate confidence scores for ranking candidates.

The work of~\cite{jurgens2015geolocation} compares different approaches in location inference attacks in social networks. There are also some other surveys discussing location inference techniques specifically in Twitter~\cite{ajao2015survey,8295255} which the reader can refer to. Note that a large portion of research is dedicated to inference attacks on geolocated data which is out of the scope of this survey~\cite{shokri2011quantifying,gambs2010show,liu2018location}. A thorough survey is also available discussing geolocation data privacy which readers can refer to it if they are interested~\cite{liu2018location}. Note that the scope of this survey is a different from ours in which we cover the location privacy issues of users based on activities in social media. 
\section{Recommendation Systems and Privacy}
Recommendation systems help individuals find information that matches with their interests by building user-interest profiles and recommending items to users based on those profiles. These profiles could be extracted from the users' interactions as they express their preferences and interests, e.g. clicks, likes/dislikes, ratings, purchases, etc~\cite{beigi2018similar,zafarani2014social}. While user profiles help recommender systems to improve the quality of the services a user receives (a.k.a utility), they also raise privacy concerns by reflecting the preferences of users~\cite{ramakrishnan2001privacy}. Many works have studied the relationship between privacy and utility and have proposed solutions to handle the trade-off. In general, these works focus on obfuscating users' interactions to hide their actual intentions and prevent accurate profiling~\cite{puglisi2015content}. Following this strategy, no third parties or external entities need to be trusted by the users to preserve their privacy.
%conclusion of ~\cite{puglisi2015content} for some introduction if needed
Existing approaches use different techniques and mechanisms and could be categorized mainly into three categories: cryptographic based techniques~\cite{aimeur2008experimental,canny2002,hoens2010private,tang2016privacy,badsha2017privacy}, differential privacy based approaches~\cite{mcsherry2009differentially,machanavajjhala2011personalized,zhu2013differential,jorgensen2014privacy,shen2014privacy,hua2015differentially,guerraoui2015d,zhu2016differential,meng2018personalized} and perturbation based techniques~\cite{parra2017pay,rebollo2011information,parra2014optimal,polat2003privacy,luo2014privacy,xin2014controlling,parameswaran2007privacy,puglisi2015content,howe2009lessons}

A group of works focus on providing cryptographic solutions to the problem of secure recommender systems. The approaches do not let the single trusted party have access to everyone's data~\cite{aimeur2008experimental,canny2002,hoens2010private,tang2016privacy,badsha2017privacy}. Instead, users’ ratings are stored as encrypted vectors and aggregates of the data are provided in the public domain. These approaches do not prevent privacy leaks through the output of recommendation systems (i.e., the recommendation themselves). Moreover, these techniques are not the scope of this survey.
\subsection{Differential Privacy Based Solutions}
Works in this group utilize a differential privacy strategy to either anonymize user data before sending it to the recommendation system or perturb the recommendation outputs. McSherry et al.~\cite{mcsherry2009differentially} modify leading algorithms for recommendation systems (i.e., SVD and $k$-nearest neighbor) for the first time so that drawing inferences about original ratings is difficult. They utilize differential privacy to construct private covariance matrices and make the collaborative filtering algorithms that use them private without having significant loss in accuracy.

In another work, Calandrino et al.~\cite{calandrino2011you} propose a new passive attack on recommender systems to infer a target user's transactions (i.e., item ratings). Their attack first monitors changes in the public outputs of a recommender system over a period of time. Public outputs may include related-items lists or an item-item covariance matrix. Then, it combines this information with a moderate amount of auxiliary information about the target user's transactions to further infer many of the target user's unknown transactions. Calandrino et al. further introduce an active inference attack on $k$-NN recommender systems. In this attack, $k$ sybil users accounts are created and the $k$ nearest neighbor of each sybil consists of $k-1$ other sybil users and the target user. The attack can then infer the target user's transactions history based on the items recommended to any of the sybils. The results of this work confirms the existence of privacy risks over the public outputs of recommender systems. The work of McSherry et al.~\cite{mcsherry2009differentially} is not effective in protecting users against this attack as it does not consider updates to the covariance matrices and cannot provide a privacy guarantee in the dynamic settings. Machanavajjhala et al.~\cite{machanavajjhala2011personalized} then quantifies the accuracy-privacy trade-off. In particular, they prove lower bounds on the minimum loss in accuracy for recommendation systems that utilize differential privacy. Moreover, they adapt two differentially private algorithms, Laplace~\cite{dwork2006calibrating} and Exponential~\cite{mcsherry2007mechanism} for the problem of recommendation without disclosing any user sensitive attributes. This work assumes that all users' attributes are sensitive. %Their results emphasize impossibility of having algorithms that are both private and accurate for all users.

Previous works~\cite{mcsherry2009differentially,machanavajjhala2011personalized} are vulnerable to $k$-nearest neighbor attack as they fail to hide similar neighbors~\cite{calandrino2011you}. Zhu et al.~\cite{zhu2013differential} also propose a private neighborhood-based collaborative filtering which protects the information of both neighbors and user ratings. The proposed work assumes that the recommender system is trusted and introduces two operations: private neighbor selection, and recommendation-aware sensitivity. The first operation seeks to protect neighbors identity by privately selecting $k$ neighbors from a list of candidates and then adopting the exponential mechanism~\cite{mcsherry2007mechanism} to arrange a probability for each candidate. The second operation is proposed in this work to enhance the performance of recommendation systems by reducing the magnitude of added noise. To do so, after selecting $k$ neighbors, the similarity of neighbors is then perturbed by adding Laplace noise to mask the ratings given by a certain neighbor. Finally, the neighborhood collaborative filtering based recommendation is performed on the private data. In another work, Jorgensen et al.~\cite{jorgensen2014privacy} assume that all users' item-rating attributes are sensitive. However, different from Machanavajjhala et al.~\cite{machanavajjhala2011personalized}, they assume that users' social relations are non-sensitive. They propose a differentially private based recommendation which incorporates social relations besides user-item ratings. To address the utility loss, this work first clusters users according to their social relations. Then, noisy averages of the user-item preferences are computed for each cluster using the differential privacy mechanism. Results of this method show that the clustering phase reduces sensitivity and the amount of added noise which further reduces the utility loss.

Shen et al.~\cite{shen2014privacy} assume that the recommender system is untrusted. They propose a user perturbation framework which anonymizes user data under a novel mechanism for differential privacy: relaxed admissible mechanism. The recommender system then utilizes users' perturbed data to perform recommendation. They provide mathematical bounds on the privacy and utility of the anonymized data. Hua et al.~\cite{hua2015differentially} also propose a matrix factorization based recommender system which is differentially private. In particular, they solve this problem for two scenarios, trusted recommender and untrusted recommender. For the first scenario, user and item profile vectors are learned via regular and private version of matrix factorization, respectively. Private version of matrix factorization adds noises to item vectors to make them differentially private. In the second scenario, item profile vectors are first differentially privately learned with private matrix factorization problem. Then, since a user's profile depends on her own ratings rather than other users, her differentially private profile vector is derived from the private item profiles.

A novel and strong form of differential privacy, namely distance-based differential privacy, has been introduced by Guerroaui et al.~\cite{guerraoui2015d}. Distance-based differential privacy ensures privacy for all the items rated by a user and the ones that are within a distance $\lambda$ from it. The distance parameter $\lambda$ controls the level of privacy and aids in tuning the recommendation privacy-utility trade-off. The proposed protocol first finds a group of similar items for each given item. Then, it creates a manipulated user profile to preserve $(\epsilon, \lambda)$-differential privacy by selecting an item and replacing it with another one. The $k$ most similar users for an active user also get updated periodically using the altered profiles generated in previous step.

Another differential privacy based recommendation by Zhu et al.~\cite{zhu2016differential} seeks to solve the privacy problem in recommendations by applying a differential privacy mechanism into the procedure of recommendation. In particular, it proposed two approaches: item-based and user-based recommendation algorithms. In the item-based one, the exponential mechanism~\cite{mcsherry2007mechanism} is applied to the selection of the related items in order to guarantee differential privacy. Such resultant differentially private items list is further used to find recommendation for a given user. Similarly, in the user-based recommendation system, a list of related users are selected for each target user. This list is further used to find the relevance score for each item by calculating the sum of ratings provided by the related users. The exponential mechanism is used in the item selection process to make the recommendation process differentially private. Another work differentiates sensitive and non-sensitive ratings to further improve the quality of recommendation systems in the long run~\cite{meng2018personalized}. Meng et al.~\cite{meng2018personalized} propose a personalized privacy preserving recommender system. Given sets of sensitive and non-sensitive ratings for each user, their approach utilizes differential privacy~\cite{dwork2008differential} to perturb users' ratings. Smaller and larger privacy budgets are considered for sensitive and non-sensitive ratings, respectively. This protects users' privacy while retaining recommendation effectiveness. In order to protect sensitive ratings from untrusted friends, Meng et al. then utilize only non-sensitive ratings to calculate social relations regularization.

\subsection{Perturbation Based Solutions}
Perturbation based techniques usually obfuscate users item ratings by adding random noise to the user data. Rebollo et al.~\cite{rebollo2011information} propose an approach which first measures the user's privacy risk as the Kullback–Leibler(KL)-divergence (a.k.a relative entropy)~\cite{cover2012elements} between user's apparent profile $s$ and average population's distribution profile $p$. The idea is that the more a user's profile diverges from the general population, the more information an attacker can learn about her. Then it seeks to find the obfuscation rate $\rho$ for generating forged user profiles so that the privacy risk is minimized. Authors then provide a closed-form solution for perturbing users interactions with a recommender system in order to optimize the privacy risk function. 

Puglisi et al.~\cite{puglisi2015content} further extend Rebollo et al.'s work~\cite{rebollo2011information} to investigate the impact of this technique on content-based recommendation in terms of privacy and the potential degradation of the recommendation utility. This work measures a user's privacy risk similar to the approach proposed in~\cite{rebollo2011information}. The utility of the service is also measured by the prediction accuracy of the recommender system. This paper evaluates three different strategies, namely: optimized tag forgery~\cite{rebollo2010optimized}, uniform tag forgery and TrackmeNot (TMN)~\cite{howe2009lessons}. The uniform tag forgery method assign forged tags according to a uniform distribution across all categories of the user profile. TMN constructed eleven categories from Open Directory Project (ODP) classification scheme\footnote{http://www.dmoz.com} and selected the tags uniformly from this set. According to this work, users tend to mimic the profile of the population distribution when larger values of obfuscation rate is considered which results in less privacy risk but lower utility rate. Moreover, the authors have found that for a small forgery rate ($\rho = 0.2$), it is possible to obtain an increase in privacy against a small degradation of utility.

Polat et al.~\cite{polat2003privacy} use a randomized perturbation technique~\cite{agrawal2000privacy} to obfuscate user generated data. Each user generates the disguised z-score for the item he has rated. The z-score for each user-item pair is based on the original item-rating, the user's average ratings and the total number of items she has rated. The proposed approach approach then passes the perturbed private data to the collaborative filtering based recommender system to perform recommendation. The reason that this technique works is because collaborative filtering works with the aggregated user data. Therefore, although information from each individual
user is scrambled, since the number of users is significantly large, the aggregate information of
these users can be estimated with decent accuracy. The accuracy of predictions with this approach depends on the amount of noise added. Another work from Parameswaran et al.~\cite{parameswaran2007privacy} obfuscate user rating information and then pass disguised information to the collaborative filtering system for further recommendation. The proposed Nearest Neighbor Data Substitution (NeNDS) obfuscation method substitutes a user's data elements with one of her neighbors in the metric space~\cite{parameswaran2005robust}. However, one drawback of NeNDS is that the value of the perturbed data could be close enough to the original value which thus makes the data vulnerable. A hybrid version of NeNDS is then proposed which provides stronger privacy by combining geometric transformations with NeNDS. In this technique, the data sets are first geometrically transformed, and then operated upon by NeNDS.

In contrast to Mcsherry et al.~\cite{mcsherry2009differentially}, Xin et al, assume that the recommender is not trusted and the onus is on the users to protect their privacy~\cite{xin2014controlling}. Their approach separates the computations that can be done by the users locally and privately and those that must be done by the recommender system. In particular, item features are learned by the system and user features are obtained locally by the users and further used for recommendation. Their approach also divides users into two groups, users who publicly share their information, and those who keep their preferences private. It then uses information of users in the first group to estimate items features. Xin et al. show theoretically and empirically that having the public information of a moderate number of users with a high number of ratings is enough to have an accurate estimation. Moreover, they propose a new privacy mechanism which privately releases second order information that is needed for estimating item features. This information is extracted from users who keep their preferences private. The main assumption behind this work is not realistic though, as in a real-world scenario it is not easy to collect ratings of a moderate number of people with a high number of ratings.

Luo et al.~\cite{luo2014privacy} propose a perturbation-based group recommendation method which assumes that similar users are grouped to each other and they are not willing to expose their preferences to anybody other than the group members. The recommendation system then recommends items to the users within the same group. Their algorithm has four steps. In the first step, users are required to exchange their rating data among users in the same group given a secret key. This key varies for different users. The output of this step is a fake preference vector for each user. The value of the rating is then obfuscated in the second step by a chaos-based scrambling method. Similar to the traditional perturbation-based scheme in Polat et al.~\cite{polat2003privacy}, randomness is added to the output of the previous step to make sure no sensitive information remains in the published data for the attacker to misuse. This information is then sent to the recommender system and it iteratively extracts information about aggregated ratings of the users. Extracted information is then used to estimate a group preference vector for collaborative filtering based recommendation.

Parra-Arnau et al.~\cite{parra2014optimal}, propose a privacy enhancing technology framework, PET, which perturbs users preferences information by combining two techniques, namely, the forgery and the suppression of ratings. In this scenario, users may avoid rating items they like and instead rate those which do not reflect their actual preferences. Therefore, the apparent profile of users will be different from their actual profile. Similar to~\cite{rebollo2011information}, the privacy risk of each user is then measured as the KL-divergence~\cite{cover2012elements} between the user's apparent profile and the average population distribution. Utility is also controlled with the forgery rate $\rho \in [0,\inf)$ and suppression rate $\sigma \in [0,1)$. Then, authors define the privacy-forgery-suppression optimization function which characterizes the optimal trade-off among privacy, forgery rate and suppression rate. In particular, the solution of the optimization problem contains information about which ratings for each user should be forged and which ones should be suppressed to achieve the minimum privacy risk while keeping the utility of the data as high as possible. Similarly, Parra-Arnau et al.~\cite{parra2017pay} propose a system which generates a perturbed version of a user rating profile according to her privacy preferences. The system has two components, 1) a profile-density model in which the user's profile will be more similar to the crowd's, and 2) a classification model in which the user will not be identified as a member of a given group of users. Their proposed framework considers the money loss for advertisement venue in the exchange of privacy and optimizes the trade-off between privacy and economic compensation. The system utilizes different privacy metrics such as KL-divergence and mutual information. The final output is a decision on whether each service provider (i.e. tracker) can have access to the user's profile, or it should be blocked, or the user should be notified about the privacy risks. 

Recently, the work of Biega et. al.~\cite{biega2017privacy} proposes a framework which scrambles the users' rating history to preserve both their privacy and utility. The main assumption of this paper is that service providers, i.e. recommender systems, do not need the complete and accurate user profiles to have a personalized recommendation. Therefore, it splits users' profiles which is consisted of pairs of user-item interactions to \textit{Mediator Accounts} in a way that coherent pieces of different users' profiles are kept intact in the MAs. The service provider will then deal with the MAs rather than real user profiles. This helps to first preserve users' privacy by scrambling the user-item interactions across various proxy accounts. Moreover, it keeps the user utility high as possible since it tries to assign an user-item interaction to a proxy account which minimizes the average coherence loss over all other objects in the account. This framework also quantifies the user's privacy-utility trade-off.

Another work from Guerraoui et. al.~\cite{guerraoui2017utility} introduces metrics for measuring the utility and privacy effect of a user's behavior such as clicks, and likes/dislikes. Then, it shows that there is not always a trade-off between utility and privacy. This paper also proposed a click-advisor platform which is an application of the utility and privacy metrics and could warn users regarding the status of their click with respect to the privacy and utility. This paper assumes that the recommender is trusted itself and the users' sensitive information could be learned by curious users who could deduce profiles through what is recommended to them. According to this work, the utility of a click by user $u$ is the difference between commonality of this user before and after that click. Commonality is defined as the closeness of the user profileto other users profiles in the system. The disclosure degree of a user is measured as the probability that the user like items. The disclosure risk of a click is accordingly defined as the difference of the disclosure degree of a user before and after the click. It then uses privacy and utility metrics to guide users in their action by telling them whether their intended action leads to privacy leakage or if it has any effect on their utility or not.

\section{Summary and Future Research Directions}
\begin{figure}[t]
	\centering
	\includegraphics[width=0.6\linewidth]{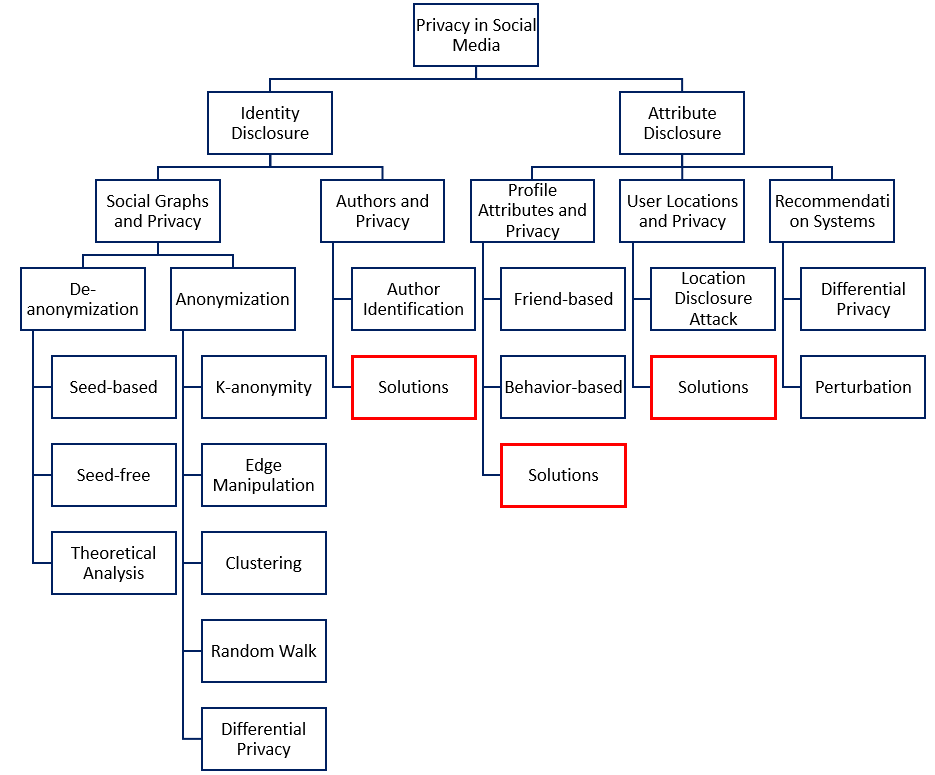}
	\captionof{figure}{\textbf{An overview of privacy attacks and corresponding defenses in social media platforms. Tasks highlighted in red have not been extensively studied.}}
	\label{outline}
\end{figure}

\begin{figure}[t]
	\centering
	\includegraphics[width=0.7\linewidth]{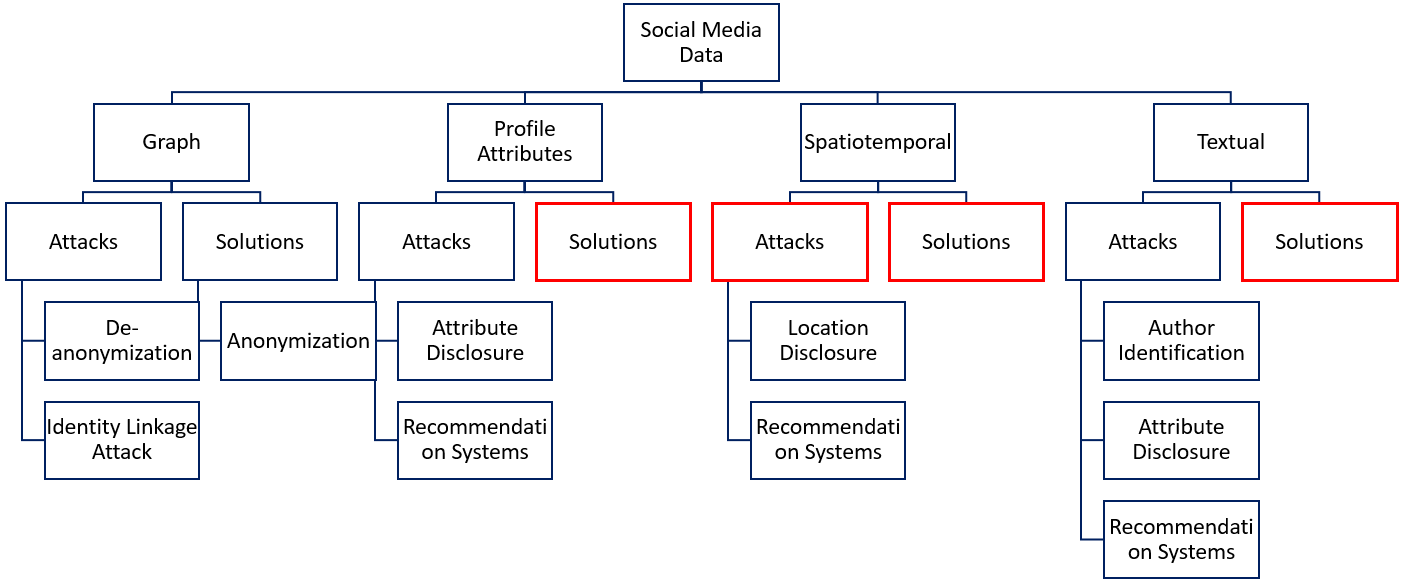}
	\captionof{figure}{\textbf{An overview of privacy issues with respect to the type of social media data. Tasks highlighted in red have not been extensively studied.}}
	\label{outline-data}
\end{figure}

Online users are increasingly sharing their personal information on social media platforms. These platforms publish and share user-generated data with third party consumers. This data is rich in content and contains sensitive information about users which risks exposing individuals' privacy. Recent research has shown the vulnerability of user-generated data against the two general types of attacks, identity disclosure and attribute disclosure. Sanitizing user-generated social media data is more challenging than structured data as it is heterogeneous, highly unstructured, noisy and inherently different from relational and tabular data. In this survey, we reviewed the recent developments in the field of privacy of social media data. We first reviewed traditional privacy models for structural data. Then, we reviewed, categorized and compared existing methods in terms of privacy models, privacy leakage attacks, and anonymization algorithms. We also reviewed privacy risks which exists in different aspects of social media such as users graph information (e.g. social relations, mobility traces, sociotemporal information, etc.), profile attributes, textual information (e.g. posts) and preferences. We categorized relevant works into five groups 1) graph data anonymization and de-anonymization, 2) author identification, 3) profile attribute disclosure, 4) user location and privacy, and 5) recommender systems and privacy issues. For each category, we discussed existing attacks and solutions (if any was proposed) and classified them based on the type of data and the used technique. We outlined the privacy attacks/solutions in Figure~\ref{outline}. Figure~\ref{outline-data} also depicts the relevant privacy issues with respect to the type of social media data.

Detecting privacy issues and proposing techniques to protect privacy of users in social media is a challenging issue. Most of the existing works focus on introducing new attacks and thus the gap between protection and detection becomes larger. Although a large body of work has emerged in recent years for investigating privacy issues for social media data, the development of tasks in each category is highly imbalanced. Some of them are well studied, whereas others need further investigation. We highlight these tasks in red in Figure \ref{outline} and Figure \ref{outline-data} based on privacy issues and user-generated data type, respectively. Below, we identify potential research directions in this field:
\begin{itemize}[leftmargin=*]
\item \textbf{\textit{Protecting privacy of textual information}}: Textual information is noisy, high-dimensional and unstructured. It is rich in content and could reveal many sensitive information that user does not originally expose such as demographic information and location. This makes textual data a very important source of information for adversaries and could be exploited in many attacks. We thus need more research for anonymizing users' textual information to preserve privacy of users against various attacks such as author identification, and profile attribute disclosure.

\item \textbf{\textit{Protecting privacy of profile attribute information}}: We also reviewed many state-of-the-art works which introduces privacy risks with respect to profile attributes. In particular, these works introduce new attacks which infer target users profile attributes considering their behavior in social media platforms. To the best of our knowledge, there is no work on introducing defense mechanism against these attacks. One research direction could be either in terms of a privacy preserving tool for users which warns them against their activities and possibility of privacy leakage. Another direction would be to propose a privacy protection technique which will be deployed before sharing users' data with third parties. Profile attributes are very similar to tabular datasets but could be easily inferred from user-generated unstructured data.

\item \textbf{\textit{Privacy of spatiotemporal social media data}}: Social media platforms support space-time indexed data and users have created a large volume of time-stamped, geo-located data. Such spatiotemporal data has an immense value for understanding users behavior better. In this survey, we reviewed the state-of-the-art re-identification attacks which incorporate this data to breach privacy of users. This information could be used to infer users' location as well as their preferences and interests in case of recommendation systems. One future research direction could be investigating the role of temporal information in privacy of online users. More research should be done to build anonymization frameworks for protecting users temporal information.

\item \textbf{\textit{Privacy of heterogeneous social media data}}: 
User-generated social media data is heterogeneous and consists of different aspects. Most of the previous works illustrate the vulnerability of each aspect of social media data against identity and attribute disclosure attacks. Existing anonymization techniques also assume that it is enough to anonymize each aspect of heterogeneous social media data independently. Beigi et al.~\cite{beigi2018securing} evaluated this assumption for two specific aspects of data, i.e. textual and graph, and showed that this is not a correct assumption due to the hidden relations between different aspects of the heterogeneous data. One potential research direction is to examine how different combinations of heterogeneous data (e.g., a combination of location
and textual information) are vulnerable to the de-anonymization attack. Another potential direction is to improve anonymization techniques to preserve the privacy of users in social media data by considering hidden relations between different components of the data due to the innate heterogeneity of user-generated data.

\item \textbf{\textit{Privacy protection against identity and attribute disclosure attacks}}: User-generated data in social media platforms such as profile information, graph data, location and interest beliefs plays an important role in helping online service providers to offer better services for their users. We reviewed many related works in this survey which show how these information make users vulnerable against privacy breaches. However, very limited research has been done to exploit effective anonymization techniques for preserving privacy of users against these attacks. More research needs to be done to develop data sanitizaion approaches specialized for social media data.
\end{itemize}
\begin{acks}
The authors would like to thank Alexander Nou for his help throughout the paper. This material is based upon the work supported in part by Army Research Office (ARO) under grant number W911NF-15-1-0328 and Office of Naval Research (ONR) under grant number N00014-17-1-2605.
\end{acks}

\bibliographystyle{ACM-Reference-Format}
\bibliography{sample-bibliography}


\begin{thebibliography}{200}

%%% the Web address.

\ifx \showCODEN    \undefined \def \showCODEN     #1{\unskip}     \fi
\ifx \showDOI      \undefined \def \showDOI       #1{#1}\fi
\ifx \showISBNx    \undefined \def \showISBNx     #1{\unskip}     \fi
\ifx \showISBNxiii \undefined \def \showISBNxiii  #1{\unskip}     \fi
\ifx \showISSN     \undefined \def \showISSN      #1{\unskip}     \fi
\ifx \showLCCN     \undefined \def \showLCCN      #1{\unskip}     \fi
\ifx \shownote     \undefined \def \shownote      #1{#1}          \fi
\ifx \showarticletitle \undefined \def \showarticletitle #1{#1}   \fi
\ifx \showURL      \undefined \def \showURL       {\relax}        \fi
% invisible to TeX
\providecommand\bibfield[2]{#2}
\providecommand\bibinfo[2]{#2}
\providecommand\natexlab[1]{#1}
\providecommand\showeprint[2][]{arXiv:#2}

\bibitem[\protect\citeauthoryear{Abawajy, Ninggal, and Herawan}{Abawajy
  et~al\mbox{.}}{2016}]%
        {abawajy2016privacy}
\bibfield{author}{\bibinfo{person}{Jemal~H Abawajy}, \bibinfo{person}{Mohd
  Izuan~Hafez Ninggal}, {and} \bibinfo{person}{Tutut Herawan}.}
  \bibinfo{year}{2016}\natexlab{}.
\newblock \showarticletitle{Privacy preserving social network data
  publication}.
\newblock \bibinfo{journal}{\emph{IEEE communications surveys \& tutorials}}
  \bibinfo{volume}{18}, \bibinfo{number}{3} (\bibinfo{year}{2016}),
  \bibinfo{pages}{1974--1997}.
\newblock


\bibitem[\protect\citeauthoryear{Abbasi and Chen}{Abbasi and Chen}{2008}]%
        {abbasi2008writeprints}
\bibfield{author}{\bibinfo{person}{Ahmed Abbasi} {and}
  \bibinfo{person}{Hsinchun Chen}.} \bibinfo{year}{2008}\natexlab{}.
\newblock \showarticletitle{Writeprints: A stylometric approach to
  identity-level identification and similarity detection in cyberspace}.
\newblock \bibinfo{journal}{\emph{ACM Transactions on Information Systems
  (TOIS)}} \bibinfo{volume}{26}, \bibinfo{number}{2} (\bibinfo{year}{2008}),
  \bibinfo{pages}{7}.
\newblock


\bibitem[\protect\citeauthoryear{Afroz, Brennan, and Greenstadt}{Afroz
  et~al\mbox{.}}{2012}]%
        {afroz2012detecting}
\bibfield{author}{\bibinfo{person}{Sadia Afroz}, \bibinfo{person}{Michael
  Brennan}, {and} \bibinfo{person}{Rachel Greenstadt}.}
  \bibinfo{year}{2012}\natexlab{}.
\newblock \showarticletitle{Detecting hoaxes, frauds, and deception in writing
  style online}. In \bibinfo{booktitle}{\emph{Security and Privacy (SP), 2012
  IEEE Symposium on}}. IEEE, \bibinfo{pages}{461--475}.
\newblock


\bibitem[\protect\citeauthoryear{Aggarwal, Feder, Kenthapadi, Motwani,
  Panigrahy, Thomas, and Zhu}{Aggarwal et~al\mbox{.}}{2005}]%
        {aggarwal2005approximation}
\bibfield{author}{\bibinfo{person}{Gagan Aggarwal}, \bibinfo{person}{Tomas
  Feder}, \bibinfo{person}{Krishnaram Kenthapadi}, \bibinfo{person}{Rajeev
  Motwani}, \bibinfo{person}{Rina Panigrahy}, \bibinfo{person}{Dilys Thomas},
  {and} \bibinfo{person}{An Zhu}.} \bibinfo{year}{2005}\natexlab{}.
\newblock \showarticletitle{Approximation algorithms for k-anonymity}.
\newblock \bibinfo{journal}{\emph{Journal of Privacy Technology (JOPT)}}
  (\bibinfo{year}{2005}).
\newblock


\bibitem[\protect\citeauthoryear{Agrawal and Srikant}{Agrawal and
  Srikant}{2000}]%
        {agrawal2000privacy}
\bibfield{author}{\bibinfo{person}{Rakesh Agrawal} {and}
  \bibinfo{person}{Ramakrishnan Srikant}.} \bibinfo{year}{2000}\natexlab{}.
\newblock \showarticletitle{Privacy-preserving data mining}. In
  \bibinfo{booktitle}{\emph{ACM Sigmod Record}}, Vol.~\bibinfo{volume}{29}.
\newblock


\bibitem[\protect\citeauthoryear{Aimeur, Brassard, Fernandez, Onana, and
  Rakowski}{Aimeur et~al\mbox{.}}{2008}]%
        {aimeur2008experimental}
\bibfield{author}{\bibinfo{person}{Esma Aimeur}, \bibinfo{person}{Gilles
  Brassard}, \bibinfo{person}{Jose~M Fernandez}, \bibinfo{person}{Flavien
  Serge~Mani Onana}, {and} \bibinfo{person}{Zbigniew Rakowski}.}
  \bibinfo{year}{2008}\natexlab{}.
\newblock \showarticletitle{Experimental demonstration of a hybrid
  privacy-preserving recommender system}. In
  \bibinfo{booktitle}{\emph{Availability, Reliability and Security, 2008. ARES
  08. Third International Conference on}}. IEEE, \bibinfo{pages}{161--170}.
\newblock


\bibitem[\protect\citeauthoryear{Ajao, Hong, and Liu}{Ajao
  et~al\mbox{.}}{2015}]%
        {ajao2015survey}
\bibfield{author}{\bibinfo{person}{Oluwaseun Ajao}, \bibinfo{person}{Jun Hong},
  {and} \bibinfo{person}{Weiru Liu}.} \bibinfo{year}{2015}\natexlab{}.
\newblock \showarticletitle{A survey of location inference techniques on
  Twitter}.
\newblock \bibinfo{journal}{\emph{Journal of Information Science}}
  \bibinfo{volume}{41}, \bibinfo{number}{6} (\bibinfo{year}{2015}),
  \bibinfo{pages}{855--864}.
\newblock


\bibitem[\protect\citeauthoryear{Al-Qurishi, Al-Rakhami, Alamri, Alrubaian,
  Rahman, and Hossain}{Al-Qurishi et~al\mbox{.}}{2017}]%
        {al2017sybil}
\bibfield{author}{\bibinfo{person}{Muhammad Al-Qurishi},
  \bibinfo{person}{Mabrook Al-Rakhami}, \bibinfo{person}{Atif Alamri},
  \bibinfo{person}{Majed Alrubaian}, \bibinfo{person}{Sk~Md~Mizanur Rahman},
  {and} \bibinfo{person}{M~Shamim Hossain}.} \bibinfo{year}{2017}\natexlab{}.
\newblock \showarticletitle{Sybil defense techniques in online social networks:
  a survey}.
\newblock \bibinfo{journal}{\emph{IEEE Access}}  \bibinfo{volume}{5}
  (\bibinfo{year}{2017}), \bibinfo{pages}{1200--1219}.
\newblock


\bibitem[\protect\citeauthoryear{Al~Zamal, Liu, and Ruths}{Al~Zamal
  et~al\mbox{.}}{2012}]%
        {al2012homophily}
\bibfield{author}{\bibinfo{person}{Faiyaz Al~Zamal}, \bibinfo{person}{Wendy
  Liu}, {and} \bibinfo{person}{Derek Ruths}.} \bibinfo{year}{2012}\natexlab{}.
\newblock \showarticletitle{Homophily and Latent Attribute Inference: Inferring
  Latent Attributes of Twitter Users from Neighbors.}
\newblock  (\bibinfo{year}{2012}).
\newblock


\bibitem[\protect\citeauthoryear{Almishari and Tsudik}{Almishari and
  Tsudik}{2012}]%
        {almishari2012exploring}
\bibfield{author}{\bibinfo{person}{Mishari Almishari} {and}
  \bibinfo{person}{Gene Tsudik}.} \bibinfo{year}{2012}\natexlab{}.
\newblock \showarticletitle{Exploring linkability of user reviews}. In
  \bibinfo{booktitle}{\emph{European Symposium on Research in Computer
  Security}}. Springer, \bibinfo{pages}{307--324}.
\newblock


\bibitem[\protect\citeauthoryear{Alufaisan, Zhou, Kantarcioglu, and
  Thuraisingham}{Alufaisan et~al\mbox{.}}{2017}]%
        {alufaisan2017hacking}
\bibfield{author}{\bibinfo{person}{Yasmeen Alufaisan}, \bibinfo{person}{Yan
  Zhou}, \bibinfo{person}{Murat Kantarcioglu}, {and} \bibinfo{person}{Bhavani
  Thuraisingham}.} \bibinfo{year}{2017}\natexlab{}.
\newblock \showarticletitle{Hacking social network data mining}. In
  \bibinfo{booktitle}{\emph{Intelligence and Security Informatics (ISI), 2017
  IEEE International Conference on}}. IEEE, \bibinfo{pages}{54--59}.
\newblock


\bibitem[\protect\citeauthoryear{Alvari, Hajibagheri, Sukthankar, and
  Lakkaraju}{Alvari et~al\mbox{.}}{2016}]%
        {alvari2016identifying}
\bibfield{author}{\bibinfo{person}{Hamidreza Alvari}, \bibinfo{person}{Alireza
  Hajibagheri}, \bibinfo{person}{Gita Sukthankar}, {and} \bibinfo{person}{Kiran
  Lakkaraju}.} \bibinfo{year}{2016}\natexlab{}.
\newblock \showarticletitle{Identifying community structures in dynamic
  networks}.
\newblock \bibinfo{journal}{\emph{Social Network Analysis and Mining}}
  \bibinfo{volume}{6}, \bibinfo{number}{1} (\bibinfo{year}{2016}),
  \bibinfo{pages}{77}.
\newblock


\bibitem[\protect\citeauthoryear{Alvari, Lakkaraju, Sukthankar, and
  Whetzel}{Alvari et~al\mbox{.}}{2014}]%
        {alvari2014predicting}
\bibfield{author}{\bibinfo{person}{Hamidreza Alvari}, \bibinfo{person}{Kiran
  Lakkaraju}, \bibinfo{person}{Gita Sukthankar}, {and} \bibinfo{person}{Jon
  Whetzel}.} \bibinfo{year}{2014}\natexlab{}.
\newblock \showarticletitle{Predicting guild membership in massively
  multiplayer online games}. In \bibinfo{booktitle}{\emph{International
  Conference on Social Computing, Behavioral-Cultural Modeling, and
  Prediction}}. Springer, \bibinfo{pages}{215--222}.
\newblock


\bibitem[\protect\citeauthoryear{Andreou, Goga, and Loiseau}{Andreou
  et~al\mbox{.}}{2017}]%
        {andreou2017identity}
\bibfield{author}{\bibinfo{person}{Athanasios Andreou}, \bibinfo{person}{Oana
  Goga}, {and} \bibinfo{person}{Patrick Loiseau}.}
  \bibinfo{year}{2017}\natexlab{}.
\newblock \showarticletitle{Identity vs. Attribute Disclosure Risks for Users
  with Multiple Social Profiles}. In \bibinfo{booktitle}{\emph{Proceedings of
  the 2017 IEEE/ACM ASONAM}}. ACM, \bibinfo{pages}{163--170}.
\newblock


\bibitem[\protect\citeauthoryear{Backes, Berrang, Goga, Gummadi, and
  Manoharan}{Backes et~al\mbox{.}}{2016}]%
        {backes2016profile}
\bibfield{author}{\bibinfo{person}{Michael Backes}, \bibinfo{person}{Pascal
  Berrang}, \bibinfo{person}{Oana Goga}, \bibinfo{person}{Krishna~P Gummadi},
  {and} \bibinfo{person}{Praveen Manoharan}.} \bibinfo{year}{2016}\natexlab{}.
\newblock \showarticletitle{On profile linkability despite anonymity in social
  media systems}. In \bibinfo{booktitle}{\emph{Proceedings of the 2016 ACM on
  Workshop on Privacy in the Electronic Society}}. ACM,
  \bibinfo{pages}{25--35}.
\newblock


\bibitem[\protect\citeauthoryear{Backes, Humbert, Pang, and Zhang}{Backes
  et~al\mbox{.}}{2017}]%
        {backes2017walk2friends}
\bibfield{author}{\bibinfo{person}{Michael Backes}, \bibinfo{person}{Mathias
  Humbert}, \bibinfo{person}{Jun Pang}, {and} \bibinfo{person}{Yang Zhang}.}
  \bibinfo{year}{2017}\natexlab{}.
\newblock \showarticletitle{walk2friends: Inferring Social Links from Mobility
  Profiles}. In \bibinfo{booktitle}{\emph{Proceedings of the 2017 ACM SIGSAC
  Conference on Computer and Communications Security}}.
\newblock


\bibitem[\protect\citeauthoryear{Backstrom, Dwork, and Kleinberg}{Backstrom
  et~al\mbox{.}}{2007}]%
        {backstrom2007wherefore}
\bibfield{author}{\bibinfo{person}{Lars Backstrom}, \bibinfo{person}{Cynthia
  Dwork}, {and} \bibinfo{person}{Jon Kleinberg}.}
  \bibinfo{year}{2007}\natexlab{}.
\newblock \showarticletitle{Wherefore art thou r3579x?: anonymized social
  networks, hidden patterns, and structural steganography}. In
  \bibinfo{booktitle}{\emph{Proceedings of the 16th international conference on
  WWW}}.
\newblock


\bibitem[\protect\citeauthoryear{Backstrom, Sun, and Marlow}{Backstrom
  et~al\mbox{.}}{2010}]%
        {backstrom2010find}
\bibfield{author}{\bibinfo{person}{Lars Backstrom}, \bibinfo{person}{Eric Sun},
  {and} \bibinfo{person}{Cameron Marlow}.} \bibinfo{year}{2010}\natexlab{}.
\newblock \showarticletitle{Find me if you can: improving geographical
  prediction with social and spatial proximity}. In
  \bibinfo{booktitle}{\emph{Proceedings of the 19th international conference on
  WWW}}.
\newblock


\bibitem[\protect\citeauthoryear{Badsha, Yi, Khalil, and Bertino}{Badsha
  et~al\mbox{.}}{2017}]%
        {badsha2017privacy}
\bibfield{author}{\bibinfo{person}{Shahriar Badsha}, \bibinfo{person}{Xun Yi},
  \bibinfo{person}{Ibrahim Khalil}, {and} \bibinfo{person}{Elisa Bertino}.}
  \bibinfo{year}{2017}\natexlab{}.
\newblock \showarticletitle{Privacy preserving user-based recommender system}.
  In \bibinfo{booktitle}{\emph{Distributed Computing Systems (ICDCS), 2017 IEEE
  37th International Conference on}}. IEEE, \bibinfo{pages}{1074--1083}.
\newblock


\bibitem[\protect\citeauthoryear{Beigi}{Beigi}{2018}]%
        {beigi2018social}
\bibfield{author}{\bibinfo{person}{Ghazaleh Beigi}.}
  \bibinfo{year}{2018}\natexlab{}.
\newblock \showarticletitle{Social Media and User Privacy}.
\newblock \bibinfo{journal}{\emph{arXiv preprint arXiv:1806.09786}}
  (\bibinfo{year}{2018}).
\newblock


\bibitem[\protect\citeauthoryear{Beigi, Jalili, Alvari, and Sukthankar}{Beigi
  et~al\mbox{.}}{2014}]%
        {gbeigi_trust}
\bibfield{author}{\bibinfo{person}{Ghazaleh Beigi}, \bibinfo{person}{Mahdi
  Jalili}, \bibinfo{person}{Hamidreza Alvari}, {and} \bibinfo{person}{Gita
  Sukthankar}.} \bibinfo{year}{2014}\natexlab{}.
\newblock \showarticletitle{Leveraging Community Detection for Accurate Trust
  Prediction}. In \bibinfo{booktitle}{\emph{ASE International Conference on
  Social Computing, Palo Alto, CA, May 2014}}.
\newblock


\bibitem[\protect\citeauthoryear{Beigi and Liu}{Beigi and Liu}{2018}]%
        {beigi2018similar}
\bibfield{author}{\bibinfo{person}{Ghazaleh Beigi} {and} \bibinfo{person}{Huan
  Liu}.} \bibinfo{year}{2018}\natexlab{}.
\newblock \showarticletitle{Similar but Different: Exploiting Users' Congruity
  for Recommendation Systems}. In \bibinfo{booktitle}{\emph{International
  Conference on Social Computing, Behavioral-Cultural Modeling, and
  Prediction}}. Springer.
\newblock


\bibitem[\protect\citeauthoryear{Beigi, Shu, Zhang, and Liu}{Beigi
  et~al\mbox{.}}{2018}]%
        {beigi2018securing}
\bibfield{author}{\bibinfo{person}{Ghazaleh Beigi}, \bibinfo{person}{Kai Shu},
  \bibinfo{person}{Yanchao Zhang}, {and} \bibinfo{person}{Huan Liu}.}
  \bibinfo{year}{2018}\natexlab{}.
\newblock \showarticletitle{Securing Social Media User Data: An Adversarial
  Approach}. In \bibinfo{booktitle}{\emph{Proceedings of the 29th on Hypertext
  and Social Media}}. \bibinfo{publisher}{ACM}, \bibinfo{pages}{165--173}.
\newblock


\bibitem[\protect\citeauthoryear{Beigi, Tang, and Liu}{Beigi
  et~al\mbox{.}}{2016a}]%
        {beigi2016signed}
\bibfield{author}{\bibinfo{person}{Ghazaleh Beigi}, \bibinfo{person}{Jiliang
  Tang}, {and} \bibinfo{person}{Huan Liu}.} \bibinfo{year}{2016}\natexlab{a}.
\newblock \showarticletitle{Signed link analysis in social media networks}. In
  \bibinfo{booktitle}{\emph{10th International Conference on Web and Social
  Media, ICWSM 2016}}. AAAI Press.
\newblock


\bibitem[\protect\citeauthoryear{Beigi, Tang, Wang, and Liu}{Beigi
  et~al\mbox{.}}{2016b}]%
        {beigi2016exploiting}
\bibfield{author}{\bibinfo{person}{Ghazaleh Beigi}, \bibinfo{person}{Jiliang
  Tang}, \bibinfo{person}{Suhang Wang}, {and} \bibinfo{person}{Huan Liu}.}
  \bibinfo{year}{2016}\natexlab{b}.
\newblock \showarticletitle{Exploiting emotional information for trust/distrust
  prediction}. In \bibinfo{booktitle}{\emph{Proceedings of the 2016 SIAM
  International Conference on Data Mining}}. SIAM, \bibinfo{pages}{81--89}.
\newblock


\bibitem[\protect\citeauthoryear{Bhagat, Cormode, Krishnamurthy, and
  Srivastava}{Bhagat et~al\mbox{.}}{2009}]%
        {bhagat2009class}
\bibfield{author}{\bibinfo{person}{Smriti Bhagat}, \bibinfo{person}{Graham
  Cormode}, \bibinfo{person}{Balachander Krishnamurthy}, {and}
  \bibinfo{person}{Divesh Srivastava}.} \bibinfo{year}{2009}\natexlab{}.
\newblock \showarticletitle{Class-based graph anonymization for social network
  data}.
\newblock \bibinfo{journal}{\emph{Proceedings of the VLDB Endowment}}
  \bibinfo{volume}{2}, \bibinfo{number}{1} (\bibinfo{year}{2009}),
  \bibinfo{pages}{766--777}.
\newblock


\bibitem[\protect\citeauthoryear{Bhagat, Weinsberg, Ioannidis, and Taft}{Bhagat
  et~al\mbox{.}}{2014}]%
        {bhagat2014recommending}
\bibfield{author}{\bibinfo{person}{Smriti Bhagat}, \bibinfo{person}{Udi
  Weinsberg}, \bibinfo{person}{Stratis Ioannidis}, {and} \bibinfo{person}{Nina
  Taft}.} \bibinfo{year}{2014}\natexlab{}.
\newblock \showarticletitle{Recommending with an agenda: Active learning of
  private attributes using matrix factorization}. In
  \bibinfo{booktitle}{\emph{Proceedings of RecSys}}. ACM.
\newblock


\bibitem[\protect\citeauthoryear{Biega, Roy, and Weikum}{Biega
  et~al\mbox{.}}{2017}]%
        {biega2017privacy}
\bibfield{author}{\bibinfo{person}{Asia~J Biega},
  \bibinfo{person}{Rishiraj~Saha Roy}, {and} \bibinfo{person}{Gerhard Weikum}.}
  \bibinfo{year}{2017}\natexlab{}.
\newblock \showarticletitle{Privacy through Solidarity: A
  User-Utility-Preserving Framework to Counter Profiling}. In
  \bibinfo{booktitle}{\emph{Proceedings of ACM SIGIR}}. ACM,
  \bibinfo{pages}{665--674}.
\newblock


\bibitem[\protect\citeauthoryear{Blei, Ng, and Jordan}{Blei
  et~al\mbox{.}}{2003}]%
        {blei2003latent}
\bibfield{author}{\bibinfo{person}{David~M Blei}, \bibinfo{person}{Andrew~Y
  Ng}, {and} \bibinfo{person}{Michael~I Jordan}.}
  \bibinfo{year}{2003}\natexlab{}.
\newblock \showarticletitle{Latent dirichlet allocation}.
\newblock \bibinfo{journal}{\emph{Journal of machine Learning research}}
  \bibinfo{volume}{3}, \bibinfo{number}{Jan} (\bibinfo{year}{2003}),
  \bibinfo{pages}{993--1022}.
\newblock


\bibitem[\protect\citeauthoryear{Bonneau, Anderson, and Danezis}{Bonneau
  et~al\mbox{.}}{2009}]%
        {bonneau2009prying}
\bibfield{author}{\bibinfo{person}{Joseph Bonneau}, \bibinfo{person}{Jonathan
  Anderson}, {and} \bibinfo{person}{George Danezis}.}
  \bibinfo{year}{2009}\natexlab{}.
\newblock \showarticletitle{Prying data out of a social network}. In
  \bibinfo{booktitle}{\emph{Social Network Analysis and Mining, 2009.
  ASONAM'09. International Conference on Advances in}}. IEEE,
  \bibinfo{pages}{249--254}.
\newblock


\bibitem[\protect\citeauthoryear{Bowers, Williams, Dozier, and Williams}{Bowers
  et~al\mbox{.}}{2015}]%
        {Bower15}
\bibfield{author}{\bibinfo{person}{Jasmine Bowers}, \bibinfo{person}{Henry
  Williams}, \bibinfo{person}{Gerry Dozier}, {and} \bibinfo{person}{R
  Williams}.} \bibinfo{year}{2015}\natexlab{}.
\newblock \showarticletitle{Mitigation deanonymization attacks via language
  translation for anonymous social networks}.
\newblock \bibinfo{journal}{\emph{Proceedings of the 7th International
  Conference on Machine Learning and Computing}} (\bibinfo{year}{2015}).
\newblock


\bibitem[\protect\citeauthoryear{Bradley, Kelley, and Roth}{Bradley
  et~al\mbox{.}}{[n. d.]}]%
        {bradley2008author}
\bibfield{author}{\bibinfo{person}{Joseph~K Bradley},
  \bibinfo{person}{Patrick~Gage Kelley}, {and} \bibinfo{person}{Aaron Roth}.}
  \bibinfo{year}{[n. d.]}\natexlab{}.
\newblock \showarticletitle{Author identification from citations}.
\newblock  (\bibinfo{year}{[n. d.]}).
\newblock


\bibitem[\protect\citeauthoryear{Bringmann, Friedrich, and Krohmer}{Bringmann
  et~al\mbox{.}}{2014}]%
        {bringmann2014anonymization}
\bibfield{author}{\bibinfo{person}{Karl Bringmann}, \bibinfo{person}{Tobias
  Friedrich}, {and} \bibinfo{person}{Anton Krohmer}.}
  \bibinfo{year}{2014}\natexlab{}.
\newblock \showarticletitle{De-anonymization of heterogeneous random graphs in
  quasilinear time}. In \bibinfo{booktitle}{\emph{European Symposium on
  Algorithms}}. Springer, \bibinfo{pages}{197--208}.
\newblock


\bibitem[\protect\citeauthoryear{Calandrino, Kilzer, Narayanan, Felten, and
  Shmatikov}{Calandrino et~al\mbox{.}}{2011}]%
        {calandrino2011you}
\bibfield{author}{\bibinfo{person}{Joseph~A Calandrino}, \bibinfo{person}{Ann
  Kilzer}, \bibinfo{person}{Arvind Narayanan}, \bibinfo{person}{Edward~W
  Felten}, {and} \bibinfo{person}{Vitaly Shmatikov}.}
  \bibinfo{year}{2011}\natexlab{}.
\newblock \showarticletitle{" You Might Also Like:" Privacy Risks of
  Collaborative Filtering}. In \bibinfo{booktitle}{\emph{Security and Privacy
  (SP)}}. IEEE.
\newblock


\bibitem[\protect\citeauthoryear{Canny}{Canny}{2002}]%
        {canny2002}
\bibfield{author}{\bibinfo{person}{John Canny}.}
  \bibinfo{year}{2002}\natexlab{}.
\newblock \showarticletitle{Collaborative filtering with privacy via factor
  analysis}. In \bibinfo{booktitle}{\emph{SIGIR}}. ACM,
  \bibinfo{pages}{238--245}.
\newblock


\bibitem[\protect\citeauthoryear{Chaabane, Acs, Kaafar, et~al\mbox{.}}{Chaabane
  et~al\mbox{.}}{2012}]%
        {chaabane2012you}
\bibfield{author}{\bibinfo{person}{Abdelberi Chaabane},
  \bibinfo{person}{Gergely Acs}, \bibinfo{person}{Mohamed~Ali Kaafar},
  {et~al\mbox{.}}} \bibinfo{year}{2012}\natexlab{}.
\newblock \showarticletitle{You are what you like! information leakage through
  users’ interests}. In \bibinfo{booktitle}{\emph{Proceedings of the 19th
  Annual Network \& Distributed System Security Symposium(NDSS)}}.
\newblock


\bibitem[\protect\citeauthoryear{Chaski}{Chaski}{2005}]%
        {chaski2005s}
\bibfield{author}{\bibinfo{person}{Carole~E Chaski}.}
  \bibinfo{year}{2005}\natexlab{}.
\newblock \showarticletitle{Who is at the keyboard? Authorship attribution in
  digital evidence investigations}.
\newblock \bibinfo{journal}{\emph{International journal of digital evidence}}
  \bibinfo{volume}{4}, \bibinfo{number}{1} (\bibinfo{year}{2005}),
  \bibinfo{pages}{1--13}.
\newblock


\bibitem[\protect\citeauthoryear{Cheng, Fu, and Liu}{Cheng
  et~al\mbox{.}}{2010b}]%
        {cheng2010k}
\bibfield{author}{\bibinfo{person}{James Cheng}, \bibinfo{person}{Ada Wai-chee
  Fu}, {and} \bibinfo{person}{Jia Liu}.} \bibinfo{year}{2010}\natexlab{b}.
\newblock \showarticletitle{K-isomorphism: privacy preserving network
  publication against structural attacks}. In
  \bibinfo{booktitle}{\emph{Proceedings of ACM SIGMOD International Conference
  on Management of data}}.
\newblock


\bibitem[\protect\citeauthoryear{Cheng, Caverlee, and Lee}{Cheng
  et~al\mbox{.}}{2010a}]%
        {cheng2010you}
\bibfield{author}{\bibinfo{person}{Zhiyuan Cheng}, \bibinfo{person}{James
  Caverlee}, {and} \bibinfo{person}{Kyumin Lee}.}
  \bibinfo{year}{2010}\natexlab{a}.
\newblock \showarticletitle{You are where you tweet: a content-based approach
  to geo-locating twitter users}. In \bibinfo{booktitle}{\emph{Proceedings of
  CIKM}}. ACM, \bibinfo{pages}{759--768}.
\newblock


\bibitem[\protect\citeauthoryear{Chiasserini, Garetto, and
  Leonardi}{Chiasserini et~al\mbox{.}}{2016}]%
        {chiasserini2016social}
\bibfield{author}{\bibinfo{person}{Carla-Fabiana Chiasserini},
  \bibinfo{person}{Michele Garetto}, {and} \bibinfo{person}{Emilio Leonardi}.}
  \bibinfo{year}{2016}\natexlab{}.
\newblock \showarticletitle{Social network de-anonymization under scale-free
  user relations}.
\newblock \bibinfo{journal}{\emph{IEEE/ACM Transactions on Networking}}
  \bibinfo{volume}{24}, \bibinfo{number}{6} (\bibinfo{year}{2016}),
  \bibinfo{pages}{3756--3769}.
\newblock


\bibitem[\protect\citeauthoryear{Chiasserini, Garetto, and
  Leonardi}{Chiasserini et~al\mbox{.}}{2018}]%
        {chiasserini2018anonymizing}
\bibfield{author}{\bibinfo{person}{Carla-Fabiana Chiasserini},
  \bibinfo{person}{Michel Garetto}, {and} \bibinfo{person}{Emili Leonardi}.}
  \bibinfo{year}{2018}\natexlab{}.
\newblock \showarticletitle{De-anonymizing clustered social networks by
  percolation graph matching}.
\newblock \bibinfo{journal}{\emph{ACM Transactions on Knowledge Discovery from
  Data (TKDD)}} \bibinfo{volume}{12}, \bibinfo{number}{2}
  (\bibinfo{year}{2018}), \bibinfo{pages}{21}.
\newblock


\bibitem[\protect\citeauthoryear{Compton, Jurgens, and Allen}{Compton
  et~al\mbox{.}}{2014}]%
        {compton2014geotagging}
\bibfield{author}{\bibinfo{person}{Ryan Compton}, \bibinfo{person}{David
  Jurgens}, {and} \bibinfo{person}{David Allen}.}
  \bibinfo{year}{2014}\natexlab{}.
\newblock \showarticletitle{Geotagging one hundred million twitter accounts
  with total variation minimization}. In \bibinfo{booktitle}{\emph{Big Data
  (Big Data), 2014 IEEE International Conference on}}. IEEE,
  \bibinfo{pages}{393--401}.
\newblock


\bibitem[\protect\citeauthoryear{Cover and Thomas}{Cover and Thomas}{2012}]%
        {cover2012elements}
\bibfield{author}{\bibinfo{person}{Thomas~M Cover} {and} \bibinfo{person}{Joy~A
  Thomas}.} \bibinfo{year}{2012}\natexlab{}.
\newblock \bibinfo{booktitle}{\emph{Elements of information theory}}.
\newblock \bibinfo{publisher}{John Wiley \& Sons}.
\newblock


\bibitem[\protect\citeauthoryear{Dey, Tang, Ross, and Saxena}{Dey
  et~al\mbox{.}}{2012}]%
        {dey2012estimating}
\bibfield{author}{\bibinfo{person}{Ratan Dey}, \bibinfo{person}{Cong Tang},
  \bibinfo{person}{Keith Ross}, {and} \bibinfo{person}{Nitesh Saxena}.}
  \bibinfo{year}{2012}\natexlab{}.
\newblock \showarticletitle{Estimating age privacy leakage in online social
  networks}. In \bibinfo{booktitle}{\emph{INFOCOM, 2012 Proceedings IEEE}}.
  IEEE, \bibinfo{pages}{2836--2840}.
\newblock


\bibitem[\protect\citeauthoryear{Dimitropoulos, Krioukov, Vahdat, and
  Riley}{Dimitropoulos et~al\mbox{.}}{2009}]%
        {dimitropoulos2009graph}
\bibfield{author}{\bibinfo{person}{Xenofontas Dimitropoulos},
  \bibinfo{person}{Dmitri Krioukov}, \bibinfo{person}{Amin Vahdat}, {and}
  \bibinfo{person}{George Riley}.} \bibinfo{year}{2009}\natexlab{}.
\newblock \showarticletitle{Graph annotations in modeling complex network
  topologies}.
\newblock \bibinfo{journal}{\emph{ACM Transactions on Modeling and Computer
  Simulation (TOMACS)}} \bibinfo{volume}{19}, \bibinfo{number}{4}
  (\bibinfo{year}{2009}), \bibinfo{pages}{17}.
\newblock


\bibitem[\protect\citeauthoryear{Dingledine, Mathewson, and
  Syverson}{Dingledine et~al\mbox{.}}{2004}]%
        {dingledine2004tor}
\bibfield{author}{\bibinfo{person}{Roger Dingledine}, \bibinfo{person}{Nick
  Mathewson}, {and} \bibinfo{person}{Paul Syverson}.}
  \bibinfo{year}{2004}\natexlab{}.
\newblock \bibinfo{booktitle}{\emph{Tor: The second-generation onion router}}.
\newblock \bibinfo{type}{{T}echnical {R}eport}. \bibinfo{institution}{Naval
  Research Lab Washington DC}.
\newblock


\bibitem[\protect\citeauthoryear{Duncan and Lambert}{Duncan and
  Lambert}{1986}]%
        {duncan1986disclosure}
\bibfield{author}{\bibinfo{person}{George~T Duncan} {and}
  \bibinfo{person}{Diane Lambert}.} \bibinfo{year}{1986}\natexlab{}.
\newblock \showarticletitle{Disclosure-limited data dissemination}.
\newblock \bibinfo{journal}{\emph{Journal of the American statistical
  association}} \bibinfo{volume}{81}, \bibinfo{number}{393}
  (\bibinfo{year}{1986}), \bibinfo{pages}{10--18}.
\newblock


\bibitem[\protect\citeauthoryear{Dwork}{Dwork}{2008}]%
        {dwork2008differential}
\bibfield{author}{\bibinfo{person}{Cynthia Dwork}.}
  \bibinfo{year}{2008}\natexlab{}.
\newblock \showarticletitle{Differential privacy: A survey of results}. In
  \bibinfo{booktitle}{\emph{International Conference on Theory and Applications
  of Models of Computation}}. Springer, \bibinfo{pages}{1--19}.
\newblock


\bibitem[\protect\citeauthoryear{Dwork, McSherry, Nissim, and Smith}{Dwork
  et~al\mbox{.}}{2006}]%
        {dwork2006calibrating}
\bibfield{author}{\bibinfo{person}{Cynthia Dwork}, \bibinfo{person}{Frank
  McSherry}, \bibinfo{person}{Kobbi Nissim}, {and} \bibinfo{person}{Adam
  Smith}.} \bibinfo{year}{2006}\natexlab{}.
\newblock \showarticletitle{Calibrating noise to sensitivity in private data
  analysis}. In \bibinfo{booktitle}{\emph{Theory of Cryptography Conference}}.
  Springer, \bibinfo{pages}{265--284}.
\newblock


\bibitem[\protect\citeauthoryear{Evfimievski, Srikant, Agrawal, and
  Gehrke}{Evfimievski et~al\mbox{.}}{2004}]%
        {evfimievski2004privacy}
\bibfield{author}{\bibinfo{person}{Alexandre Evfimievski},
  \bibinfo{person}{Ramakrishnan Srikant}, \bibinfo{person}{Rakesh Agrawal},
  {and} \bibinfo{person}{Johannes Gehrke}.} \bibinfo{year}{2004}\natexlab{}.
\newblock \showarticletitle{Privacy preserving mining of association rules}.
\newblock \bibinfo{journal}{\emph{Information Systems}} \bibinfo{volume}{29},
  \bibinfo{number}{4} (\bibinfo{year}{2004}), \bibinfo{pages}{343--364}.
\newblock


\bibitem[\protect\citeauthoryear{Fabiana, Garetto, and Leonardi}{Fabiana
  et~al\mbox{.}}{2015}]%
        {fabiana2015anonymizing}
\bibfield{author}{\bibinfo{person}{Carla Fabiana}, \bibinfo{person}{Michele
  Garetto}, {and} \bibinfo{person}{Emilio Leonardi}.}
  \bibinfo{year}{2015}\natexlab{}.
\newblock \showarticletitle{De-anonymizing scale-free social networks by
  percolation graph matching}. In \bibinfo{booktitle}{\emph{Computer
  Communications (INFOCOM), 2015 IEEE Conference on}}. IEEE,
  \bibinfo{pages}{1571--1579}.
\newblock


\bibitem[\protect\citeauthoryear{Fu, Zhang, and Xie}{Fu et~al\mbox{.}}{2014}]%
        {fu2014anonymizing}
\bibfield{author}{\bibinfo{person}{Hao Fu}, \bibinfo{person}{Aston Zhang},
  {and} \bibinfo{person}{Xing Xie}.} \bibinfo{year}{2014}\natexlab{}.
\newblock \showarticletitle{De-anonymizing social graphs via node similarity}.
  In \bibinfo{booktitle}{\emph{Proceedings of the 23rd International Conference
  on World Wide Web}}. ACM, \bibinfo{pages}{263--264}.
\newblock


\bibitem[\protect\citeauthoryear{Fu, Zhang, and Xie}{Fu et~al\mbox{.}}{2015}]%
        {fu2015effective}
\bibfield{author}{\bibinfo{person}{Hao Fu}, \bibinfo{person}{Aston Zhang},
  {and} \bibinfo{person}{Xing Xie}.} \bibinfo{year}{2015}\natexlab{}.
\newblock \showarticletitle{Effective social graph deanonymization based on
  graph structure and descriptive information}.
\newblock \bibinfo{journal}{\emph{ACM Transactions on Intelligent Systems and
  Technology (TIST)}} \bibinfo{volume}{6}, \bibinfo{number}{4}
  (\bibinfo{year}{2015}), \bibinfo{pages}{49}.
\newblock


\bibitem[\protect\citeauthoryear{Fu, Hu, Xu, Fu, and Wang}{Fu
  et~al\mbox{.}}{2017}]%
        {fu2017anonymization2}
\bibfield{author}{\bibinfo{person}{Xinzhe Fu}, \bibinfo{person}{Zhongzhao Hu},
  \bibinfo{person}{Zhiying Xu}, \bibinfo{person}{Luoyi Fu}, {and}
  \bibinfo{person}{Xinbing Wang}.} \bibinfo{year}{2017}\natexlab{}.
\newblock \showarticletitle{De-anonymization of Networks with Communities: When
  Quantifications Meet Algorithms}. In \bibinfo{booktitle}{\emph{IEEE Global
  Communications Conference}}.
\newblock


\bibitem[\protect\citeauthoryear{Fung, Wang, Chen, and Philip}{Fung
  et~al\mbox{.}}{2010}]%
        {fungprivacy}
\bibfield{author}{\bibinfo{person}{Benjamin~CM Fung}, \bibinfo{person}{K Wang},
  \bibinfo{person}{R Chen}, {and} \bibinfo{person}{S~Yu Philip}.}
  \bibinfo{year}{2010}\natexlab{}.
\newblock \showarticletitle{Privacy-Preserving Data Publishing: A Survey on
  Recent Developments}.
\newblock \bibinfo{journal}{\emph{ACM Computations Survey}}
  \bibinfo{volume}{42}, \bibinfo{number}{4} (\bibinfo{year}{2010}),
  \bibinfo{pages}{1--53}.
\newblock


\bibitem[\protect\citeauthoryear{Gambs, Killijian, and del Prado~Cortez}{Gambs
  et~al\mbox{.}}{2010}]%
        {gambs2010show}
\bibfield{author}{\bibinfo{person}{S{\'e}bastien Gambs},
  \bibinfo{person}{Marc-Olivier Killijian}, {and}
  \bibinfo{person}{Miguel~N{\'u}{\~n}ez del Prado~Cortez}.}
  \bibinfo{year}{2010}\natexlab{}.
\newblock \showarticletitle{Show me how you move and I will tell you who you
  are}. In \bibinfo{booktitle}{\emph{Proceedings of SIGSPATIAL International
  Workshop on Security and Privacy in GIS and LBS}}.
\newblock


\bibitem[\protect\citeauthoryear{Gayo~Avello}{Gayo~Avello}{2011}]%
        {gayo2011all}
\bibfield{author}{\bibinfo{person}{Daniel Gayo~Avello}.}
  \bibinfo{year}{2011}\natexlab{}.
\newblock \showarticletitle{All liaisons are dangerous when all your friends
  are known to us}. In \bibinfo{booktitle}{\emph{Proceedings of the 22nd ACM
  conference on Hypertext and hypermedia}}. ACM, \bibinfo{pages}{171--180}.
\newblock


\bibitem[\protect\citeauthoryear{Goga, Lei, Parthasarathi, Friedland, Sommer,
  and Teixeira}{Goga et~al\mbox{.}}{2013}]%
        {goga2013exploiting}
\bibfield{author}{\bibinfo{person}{Oana Goga}, \bibinfo{person}{Howard Lei},
  \bibinfo{person}{Sree Hari~Krishnan Parthasarathi}, \bibinfo{person}{Gerald
  Friedland}, \bibinfo{person}{Robin Sommer}, {and} \bibinfo{person}{Renata
  Teixeira}.} \bibinfo{year}{2013}\natexlab{}.
\newblock \showarticletitle{Exploiting innocuous activity for correlating users
  across sites}. In \bibinfo{booktitle}{\emph{Proceedings of WWW}}.
\newblock


\bibitem[\protect\citeauthoryear{Gong and Liu}{Gong and Liu}{2016}]%
        {gong2016you}
\bibfield{author}{\bibinfo{person}{Neil~Zhenqiang Gong} {and}
  \bibinfo{person}{Bin Liu}.} \bibinfo{year}{2016}\natexlab{}.
\newblock \showarticletitle{You Are Who You Know and How You Behave: Attribute
  Inference Attacks via Users' Social Friends and Behaviors.}. In
  \bibinfo{booktitle}{\emph{USENIX Security Symposium}}.
  \bibinfo{pages}{979--995}.
\newblock


\bibitem[\protect\citeauthoryear{Gong and Liu}{Gong and Liu}{2018}]%
        {gong2018attribute}
\bibfield{author}{\bibinfo{person}{Neil~Zhenqiang Gong} {and}
  \bibinfo{person}{Bin Liu}.} \bibinfo{year}{2018}\natexlab{}.
\newblock \showarticletitle{Attribute Inference Attacks in Online Social
  Networks}.
\newblock \bibinfo{journal}{\emph{ACM Transactions on Privacy and Security
  (TOPS)}} \bibinfo{volume}{21}, \bibinfo{number}{1} (\bibinfo{year}{2018}),
  \bibinfo{pages}{3}.
\newblock


\bibitem[\protect\citeauthoryear{Gong, Talwalkar, Mackey, Huang, Shin,
  Stefanov, Shi, and Song}{Gong et~al\mbox{.}}{2014}]%
        {gong2014joint}
\bibfield{author}{\bibinfo{person}{Neil~Zhenqiang Gong}, \bibinfo{person}{Ameet
  Talwalkar}, \bibinfo{person}{Lester Mackey}, \bibinfo{person}{Ling Huang},
  \bibinfo{person}{Eui Chul~Richard Shin}, \bibinfo{person}{Emil Stefanov},
  \bibinfo{person}{Elaine~Runting Shi}, {and} \bibinfo{person}{Dawn Song}.}
  \bibinfo{year}{2014}\natexlab{}.
\newblock \showarticletitle{Joint link prediction and attribute inference using
  a social-attribute network}.
\newblock \bibinfo{journal}{\emph{ACM Transactions on Intelligent Systems and
  Technology (TIST)}} \bibinfo{volume}{5}, \bibinfo{number}{2}
  (\bibinfo{year}{2014}), \bibinfo{pages}{27}.
\newblock


\bibitem[\protect\citeauthoryear{Guerraoui, Kermarrec, Patra, and
  Taziki}{Guerraoui et~al\mbox{.}}{2015}]%
        {guerraoui2015d}
\bibfield{author}{\bibinfo{person}{Rachid Guerraoui},
  \bibinfo{person}{Anne-Marie Kermarrec}, \bibinfo{person}{Rhicheek Patra},
  {and} \bibinfo{person}{Mahsa Taziki}.} \bibinfo{year}{2015}\natexlab{}.
\newblock \showarticletitle{D 2 p: distance-based differential privacy in
  recommenders}.
\newblock \bibinfo{journal}{\emph{Proceedings of the VLDB Endowment}}
  \bibinfo{volume}{8}, \bibinfo{number}{8} (\bibinfo{year}{2015}),
  \bibinfo{pages}{862--873}.
\newblock


\bibitem[\protect\citeauthoryear{Guerraoui, Kermarrec, and Taziki}{Guerraoui
  et~al\mbox{.}}{2017}]%
        {guerraoui2017utility}
\bibfield{author}{\bibinfo{person}{Rachid Guerraoui},
  \bibinfo{person}{Anne-Marie Kermarrec}, {and} \bibinfo{person}{Mahsa
  Taziki}.} \bibinfo{year}{2017}\natexlab{}.
\newblock \showarticletitle{The Utility and Privacy Effects of a Click}. In
  \bibinfo{booktitle}{\emph{Proceedings of ACM SIGIR Conference on Research and
  Development in Information Retrieval}}. ACM.
\newblock


\bibitem[\protect\citeauthoryear{Gupta, Gottipati, Jiang, and Gao}{Gupta
  et~al\mbox{.}}{2013}]%
        {gupta2013your}
\bibfield{author}{\bibinfo{person}{Payas Gupta}, \bibinfo{person}{Swapna
  Gottipati}, \bibinfo{person}{Jing Jiang}, {and} \bibinfo{person}{Debin Gao}.}
  \bibinfo{year}{2013}\natexlab{}.
\newblock \showarticletitle{Your love is public now: Questioning the use of
  personal information in authentication}. In
  \bibinfo{booktitle}{\emph{Proceedings of ACM SIGSAC}}. ACM.
\newblock


\bibitem[\protect\citeauthoryear{Hajibagheri, Sukthankar, Lakkaraju, Alvari,
  Wigand, and Agarwal}{Hajibagheri et~al\mbox{.}}{2018}]%
        {hajibagheri2018using}
\bibfield{author}{\bibinfo{person}{Alireza Hajibagheri}, \bibinfo{person}{Gita
  Sukthankar}, \bibinfo{person}{Kiran Lakkaraju}, \bibinfo{person}{Hamidreza
  Alvari}, \bibinfo{person}{Rolf~T Wigand}, {and} \bibinfo{person}{Nitin
  Agarwal}.} \bibinfo{year}{2018}\natexlab{}.
\newblock \showarticletitle{Using Massively Multiplayer Online Game Data to
  Analyze the Dynamics of Social Interactions}.
\newblock \bibinfo{journal}{\emph{Social Interactions in Virtual Worlds: An
  Interdisciplinary Perspective}} (\bibinfo{year}{2018}).
\newblock


\bibitem[\protect\citeauthoryear{Hay, Miklau, Jensen, Towsley, and Weis}{Hay
  et~al\mbox{.}}{2008}]%
        {hay2008resisting}
\bibfield{author}{\bibinfo{person}{Michael Hay}, \bibinfo{person}{Gerome
  Miklau}, \bibinfo{person}{David Jensen}, \bibinfo{person}{Don Towsley}, {and}
  \bibinfo{person}{Philipp Weis}.} \bibinfo{year}{2008}\natexlab{}.
\newblock \showarticletitle{Resisting structural re-identification in
  anonymized social networks}.
\newblock \bibinfo{journal}{\emph{Proceedings of the VLDB Endowment}}
  \bibinfo{volume}{1}, \bibinfo{number}{1} (\bibinfo{year}{2008}),
  \bibinfo{pages}{102--114}.
\newblock


\bibitem[\protect\citeauthoryear{He, Chu, and Liu}{He et~al\mbox{.}}{2006}]%
        {he2006inferring}
\bibfield{author}{\bibinfo{person}{Jianming He}, \bibinfo{person}{Wesley~W
  Chu}, {and} \bibinfo{person}{Zhenyu~Victor Liu}.}
  \bibinfo{year}{2006}\natexlab{}.
\newblock \showarticletitle{Inferring privacy information from social
  networks}. In \bibinfo{booktitle}{\emph{International Conference on
  Intelligence and Security Informatics}}. Springer, \bibinfo{pages}{154--165}.
\newblock


\bibitem[\protect\citeauthoryear{Hecht, Hong, Suh, and Chi}{Hecht
  et~al\mbox{.}}{2011}]%
        {hecht2011tweets}
\bibfield{author}{\bibinfo{person}{Brent Hecht}, \bibinfo{person}{Lichan Hong},
  \bibinfo{person}{Bongwon Suh}, {and} \bibinfo{person}{Ed~H Chi}.}
  \bibinfo{year}{2011}\natexlab{}.
\newblock \showarticletitle{Tweets from Justin Bieber's heart: the dynamics of
  the location field in user profiles}. In
  \bibinfo{booktitle}{\emph{Proceedings of the SIGCHI}}. ACM,
  \bibinfo{pages}{237--246}.
\newblock


\bibitem[\protect\citeauthoryear{Hill and Provost}{Hill and Provost}{2003}]%
        {hill2003myth}
\bibfield{author}{\bibinfo{person}{Shawndra Hill} {and} \bibinfo{person}{Foster
  Provost}.} \bibinfo{year}{2003}\natexlab{}.
\newblock \showarticletitle{The myth of the double-blind review?: author
  identification using only citations}.
\newblock \bibinfo{journal}{\emph{Acm Sigkdd Explorations Newsletter}}
  \bibinfo{volume}{5}, \bibinfo{number}{2} (\bibinfo{year}{2003}),
  \bibinfo{pages}{179--184}.
\newblock


\bibitem[\protect\citeauthoryear{Hoens, Blanton, and Chawla}{Hoens
  et~al\mbox{.}}{2010}]%
        {hoens2010private}
\bibfield{author}{\bibinfo{person}{T~Ryan Hoens}, \bibinfo{person}{Marina
  Blanton}, {and} \bibinfo{person}{Nitesh~V Chawla}.}
  \bibinfo{year}{2010}\natexlab{}.
\newblock \showarticletitle{A private and reliable recommendation system for
  social networks}. In \bibinfo{booktitle}{\emph{Social Computing (SocialCom),
  2010 IEEE Second International Conference on}}. IEEE,
  \bibinfo{pages}{816--825}.
\newblock


\bibitem[\protect\citeauthoryear{Howe and Nissenbaum}{Howe and
  Nissenbaum}{2009}]%
        {howe2009lessons}
\bibfield{author}{\bibinfo{person}{DC Howe} {and} \bibinfo{person}{H
  Nissenbaum}.} \bibinfo{year}{2009}\natexlab{}.
\newblock \showarticletitle{Lessons from the identity trail: privacy, anonymity
  and identity in a networked society}.
\newblock \bibinfo{journal}{\emph{ch. TrackMeNot: Resisting surveillance in web
  search, Oxford Univ. Press, NY}} (\bibinfo{year}{2009}),
  \bibinfo{pages}{417--436}.
\newblock


\bibitem[\protect\citeauthoryear{Hua, Xia, and Zhong}{Hua
  et~al\mbox{.}}{2015}]%
        {hua2015differentially}
\bibfield{author}{\bibinfo{person}{Jingyu Hua}, \bibinfo{person}{Chang Xia},
  {and} \bibinfo{person}{Sheng Zhong}.} \bibinfo{year}{2015}\natexlab{}.
\newblock \showarticletitle{Differentially Private Matrix Factorization.}. In
  \bibinfo{booktitle}{\emph{IJCAI}}.
\newblock


\bibitem[\protect\citeauthoryear{Humbert, Studer, Grossglauser, and
  Hubaux}{Humbert et~al\mbox{.}}{2013}]%
        {humbert2013nowhere}
\bibfield{author}{\bibinfo{person}{Mathias Humbert},
  \bibinfo{person}{Th{\'e}ophile Studer}, \bibinfo{person}{Matthias
  Grossglauser}, {and} \bibinfo{person}{Jean-Pierre Hubaux}.}
  \bibinfo{year}{2013}\natexlab{}.
\newblock \showarticletitle{Nowhere to hide: Navigating around privacy in
  online social networks}. In \bibinfo{booktitle}{\emph{European Symposium on
  Research in Computer Security}}.
\newblock


\bibitem[\protect\citeauthoryear{Indyk and Motwani}{Indyk and Motwani}{1998}]%
        {indyk1998approximate}
\bibfield{author}{\bibinfo{person}{Piotr Indyk} {and} \bibinfo{person}{Rajeev
  Motwani}.} \bibinfo{year}{1998}\natexlab{}.
\newblock \showarticletitle{Approximate nearest neighbors: towards removing the
  curse of dimensionality}. In \bibinfo{booktitle}{\emph{Proceedings of the
  thirtieth annual ACM symposium on Theory of computing}}. ACM,
  \bibinfo{pages}{604--613}.
\newblock


\bibitem[\protect\citeauthoryear{James}{James}{1992}]%
        {knowledgegraphs}
\bibfield{author}{\bibinfo{person}{P James}.} \bibinfo{year}{1992}\natexlab{}.
\newblock \showarticletitle{Knowledge Graphs}. In
  \bibinfo{booktitle}{\emph{Order 501}}.
\newblock


\bibitem[\protect\citeauthoryear{Ji, Li, Gong, Mittal, and Beyah}{Ji
  et~al\mbox{.}}{2015a}]%
        {ji2015your}
\bibfield{author}{\bibinfo{person}{Shouling Ji}, \bibinfo{person}{Weiqing Li},
  \bibinfo{person}{Neil~Zhenqiang Gong}, \bibinfo{person}{Prateek Mittal},
  {and} \bibinfo{person}{Raheem~A Beyah}.} \bibinfo{year}{2015}\natexlab{a}.
\newblock \showarticletitle{On Your Social Network De-anonymizablity:
  Quantification and Large Scale Evaluation with Seed Knowledge.}
\newblock \bibinfo{journal}{\emph{NDSS}}.
\newblock


\bibitem[\protect\citeauthoryear{Ji, Li, Gong, Mittal, and Beyah}{Ji
  et~al\mbox{.}}{2016a}]%
        {ji2016your}
\bibfield{author}{\bibinfo{person}{Shouling Ji}, \bibinfo{person}{Weiqing Li},
  \bibinfo{person}{Neil~Zhenqiang Gong}, \bibinfo{person}{Prateek Mittal},
  {and} \bibinfo{person}{Raheem~A Beyah}.} \bibinfo{year}{2016}\natexlab{a}.
\newblock \showarticletitle{Seed based Deanonymizability Quantification of
  Social Networks.}
\newblock \bibinfo{journal}{\emph{IEEE Transactions on Information Forensics \&
  Security (TIFS)}} \bibinfo{volume}{11}, \bibinfo{number}{7},
  \bibinfo{pages}{1398--1411}.
\newblock


\bibitem[\protect\citeauthoryear{Ji, Li, Mittal, and Beyah}{Ji
  et~al\mbox{.}}{2015b}]%
        {secgraph}
\bibfield{author}{\bibinfo{person}{Shouling Ji}, \bibinfo{person}{Weiqing Li},
  \bibinfo{person}{Prateek Mittal}, {and} \bibinfo{person}{Raheem Beyah}.}
  \bibinfo{year}{2015}\natexlab{b}.
\newblock \showarticletitle{SecGraph: A Uniform and Open-source Evaluation
  System for Graph Data Anonymization and De-anonymization}. In
  \bibinfo{booktitle}{\emph{USENIX Security Symposium}}.
  \bibinfo{pages}{303--318}.
\newblock


\bibitem[\protect\citeauthoryear{Ji, Li, Srivatsa, and Beyah}{Ji
  et~al\mbox{.}}{2014a}]%
        {ji2014structural}
\bibfield{author}{\bibinfo{person}{Shouling Ji}, \bibinfo{person}{Weiqing Li},
  \bibinfo{person}{Mudhakar Srivatsa}, {and} \bibinfo{person}{Raheem Beyah}.}
  \bibinfo{year}{2014}\natexlab{a}.
\newblock \showarticletitle{Structural data de-anonymization: Quantification,
  practice, and implications}. In \bibinfo{booktitle}{\emph{Proceedings of the
  2014 ACM SIGSAC}}. ACM, \bibinfo{pages}{1040--1053}.
\newblock


\bibitem[\protect\citeauthoryear{Ji, Li, Srivatsa, and Beyah}{Ji
  et~al\mbox{.}}{2016b}]%
        {ji2016structural}
\bibfield{author}{\bibinfo{person}{Shouling Ji}, \bibinfo{person}{Weiqing Li},
  \bibinfo{person}{Mudhakar Srivatsa}, {and} \bibinfo{person}{Raheem Beyah}.}
  \bibinfo{year}{2016}\natexlab{b}.
\newblock \showarticletitle{Structural data de-anonymization: theory and
  practice}.
\newblock \bibinfo{journal}{\emph{IEEE/ACM Transactions on Networking}}
  \bibinfo{volume}{24}, \bibinfo{number}{6} (\bibinfo{year}{2016}),
  \bibinfo{pages}{3523--3536}.
\newblock


\bibitem[\protect\citeauthoryear{Ji, Li, Srivatsa, He, and Beyah}{Ji
  et~al\mbox{.}}{2014b}]%
        {ji2014structure}
\bibfield{author}{\bibinfo{person}{Shouling Ji}, \bibinfo{person}{Weiqing Li},
  \bibinfo{person}{Mudhakar Srivatsa}, \bibinfo{person}{Jing~Selena He}, {and}
  \bibinfo{person}{Raheem Beyah}.} \bibinfo{year}{2014}\natexlab{b}.
\newblock \showarticletitle{Structure based data de-anonymization of social
  networks and mobility traces}. In \bibinfo{booktitle}{\emph{International
  Conference on Information Security}}. Springer.
\newblock


\bibitem[\protect\citeauthoryear{Ji, Li, Srivatsa, He, and Beyah}{Ji
  et~al\mbox{.}}{2016c}]%
        {ji2016general}
\bibfield{author}{\bibinfo{person}{Shouling Ji}, \bibinfo{person}{Weiqing Li},
  \bibinfo{person}{Mudhakar Srivatsa}, \bibinfo{person}{Jing~Selena He}, {and}
  \bibinfo{person}{Raheem Beyah}.} \bibinfo{year}{2016}\natexlab{c}.
\newblock \showarticletitle{General graph data de-anonymization: From mobility
  traces to social networks}.
\newblock \bibinfo{journal}{\emph{ACM TISS}} \bibinfo{volume}{18},
  \bibinfo{number}{4} (\bibinfo{year}{2016}).
\newblock


\bibitem[\protect\citeauthoryear{Ji, Mittal, and Beyah}{Ji
  et~al\mbox{.}}{2016d}]%
        {ji2016graph}
\bibfield{author}{\bibinfo{person}{Shouling Ji}, \bibinfo{person}{Prateek
  Mittal}, {and} \bibinfo{person}{Raheem Beyah}.}
  \bibinfo{year}{2016}\natexlab{d}.
\newblock \showarticletitle{Graph data anonymization, de-anonymization attacks,
  and de-anonymizability quantification: A survey}.
\newblock \bibinfo{journal}{\emph{IEEE Communications Surveys \& Tutorials}}
  \bibinfo{volume}{19}, \bibinfo{number}{2} (\bibinfo{year}{2016}),
  \bibinfo{pages}{1305--1326}.
\newblock


\bibitem[\protect\citeauthoryear{Ji, Wang, Chen, Li, Mittal, and Beyah}{Ji
  et~al\mbox{.}}{2017}]%
        {ji2017sag}
\bibfield{author}{\bibinfo{person}{Shouling Ji}, \bibinfo{person}{Ting Wang},
  \bibinfo{person}{Jianhai Chen}, \bibinfo{person}{Weiqing Li},
  \bibinfo{person}{Prateek Mittal}, {and} \bibinfo{person}{Raheem Beyah}.}
  \bibinfo{year}{2017}\natexlab{}.
\newblock \showarticletitle{De-SAG: On the De-anonymization of
  Structure-Attribute Graph Data}.
\newblock \bibinfo{journal}{\emph{IEEE Transactions on Dependable and Secure
  Computing}} (\bibinfo{year}{2017}).
\newblock


\bibitem[\protect\citeauthoryear{Jia, Wang, Zhang, and Gong}{Jia
  et~al\mbox{.}}{2017}]%
        {jia2017attriinfer}
\bibfield{author}{\bibinfo{person}{Jinyuan Jia}, \bibinfo{person}{Binghui
  Wang}, \bibinfo{person}{Le Zhang}, {and} \bibinfo{person}{Neil~Zhenqiang
  Gong}.} \bibinfo{year}{2017}\natexlab{}.
\newblock \showarticletitle{AttriInfer: Inferring user attributes in online
  social networks using markov random fields}. In
  \bibinfo{booktitle}{\emph{Proceedings of the WWW}}.
  \bibinfo{pages}{1561--1569}.
\newblock


\bibitem[\protect\citeauthoryear{Jorgensen and Yu}{Jorgensen and Yu}{2014}]%
        {jorgensen2014privacy}
\bibfield{author}{\bibinfo{person}{Zach Jorgensen} {and} \bibinfo{person}{Ting
  Yu}.} \bibinfo{year}{2014}\natexlab{}.
\newblock \showarticletitle{A Privacy-Preserving Framework for Personalized,
  Social Recommendations.}
\newblock \bibinfo{journal}{\emph{EDBT}}  \bibinfo{volume}{582}.
\newblock


\bibitem[\protect\citeauthoryear{Jurgens}{Jurgens}{2013}]%
        {jurgens2013s}
\bibfield{author}{\bibinfo{person}{David Jurgens}.}
  \bibinfo{year}{2013}\natexlab{}.
\newblock \showarticletitle{That's What Friends Are For: Inferring Location in
  Online Social Media Platforms Based on Social Relationships.}
\newblock  (\bibinfo{year}{2013}).
\newblock


\bibitem[\protect\citeauthoryear{Jurgens, Finethy, McCorriston, Xu, and
  Ruths}{Jurgens et~al\mbox{.}}{2015}]%
        {jurgens2015geolocation}
\bibfield{author}{\bibinfo{person}{David Jurgens}, \bibinfo{person}{Tyler
  Finethy}, \bibinfo{person}{James McCorriston}, \bibinfo{person}{Yi~Tian Xu},
  {and} \bibinfo{person}{Derek Ruths}.} \bibinfo{year}{2015}\natexlab{}.
\newblock \showarticletitle{Geolocation Prediction in Twitter Using Social
  Networks: A Critical Analysis and Review of Current Practice.}
\newblock  (\bibinfo{year}{2015}).
\newblock


\bibitem[\protect\citeauthoryear{Khairnar and Bajpai}{Khairnar and
  Bajpai}{2014}]%
        {khairnar2014anonymization}
\bibfield{author}{\bibinfo{person}{Ms~Sonali~M Khairnar} {and}
  \bibinfo{person}{Sanchika Bajpai}.} \bibinfo{year}{2014}\natexlab{}.
\newblock \showarticletitle{Anonymization of Centralized and Distributed Social
  Networks by Incremental Clustering}.
\newblock \bibinfo{journal}{\emph{International Journal of Computer Science and
  Information Technologies}} \bibinfo{volume}{5}, \bibinfo{number}{5}
  (\bibinfo{year}{2014}), \bibinfo{pages}{6724--6727}.
\newblock


\bibitem[\protect\citeauthoryear{Kifer and Machanavajjhala}{Kifer and
  Machanavajjhala}{2011}]%
        {kifer2011no}
\bibfield{author}{\bibinfo{person}{Daniel Kifer} {and} \bibinfo{person}{Ashwin
  Machanavajjhala}.} \bibinfo{year}{2011}\natexlab{}.
\newblock \showarticletitle{No free lunch in data privacy}. In
  \bibinfo{booktitle}{\emph{Proceedings of the 2011 ACM SIGMOD International
  Conference on Management of data}}. ACM, \bibinfo{pages}{193--204}.
\newblock


\bibitem[\protect\citeauthoryear{Klein and Manning}{Klein and Manning}{2003}]%
        {klein2003accurate}
\bibfield{author}{\bibinfo{person}{Dan Klein} {and}
  \bibinfo{person}{Christopher~D Manning}.} \bibinfo{year}{2003}\natexlab{}.
\newblock \showarticletitle{Accurate unlexicalized parsing}. In
  \bibinfo{booktitle}{\emph{Proceedings of the 41st annual meeting of the
  association for computational linguistics}}.
\newblock


\bibitem[\protect\citeauthoryear{Kong, Liu, and Huang}{Kong
  et~al\mbox{.}}{2014}]%
        {kong2014spot}
\bibfield{author}{\bibinfo{person}{Longbo Kong}, \bibinfo{person}{Zhi Liu},
  {and} \bibinfo{person}{Yan Huang}.} \bibinfo{year}{2014}\natexlab{}.
\newblock \showarticletitle{Spot: Locating social media users based on social
  network context}.
\newblock \bibinfo{journal}{\emph{Proceedings of the VLDB Endowment}}
  \bibinfo{volume}{7}, \bibinfo{number}{13} (\bibinfo{year}{2014}),
  \bibinfo{pages}{1681--1684}.
\newblock


\bibitem[\protect\citeauthoryear{Koppel, Schler, and Argamon}{Koppel
  et~al\mbox{.}}{2009}]%
        {koppel2009computational}
\bibfield{author}{\bibinfo{person}{Moshe Koppel}, \bibinfo{person}{Jonathan
  Schler}, {and} \bibinfo{person}{Shlomo Argamon}.}
  \bibinfo{year}{2009}\natexlab{}.
\newblock \showarticletitle{Computational methods in authorship attribution}.
\newblock \bibinfo{journal}{\emph{Journal of the Association for Information
  Science and Technology}} \bibinfo{volume}{60}, \bibinfo{number}{1}
  (\bibinfo{year}{2009}), \bibinfo{pages}{9--26}.
\newblock


\bibitem[\protect\citeauthoryear{Koppel, Schler, and Argamon}{Koppel
  et~al\mbox{.}}{2011}]%
        {koppel2011authorship}
\bibfield{author}{\bibinfo{person}{Moshe Koppel}, \bibinfo{person}{Jonathan
  Schler}, {and} \bibinfo{person}{Shlomo Argamon}.}
  \bibinfo{year}{2011}\natexlab{}.
\newblock \showarticletitle{Authorship attribution in the wild}.
\newblock \bibinfo{journal}{\emph{Language Resources and Evaluation}}
  \bibinfo{volume}{45}, \bibinfo{number}{1} (\bibinfo{year}{2011}),
  \bibinfo{pages}{83--94}.
\newblock


\bibitem[\protect\citeauthoryear{Koppel, Schler, Argamon, and Messeri}{Koppel
  et~al\mbox{.}}{2006}]%
        {koppel2006authorship}
\bibfield{author}{\bibinfo{person}{Moshe Koppel}, \bibinfo{person}{Jonathan
  Schler}, \bibinfo{person}{Shlomo Argamon}, {and} \bibinfo{person}{Eran
  Messeri}.} \bibinfo{year}{2006}\natexlab{}.
\newblock \showarticletitle{Authorship attribution with thousands of candidate
  authors}. In \bibinfo{booktitle}{\emph{Proceedings of ACM SIGIR}}. ACM,
  \bibinfo{pages}{659--660}.
\newblock


\bibitem[\protect\citeauthoryear{Korolova, Motwani, Nabar, and Xu}{Korolova
  et~al\mbox{.}}{2008}]%
        {korolova2008link}
\bibfield{author}{\bibinfo{person}{Aleksandra Korolova},
  \bibinfo{person}{Rajeev Motwani}, \bibinfo{person}{Shubha~U Nabar}, {and}
  \bibinfo{person}{Ying Xu}.} \bibinfo{year}{2008}\natexlab{}.
\newblock \showarticletitle{Link privacy in social networks}. In
  \bibinfo{booktitle}{\emph{Proceedings of the 17th ACM conference on
  Information and knowledge management}}. ACM, \bibinfo{pages}{289--298}.
\newblock


\bibitem[\protect\citeauthoryear{Korula and Lattanzi}{Korula and
  Lattanzi}{2014}]%
        {korula2014efficient}
\bibfield{author}{\bibinfo{person}{Nitish Korula} {and} \bibinfo{person}{Silvio
  Lattanzi}.} \bibinfo{year}{2014}\natexlab{}.
\newblock \showarticletitle{An efficient reconciliation algorithm for social
  networks}.
\newblock \bibinfo{journal}{\emph{Proceedings of the VLDB Endowment}}
  \bibinfo{volume}{7}, \bibinfo{number}{5} (\bibinfo{year}{2014}),
  \bibinfo{pages}{377--388}.
\newblock


\bibitem[\protect\citeauthoryear{Kosinski, Stillwell, and Graepel}{Kosinski
  et~al\mbox{.}}{2013}]%
        {kosinski2013private}
\bibfield{author}{\bibinfo{person}{Michal Kosinski}, \bibinfo{person}{David
  Stillwell}, {and} \bibinfo{person}{Thore Graepel}.}
  \bibinfo{year}{2013}\natexlab{}.
\newblock \showarticletitle{Private traits and attributes are predictable from
  digital records of human behavior}.
\newblock \bibinfo{journal}{\emph{Proceedings of the National Academy of
  Sciences}} \bibinfo{volume}{110}, \bibinfo{number}{15}
  (\bibinfo{year}{2013}), \bibinfo{pages}{5802--5805}.
\newblock


\bibitem[\protect\citeauthoryear{Kuhn}{Kuhn}{2010}]%
        {kuhn2010hungarian}
\bibfield{author}{\bibinfo{person}{Harold~W Kuhn}.}
  \bibinfo{year}{2010}\natexlab{}.
\newblock \showarticletitle{The hungarian method for the assignment problem}.
\newblock In \bibinfo{booktitle}{\emph{50 Years of Integer Programming
  1958-2008}}. \bibinfo{publisher}{Springer}, \bibinfo{pages}{29--47}.
\newblock


\bibitem[\protect\citeauthoryear{Labitzke, Werling, Mittag, and
  Hartenstein}{Labitzke et~al\mbox{.}}{2013}]%
        {labitzke2013online}
\bibfield{author}{\bibinfo{person}{Sebastian Labitzke},
  \bibinfo{person}{Florian Werling}, \bibinfo{person}{Jens Mittag}, {and}
  \bibinfo{person}{Hannes Hartenstein}.} \bibinfo{year}{2013}\natexlab{}.
\newblock \showarticletitle{Do online social network friends still threaten my
  privacy?}. In \bibinfo{booktitle}{\emph{Proceedings of the ACM conference on
  Data and application security and privacy}}.
\newblock


\bibitem[\protect\citeauthoryear{Lambert}{Lambert}{1993}]%
        {lambert1993measures}
\bibfield{author}{\bibinfo{person}{Diane Lambert}.}
  \bibinfo{year}{1993}\natexlab{}.
\newblock \showarticletitle{Measures of disclosure risk and harm}.
\newblock \bibinfo{journal}{\emph{Journal of Official Statistics}}
  \bibinfo{volume}{9}, \bibinfo{number}{2} (\bibinfo{year}{1993}),
  \bibinfo{pages}{313}.
\newblock


\bibitem[\protect\citeauthoryear{Lee, Liu, Ji, Mittal, and Lee}{Lee
  et~al\mbox{.}}{2017a}]%
        {lee2017quantify}
\bibfield{author}{\bibinfo{person}{Wei-Han Lee}, \bibinfo{person}{Changchang
  Liu}, \bibinfo{person}{Shouling Ji}, \bibinfo{person}{Prateek Mittal}, {and}
  \bibinfo{person}{Ruby Lee}.} \bibinfo{year}{2017}\natexlab{a}.
\newblock \showarticletitle{How to Quantify Graph De-anonymization Risks}.
\newblock  (\bibinfo{year}{2017}).
\newblock


\bibitem[\protect\citeauthoryear{Lee, Liu, Ji, Mittal, and Lee}{Lee
  et~al\mbox{.}}{2017b}]%
        {lee2017blind}
\bibfield{author}{\bibinfo{person}{Wei-Han Lee}, \bibinfo{person}{Changchang
  Liu}, \bibinfo{person}{Shouling Ji}, \bibinfo{person}{Prateek Mittal}, {and}
  \bibinfo{person}{Ruby~B Lee}.} \bibinfo{year}{2017}\natexlab{b}.
\newblock \showarticletitle{Blind De-anonymization Attacks using Social
  Networks}. In \bibinfo{booktitle}{\emph{Proceedings of the 2017 on Workshop
  on Privacy in the Electronic Society}}. ACM, \bibinfo{pages}{1--4}.
\newblock


\bibitem[\protect\citeauthoryear{Lewis, Kaufman, Gonzalez, Wimmer, and
  Christakis}{Lewis et~al\mbox{.}}{2008}]%
        {lewis2008tastes}
\bibfield{author}{\bibinfo{person}{Kevin Lewis}, \bibinfo{person}{Jason
  Kaufman}, \bibinfo{person}{Marco Gonzalez}, \bibinfo{person}{Andreas Wimmer},
  {and} \bibinfo{person}{Nicholas Christakis}.}
  \bibinfo{year}{2008}\natexlab{}.
\newblock \showarticletitle{Tastes, ties, and time: A new social network
  dataset using Facebook. com}.
\newblock \bibinfo{journal}{\emph{Social networks}} \bibinfo{volume}{30},
  \bibinfo{number}{4} (\bibinfo{year}{2008}), \bibinfo{pages}{330--342}.
\newblock


\bibitem[\protect\citeauthoryear{Li, Li, and Venkatasubramanian}{Li
  et~al\mbox{.}}{2007}]%
        {li2007t}
\bibfield{author}{\bibinfo{person}{Ninghui Li}, \bibinfo{person}{Tiancheng Li},
  {and} \bibinfo{person}{Suresh Venkatasubramanian}.}
  \bibinfo{year}{2007}\natexlab{}.
\newblock \showarticletitle{t-closeness: Privacy beyond k-anonymity and
  l-diversity}. In \bibinfo{booktitle}{\emph{Data Engineering, 2007. ICDE 2007.
  IEEE 23rd International Conference on}}. IEEE, \bibinfo{pages}{106--115}.
\newblock


\bibitem[\protect\citeauthoryear{Li, Wang, and Chang}{Li
  et~al\mbox{.}}{2012a}]%
        {li2012multiple}
\bibfield{author}{\bibinfo{person}{Rui Li}, \bibinfo{person}{Shengjie Wang},
  {and} \bibinfo{person}{Kevin Chen-Chuan Chang}.}
  \bibinfo{year}{2012}\natexlab{a}.
\newblock \showarticletitle{Multiple location profiling for users and
  relationships from social network and content}.
\newblock \bibinfo{journal}{\emph{Proceedings of the VLDB Endowment}}
  \bibinfo{volume}{5}, \bibinfo{number}{11} (\bibinfo{year}{2012}),
  \bibinfo{pages}{1603--1614}.
\newblock


\bibitem[\protect\citeauthoryear{Li, Wang, Deng, Wang, and Chang}{Li
  et~al\mbox{.}}{2012b}]%
        {li2012towards}
\bibfield{author}{\bibinfo{person}{Rui Li}, \bibinfo{person}{Shengjie Wang},
  \bibinfo{person}{Hongbo Deng}, \bibinfo{person}{Rui Wang}, {and}
  \bibinfo{person}{Kevin Chen-Chuan Chang}.} \bibinfo{year}{2012}\natexlab{b}.
\newblock \showarticletitle{Towards social user profiling: unified and
  discriminative influence model for inferring home locations}. In
  \bibinfo{booktitle}{\emph{Proceedings of the 18th ACM SIGKDD international
  conference on Knowledge discovery and data mining}}. ACM,
  \bibinfo{pages}{1023--1031}.
\newblock


\bibitem[\protect\citeauthoryear{Li, Cao, Shang, Liu, Tan, and Guo}{Li
  et~al\mbox{.}}{2017}]%
        {li2017inferring}
\bibfield{author}{\bibinfo{person}{Xiaoxue Li}, \bibinfo{person}{Yanan Cao},
  \bibinfo{person}{Yanmin Shang}, \bibinfo{person}{Yanbing Liu},
  \bibinfo{person}{Jianlong Tan}, {and} \bibinfo{person}{Li Guo}.}
  \bibinfo{year}{2017}\natexlab{}.
\newblock \showarticletitle{Inferring User Profiles in Online Social Networks
  Based on Convolutional Neural Network}. In \bibinfo{booktitle}{\emph{CIKM}}.
  Springer, \bibinfo{pages}{274--286}.
\newblock


\bibitem[\protect\citeauthoryear{Lindamood, Heatherly, Kantarcioglu, and
  Thuraisingham}{Lindamood et~al\mbox{.}}{2009}]%
        {lindamood2009inferring}
\bibfield{author}{\bibinfo{person}{Jack Lindamood}, \bibinfo{person}{Raymond
  Heatherly}, \bibinfo{person}{Murat Kantarcioglu}, {and}
  \bibinfo{person}{Bhavani Thuraisingham}.} \bibinfo{year}{2009}\natexlab{}.
\newblock \showarticletitle{Inferring private information using social network
  data}. In \bibinfo{booktitle}{\emph{Proceedings of WWW}}. ACM,
  \bibinfo{pages}{1145--1146}.
\newblock


\bibitem[\protect\citeauthoryear{Liu, Zhou, Zhu, Gao, and Xiang}{Liu
  et~al\mbox{.}}{2018}]%
        {liu2018location}
\bibfield{author}{\bibinfo{person}{Bo Liu}, \bibinfo{person}{Wanlei Zhou},
  \bibinfo{person}{Tianqing Zhu}, \bibinfo{person}{Longxiang Gao}, {and}
  \bibinfo{person}{Yong Xiang}.} \bibinfo{year}{2018}\natexlab{}.
\newblock \showarticletitle{Location Privacy and Its Applications: A Systematic
  Study}.
\newblock \bibinfo{journal}{\emph{IEEE Access}}  \bibinfo{volume}{6}
  (\bibinfo{year}{2018}), \bibinfo{pages}{17606--17624}.
\newblock


\bibitem[\protect\citeauthoryear{Liu, Chakraborty, and Mittal}{Liu
  et~al\mbox{.}}{2016a}]%
        {liu2016dependence}
\bibfield{author}{\bibinfo{person}{Changchang Liu}, \bibinfo{person}{Supriyo
  Chakraborty}, {and} \bibinfo{person}{Prateek Mittal}.}
  \bibinfo{year}{2016}\natexlab{a}.
\newblock \showarticletitle{Dependence Makes You Vulnberable: Differential
  Privacy Under Dependent Tuples.}. In \bibinfo{booktitle}{\emph{NDSS}},
  Vol.~\bibinfo{volume}{16}. \bibinfo{pages}{21--24}.
\newblock


\bibitem[\protect\citeauthoryear{Liu and Mittal}{Liu and Mittal}{2016}]%
        {liu2016linkmirage}
\bibfield{author}{\bibinfo{person}{Changchang Liu} {and}
  \bibinfo{person}{Prateek Mittal}.} \bibinfo{year}{2016}\natexlab{}.
\newblock \showarticletitle{LinkMirage: Enabling Privacy-preserving Analytics
  on Social Relationships.}. In \bibinfo{booktitle}{\emph{NDSS}}.
\newblock


\bibitem[\protect\citeauthoryear{Liu and Terzi}{Liu and Terzi}{2008}]%
        {liu2008towards}
\bibfield{author}{\bibinfo{person}{Kun Liu} {and} \bibinfo{person}{Evimaria
  Terzi}.} \bibinfo{year}{2008}\natexlab{}.
\newblock \showarticletitle{Towards identity anonymization on graphs}. In
  \bibinfo{booktitle}{\emph{Proceedings of the 2008 ACM SIGMOD international
  conference on Management of data}}. ACM, \bibinfo{pages}{93--106}.
\newblock


\bibitem[\protect\citeauthoryear{Liu, Ji, and Mittal}{Liu
  et~al\mbox{.}}{2016b}]%
        {liu2016smartwalk}
\bibfield{author}{\bibinfo{person}{Yushan Liu}, \bibinfo{person}{Shouling Ji},
  {and} \bibinfo{person}{Prateek Mittal}.} \bibinfo{year}{2016}\natexlab{b}.
\newblock \showarticletitle{SmartWalk: Enhancing social network security via
  adaptive random walks}. In \bibinfo{booktitle}{\emph{Proceedings of the 2016
  ACM SIGSAC Conference on Computer and Communications Security}}. ACM,
  \bibinfo{pages}{492--503}.
\newblock


\bibitem[\protect\citeauthoryear{Luo, Xu, Zha, Du, Xie, Yang, and Zhang}{Luo
  et~al\mbox{.}}{2014}]%
        {luo2014you}
\bibfield{author}{\bibinfo{person}{Dixin Luo}, \bibinfo{person}{Hongteng Xu},
  \bibinfo{person}{Hongyuan Zha}, \bibinfo{person}{Jun Du},
  \bibinfo{person}{Rong Xie}, \bibinfo{person}{Xiaokang Yang}, {and}
  \bibinfo{person}{Wenjun Zhang}.} \bibinfo{year}{2014}\natexlab{}.
\newblock \showarticletitle{You are what you watch and when you watch:
  Inferring household structures from iptv viewing data}.
\newblock \bibinfo{journal}{\emph{IEEE Transactions on Broadcasting}}
  \bibinfo{volume}{60}, \bibinfo{number}{1} (\bibinfo{year}{2014}),
  \bibinfo{pages}{61--72}.
\newblock


\bibitem[\protect\citeauthoryear{Luo and Chen}{Luo and Chen}{2014}]%
        {luo2014privacy}
\bibfield{author}{\bibinfo{person}{Zhifeng Luo} {and} \bibinfo{person}{Zhanli
  Chen}.} \bibinfo{year}{2014}\natexlab{}.
\newblock \showarticletitle{A privacy preserving group recommender based on
  cooperative perturbation}. In \bibinfo{booktitle}{\emph{International
  Conference on Cyber-Enabled Distributed Computing and Knowledge Discovery}}.
  IEEE.
\newblock


\bibitem[\protect\citeauthoryear{Machanavajjhala, Gehrke, Kifer, and
  Venkitasubramaniam}{Machanavajjhala et~al\mbox{.}}{2006}]%
        {machanavajjhala2006diversity}
\bibfield{author}{\bibinfo{person}{Ashwin Machanavajjhala},
  \bibinfo{person}{Johannes Gehrke}, \bibinfo{person}{Daniel Kifer}, {and}
  \bibinfo{person}{Muthuramakrishnan Venkitasubramaniam}.}
  \bibinfo{year}{2006}\natexlab{}.
\newblock \showarticletitle{l-diversity: Privacy beyond k-anonymity}. In
  \bibinfo{booktitle}{\emph{Proceedings of ICDE}}. IEEE,
  \bibinfo{pages}{24--24}.
\newblock


\bibitem[\protect\citeauthoryear{Machanavajjhala, Korolova, and
  Sarma}{Machanavajjhala et~al\mbox{.}}{2011}]%
        {machanavajjhala2011personalized}
\bibfield{author}{\bibinfo{person}{Ashwin Machanavajjhala},
  \bibinfo{person}{Aleksandra Korolova}, {and} \bibinfo{person}{Atish~Das
  Sarma}.} \bibinfo{year}{2011}\natexlab{}.
\newblock \showarticletitle{Personalized social recommendations: accurate or
  private}.
\newblock \bibinfo{journal}{\emph{Proceedings of the VLDB Endowment}}
  \bibinfo{volume}{4}, \bibinfo{number}{7} (\bibinfo{year}{2011}),
  \bibinfo{pages}{440--450}.
\newblock


\bibitem[\protect\citeauthoryear{Mack, Bowers, Williams, Dozier, and
  Shelton}{Mack et~al\mbox{.}}{2015}]%
        {mack2015best}
\bibfield{author}{\bibinfo{person}{Nathan Mack}, \bibinfo{person}{Jasmine
  Bowers}, \bibinfo{person}{Henry Williams}, \bibinfo{person}{Gerry Dozier},
  {and} \bibinfo{person}{Joseph Shelton}.} \bibinfo{year}{2015}\natexlab{}.
\newblock \showarticletitle{The Best Way to a Strong Defense is a Strong
  Offense: Mitigating Deanonymization Attacks via Iterative Language
  Translation}.
\newblock \bibinfo{journal}{\emph{International Journal of Machine Learning and
  Computing}} \bibinfo{volume}{5}, \bibinfo{number}{5} (\bibinfo{year}{2015}),
  \bibinfo{pages}{409}.
\newblock


\bibitem[\protect\citeauthoryear{Mahadevan, Krioukov, Fall, and
  Vahdat}{Mahadevan et~al\mbox{.}}{2006}]%
        {mahadevan2006systematic}
\bibfield{author}{\bibinfo{person}{Priya Mahadevan}, \bibinfo{person}{Dmitri
  Krioukov}, \bibinfo{person}{Kevin Fall}, {and} \bibinfo{person}{Amin
  Vahdat}.} \bibinfo{year}{2006}\natexlab{}.
\newblock \showarticletitle{Systematic topology analysis and generation using
  degree correlations}. In \bibinfo{booktitle}{\emph{ACM SIGCOMM Computer
  Communication Review}}, Vol.~\bibinfo{volume}{36}. ACM,
  \bibinfo{pages}{135--146}.
\newblock


\bibitem[\protect\citeauthoryear{Mahmud, Nichols, and Drews}{Mahmud
  et~al\mbox{.}}{2014}]%
        {mahmud2014home}
\bibfield{author}{\bibinfo{person}{Jalal Mahmud}, \bibinfo{person}{Jeffrey
  Nichols}, {and} \bibinfo{person}{Clemens Drews}.}
  \bibinfo{year}{2014}\natexlab{}.
\newblock \showarticletitle{Home location identification of twitter users}.
\newblock \bibinfo{journal}{\emph{ACM Transactions on Intelligent Systems and
  Technology (TIST)}} \bibinfo{volume}{5}, \bibinfo{number}{3}
  (\bibinfo{year}{2014}), \bibinfo{pages}{47}.
\newblock


\bibitem[\protect\citeauthoryear{Mao, Shuai, and Kapadia}{Mao
  et~al\mbox{.}}{2011}]%
        {mao2011loose}
\bibfield{author}{\bibinfo{person}{Huina Mao}, \bibinfo{person}{Xin Shuai},
  {and} \bibinfo{person}{Apu Kapadia}.} \bibinfo{year}{2011}\natexlab{}.
\newblock \showarticletitle{Loose tweets: an analysis of privacy leaks on
  twitter}. In \bibinfo{booktitle}{\emph{Proceedings of the 10th annual ACM
  workshop on Privacy in the electronic society}}. ACM, \bibinfo{pages}{1--12}.
\newblock


\bibitem[\protect\citeauthoryear{McGee, Caverlee, and Cheng}{McGee
  et~al\mbox{.}}{2013}]%
        {mcgee2013location}
\bibfield{author}{\bibinfo{person}{Jeffrey McGee}, \bibinfo{person}{James
  Caverlee}, {and} \bibinfo{person}{Zhiyuan Cheng}.}
  \bibinfo{year}{2013}\natexlab{}.
\newblock \showarticletitle{Location prediction in social media based on tie
  strength}. In \bibinfo{booktitle}{\emph{Proceedings of CIKM}}. ACM.
\newblock


\bibitem[\protect\citeauthoryear{McGee, Caverlee, and Cheng}{McGee
  et~al\mbox{.}}{2011}]%
        {mcgee2011geographic}
\bibfield{author}{\bibinfo{person}{Jeffrey McGee}, \bibinfo{person}{James~A
  Caverlee}, {and} \bibinfo{person}{Zhiyuan Cheng}.}
  \bibinfo{year}{2011}\natexlab{}.
\newblock \showarticletitle{A geographic study of tie strength in social
  media}. In \bibinfo{booktitle}{\emph{Proceedings of CIKM}}. ACM,
  \bibinfo{pages}{2333--2336}.
\newblock


\bibitem[\protect\citeauthoryear{McPherson, Smith-Lovin, and Cook}{McPherson
  et~al\mbox{.}}{2001}]%
        {mcpherson2001birds}
\bibfield{author}{\bibinfo{person}{Miller McPherson}, \bibinfo{person}{Lynn
  Smith-Lovin}, {and} \bibinfo{person}{James~M Cook}.}
  \bibinfo{year}{2001}\natexlab{}.
\newblock \showarticletitle{Birds of a feather: Homophily in social networks}.
\newblock \bibinfo{journal}{\emph{Annual review of sociology}}
  \bibinfo{volume}{27}, \bibinfo{number}{1} (\bibinfo{year}{2001}),
  \bibinfo{pages}{415--444}.
\newblock


\bibitem[\protect\citeauthoryear{McSherry and Mironov}{McSherry and
  Mironov}{2009}]%
        {mcsherry2009differentially}
\bibfield{author}{\bibinfo{person}{Frank McSherry} {and} \bibinfo{person}{Ilya
  Mironov}.} \bibinfo{year}{2009}\natexlab{}.
\newblock \showarticletitle{Differentially private recommender systems:
  Building privacy into the netflix prize contenders}. In
  \bibinfo{booktitle}{\emph{Proceedings of the ACM SIGKDD}}. ACM.
\newblock


\bibitem[\protect\citeauthoryear{McSherry and Talwar}{McSherry and
  Talwar}{2007}]%
        {mcsherry2007mechanism}
\bibfield{author}{\bibinfo{person}{Frank McSherry} {and} \bibinfo{person}{Kunal
  Talwar}.} \bibinfo{year}{2007}\natexlab{}.
\newblock \showarticletitle{Mechanism design via differential privacy}. In
  \bibinfo{booktitle}{\emph{Foundations of Computer Science, 2007. FOCS'07.
  48th Annual IEEE Symposium on}}. IEEE, \bibinfo{pages}{94--103}.
\newblock


\bibitem[\protect\citeauthoryear{Mendenhall}{Mendenhall}{1887}]%
        {mendenhall1887characteristic}
\bibfield{author}{\bibinfo{person}{Thomas~Corwin Mendenhall}.}
  \bibinfo{year}{1887}\natexlab{}.
\newblock \showarticletitle{The characteristic curves of composition}.
\newblock \bibinfo{journal}{\emph{Science}} \bibinfo{volume}{9},
  \bibinfo{number}{214} (\bibinfo{year}{1887}), \bibinfo{pages}{237--249}.
\newblock


\bibitem[\protect\citeauthoryear{Meng, Wang, Shu, Li, Chen, Liu, and
  Zhang}{Meng et~al\mbox{.}}{2018}]%
        {meng2018personalized}
\bibfield{author}{\bibinfo{person}{Xuying Meng}, \bibinfo{person}{Suhang Wang},
  \bibinfo{person}{Kai Shu}, \bibinfo{person}{Jundong Li}, \bibinfo{person}{Bo
  Chen}, \bibinfo{person}{Huan Liu}, {and} \bibinfo{person}{Yujun Zhang}.}
  \bibinfo{year}{2018}\natexlab{}.
\newblock \showarticletitle{Personalized privacy-preserving social
  recommendation}. In \bibinfo{booktitle}{\emph{AAAI}}.
\newblock


\bibitem[\protect\citeauthoryear{Mikolov, Sutskever, Chen, Corrado, and
  Dean}{Mikolov et~al\mbox{.}}{2013}]%
        {mikolov2013distributed}
\bibfield{author}{\bibinfo{person}{Tomas Mikolov}, \bibinfo{person}{Ilya
  Sutskever}, \bibinfo{person}{Kai Chen}, \bibinfo{person}{Greg~S Corrado},
  {and} \bibinfo{person}{Jeff Dean}.} \bibinfo{year}{2013}\natexlab{}.
\newblock \showarticletitle{Distributed representations of words and phrases
  and their compositionality}. In \bibinfo{booktitle}{\emph{Advances in neural
  information processing systems}}. \bibinfo{pages}{3111--3119}.
\newblock


\bibitem[\protect\citeauthoryear{Minkus, Ding, Dey, and Ross}{Minkus
  et~al\mbox{.}}{2015}]%
        {minkus2015city}
\bibfield{author}{\bibinfo{person}{Tehila Minkus}, \bibinfo{person}{Yuan Ding},
  \bibinfo{person}{Ratan Dey}, {and} \bibinfo{person}{Keith~W Ross}.}
  \bibinfo{year}{2015}\natexlab{}.
\newblock \showarticletitle{The city privacy attack: Combining social media and
  public records for detailed profiles of adults and children}. In
  \bibinfo{booktitle}{\emph{Proceedings of the 2015 ACM on Conference on Online
  Social Networks}}. ACM, \bibinfo{pages}{71--81}.
\newblock


\bibitem[\protect\citeauthoryear{Mislove, Viswanath, Gummadi, and
  Druschel}{Mislove et~al\mbox{.}}{2010}]%
        {mislove2010you}
\bibfield{author}{\bibinfo{person}{Alan Mislove}, \bibinfo{person}{Bimal
  Viswanath}, \bibinfo{person}{Krishna~P Gummadi}, {and} \bibinfo{person}{Peter
  Druschel}.} \bibinfo{year}{2010}\natexlab{}.
\newblock \showarticletitle{You are who you know: inferring user profiles in
  online social networks}. In \bibinfo{booktitle}{\emph{Proceedings of WSDM}}.
  ACM, \bibinfo{pages}{251--260}.
\newblock


\bibitem[\protect\citeauthoryear{Mittal, Papamanthou, and Song}{Mittal
  et~al\mbox{.}}{2012}]%
        {mittal2012preserving}
\bibfield{author}{\bibinfo{person}{Prateek Mittal},
  \bibinfo{person}{Charalampos Papamanthou}, {and} \bibinfo{person}{Dawn
  Song}.} \bibinfo{year}{2012}\natexlab{}.
\newblock \showarticletitle{Preserving link privacy in social network based
  systems}.
\newblock \bibinfo{journal}{\emph{arXiv preprint arXiv:1208.6189}}
  (\bibinfo{year}{2012}).
\newblock


\bibitem[\protect\citeauthoryear{Mosteller and Wallace}{Mosteller and
  Wallace}{1964}]%
        {mosteller1964inference}
\bibfield{author}{\bibinfo{person}{Frederick Mosteller} {and}
  \bibinfo{person}{David Wallace}.} \bibinfo{year}{1964}\natexlab{}.
\newblock \showarticletitle{Inference and disputed authorship: The Federalist}.
\newblock  (\bibinfo{year}{1964}).
\newblock


\bibitem[\protect\citeauthoryear{Nanavati, Taylor, Aiello, and
  Warfield}{Nanavati et~al\mbox{.}}{2011}]%
        {nanavati2011herbert}
\bibfield{author}{\bibinfo{person}{Mihir Nanavati}, \bibinfo{person}{Nathan
  Taylor}, \bibinfo{person}{William Aiello}, {and} \bibinfo{person}{Andrew
  Warfield}.} \bibinfo{year}{2011}\natexlab{}.
\newblock \showarticletitle{Herbert West-Deanonymizer.}
  \bibinfo{publisher}{HotSec}.
\newblock


\bibitem[\protect\citeauthoryear{Narayanan, Paskov, Gong, Bethencourt,
  Stefanov, Shin, and Song}{Narayanan et~al\mbox{.}}{2012}]%
        {narayanan2012feasibility}
\bibfield{author}{\bibinfo{person}{Arvind Narayanan}, \bibinfo{person}{Hristo
  Paskov}, \bibinfo{person}{Neil~Zhenqiang Gong}, \bibinfo{person}{John
  Bethencourt}, \bibinfo{person}{Emil Stefanov}, \bibinfo{person}{Eui
  Chul~Richard Shin}, {and} \bibinfo{person}{Dawn Song}.}
  \bibinfo{year}{2012}\natexlab{}.
\newblock \showarticletitle{On the feasibility of internet-scale author
  identification}. In \bibinfo{booktitle}{\emph{Security and Privacy (SP)}}.
  IEEE.
\newblock


\bibitem[\protect\citeauthoryear{Narayanan, Shi, and Rubinstein}{Narayanan
  et~al\mbox{.}}{2011}]%
        {narayanan2011link}
\bibfield{author}{\bibinfo{person}{Arvind Narayanan}, \bibinfo{person}{Elaine
  Shi}, {and} \bibinfo{person}{Benjamin~IP Rubinstein}.}
  \bibinfo{year}{2011}\natexlab{}.
\newblock \showarticletitle{Link prediction by de-anonymization: How we won the
  kaggle social network challenge}. In \bibinfo{booktitle}{\emph{International
  Joint Conference on Neural Networks}}. IEEE.
\newblock


\bibitem[\protect\citeauthoryear{Narayanan and Shmatikov}{Narayanan and
  Shmatikov}{2008}]%
        {narayanan2008robust}
\bibfield{author}{\bibinfo{person}{Arvind Narayanan} {and}
  \bibinfo{person}{Vitaly Shmatikov}.} \bibinfo{year}{2008}\natexlab{}.
\newblock \showarticletitle{Robust de-anonymization of large sparse datasets}.
  In \bibinfo{booktitle}{\emph{Security and Privacy}}. IEEE.
\newblock


\bibitem[\protect\citeauthoryear{Narayanan and Shmatikov}{Narayanan and
  Shmatikov}{2009}]%
        {narayanan2009anonymizing}
\bibfield{author}{\bibinfo{person}{Arvind Narayanan} {and}
  \bibinfo{person}{Vitaly Shmatikov}.} \bibinfo{year}{2009}\natexlab{}.
\newblock \showarticletitle{De-anonymizing social networks}. In
  \bibinfo{booktitle}{\emph{Security and Privacy}}. IEEE.
\newblock


\bibitem[\protect\citeauthoryear{Newman}{Newman}{2003}]%
        {CMMOdel}
\bibfield{author}{\bibinfo{person}{M.E.J Newman}.}
  \bibinfo{year}{2003}\natexlab{}.
\newblock \showarticletitle{The structure and function of complex networks}. In
  \bibinfo{booktitle}{\emph{SIAM Review}}, Vol.~\bibinfo{volume}{45}.
  \bibinfo{pages}{167--256}.
\newblock


\bibitem[\protect\citeauthoryear{Nilizadeh, Kapadia, and Ahn}{Nilizadeh
  et~al\mbox{.}}{2014}]%
        {nilizadeh2014community}
\bibfield{author}{\bibinfo{person}{Shirin Nilizadeh}, \bibinfo{person}{Apu
  Kapadia}, {and} \bibinfo{person}{Yong-Yeol Ahn}.}
  \bibinfo{year}{2014}\natexlab{}.
\newblock \showarticletitle{Community-enhanced de-anonymization of online
  social networks}. In \bibinfo{booktitle}{\emph{Proceedings of the 2014 acm
  sigsac conference on computer and communications security}}. ACM,
  \bibinfo{pages}{537--548}.
\newblock


\bibitem[\protect\citeauthoryear{Nissim, Raskhodnikova, and Smith}{Nissim
  et~al\mbox{.}}{2007}]%
        {nissim2007smooth}
\bibfield{author}{\bibinfo{person}{Kobbi Nissim}, \bibinfo{person}{Sofya
  Raskhodnikova}, {and} \bibinfo{person}{Adam Smith}.}
  \bibinfo{year}{2007}\natexlab{}.
\newblock \showarticletitle{Smooth sensitivity and sampling in private data
  analysis}. In \bibinfo{booktitle}{\emph{Proceedings of the thirty-ninth
  annual ACM symposium on Theory of computing}}. ACM, \bibinfo{pages}{75--84}.
\newblock


\bibitem[\protect\citeauthoryear{Otterbacher}{Otterbacher}{2010}]%
        {otterbacher2010inferring}
\bibfield{author}{\bibinfo{person}{Jahna Otterbacher}.}
  \bibinfo{year}{2010}\natexlab{}.
\newblock \showarticletitle{Inferring gender of movie reviewers: exploiting
  writing style, content and metadata}. In
  \bibinfo{booktitle}{\emph{Proceedings of the 19th ACM international
  conference on Information and knowledge management}}. ACM,
  \bibinfo{pages}{369--378}.
\newblock


\bibitem[\protect\citeauthoryear{Parameswaran and Blough}{Parameswaran and
  Blough}{2005}]%
        {parameswaran2005robust}
\bibfield{author}{\bibinfo{person}{Rupa Parameswaran} {and} \bibinfo{person}{D
  Blough}.} \bibinfo{year}{2005}\natexlab{}.
\newblock \showarticletitle{A robust data obfuscation approach for privacy
  preservation of clustered data}. In \bibinfo{booktitle}{\emph{In Workshop on
  Privacy and Security Aspects of Data Mining}}. \bibinfo{pages}{18--25}.
\newblock


\bibitem[\protect\citeauthoryear{Parameswaran and Blough}{Parameswaran and
  Blough}{2007}]%
        {parameswaran2007privacy}
\bibfield{author}{\bibinfo{person}{Rupa Parameswaran} {and}
  \bibinfo{person}{Douglas~M Blough}.} \bibinfo{year}{2007}\natexlab{}.
\newblock \showarticletitle{Privacy preserving collaborative filtering using
  data obfuscation}. In \bibinfo{booktitle}{\emph{IEEE International Conference
  on Granular Computing}}.
\newblock


\bibitem[\protect\citeauthoryear{Parra-Arnau}{Parra-Arnau}{2017}]%
        {parra2017pay}
\bibfield{author}{\bibinfo{person}{Javier Parra-Arnau}.}
  \bibinfo{year}{2017}\natexlab{}.
\newblock \showarticletitle{Pay-per-tracking: A collaborative masking model for
  web browsing}.
\newblock \bibinfo{journal}{\emph{Information Sciences}}  \bibinfo{volume}{385}
  (\bibinfo{year}{2017}), \bibinfo{pages}{96--124}.
\newblock


\bibitem[\protect\citeauthoryear{Parra-Arnau, Rebollo-Monedero, and
  Forn{\'e}}{Parra-Arnau et~al\mbox{.}}{2014}]%
        {parra2014optimal}
\bibfield{author}{\bibinfo{person}{Javier Parra-Arnau}, \bibinfo{person}{David
  Rebollo-Monedero}, {and} \bibinfo{person}{Jordi Forn{\'e}}.}
  \bibinfo{year}{2014}\natexlab{}.
\newblock \showarticletitle{Optimal forgery and suppression of ratings for
  privacy enhancement in recommendation systems}.
\newblock \bibinfo{journal}{\emph{Entropy}} \bibinfo{volume}{16},
  \bibinfo{number}{3} (\bibinfo{year}{2014}), \bibinfo{pages}{1586--1631}.
\newblock


\bibitem[\protect\citeauthoryear{Pedarsani and Grossglauser}{Pedarsani and
  Grossglauser}{2011}]%
        {pedarsani2011privacy}
\bibfield{author}{\bibinfo{person}{Pedram Pedarsani} {and}
  \bibinfo{person}{Matthias Grossglauser}.} \bibinfo{year}{2011}\natexlab{}.
\newblock \showarticletitle{On the privacy of anonymized networks}. In
  \bibinfo{booktitle}{\emph{Proceedings of the 17th ACM SIGKDD international
  conference on Knowledge discovery and data mining}}. ACM,
  \bibinfo{pages}{1235--1243}.
\newblock


\bibitem[\protect\citeauthoryear{Peng, Li, Zou, and Wu}{Peng
  et~al\mbox{.}}{2014}]%
        {peng2014two}
\bibfield{author}{\bibinfo{person}{Wei Peng}, \bibinfo{person}{Feng Li},
  \bibinfo{person}{Xukai Zou}, {and} \bibinfo{person}{Jie Wu}.}
  \bibinfo{year}{2014}\natexlab{}.
\newblock \showarticletitle{A two-stage deanonymization attack against
  anonymized social networks}.
\newblock \bibinfo{journal}{\emph{IEEE Trans. Comput.}} \bibinfo{volume}{63},
  \bibinfo{number}{2} (\bibinfo{year}{2014}), \bibinfo{pages}{290--303}.
\newblock


\bibitem[\protect\citeauthoryear{Polat and Du}{Polat and Du}{2003}]%
        {polat2003privacy}
\bibfield{author}{\bibinfo{person}{Huseyin Polat} {and}
  \bibinfo{person}{Wenliang Du}.} \bibinfo{year}{2003}\natexlab{}.
\newblock \showarticletitle{Privacy-preserving collaborative filtering using
  randomized perturbation techniques}. In \bibinfo{booktitle}{\emph{Data
  Mining, 2003. ICDM 2003. Third IEEE International Conference on}}. IEEE,
  \bibinfo{pages}{625--628}.
\newblock


\bibitem[\protect\citeauthoryear{Proserpio, Goldberg, and McSherry}{Proserpio
  et~al\mbox{.}}{2014}]%
        {proserpio2014calibrating}
\bibfield{author}{\bibinfo{person}{Davide Proserpio}, \bibinfo{person}{Sharon
  Goldberg}, {and} \bibinfo{person}{Frank McSherry}.}
  \bibinfo{year}{2014}\natexlab{}.
\newblock \showarticletitle{Calibrating data to sensitivity in private data
  analysis: a platform for differentially-private analysis of weighted
  datasets}.
\newblock \bibinfo{journal}{\emph{Proceedings of the VLDB Endowment}}
  \bibinfo{volume}{7}, \bibinfo{number}{8} (\bibinfo{year}{2014}).
\newblock


\bibitem[\protect\citeauthoryear{Puglisi, Parra-Arnau, Forn{\'e}, and
  Rebollo-Monedero}{Puglisi et~al\mbox{.}}{2015}]%
        {puglisi2015content}
\bibfield{author}{\bibinfo{person}{Silvia Puglisi}, \bibinfo{person}{Javier
  Parra-Arnau}, \bibinfo{person}{Jordi Forn{\'e}}, {and} \bibinfo{person}{David
  Rebollo-Monedero}.} \bibinfo{year}{2015}\natexlab{}.
\newblock \showarticletitle{On content-based recommendation and user privacy in
  social-tagging systems}.
\newblock \bibinfo{journal}{\emph{Computer Standards \& Interfaces}}
  \bibinfo{volume}{41} (\bibinfo{year}{2015}), \bibinfo{pages}{17--27}.
\newblock


\bibitem[\protect\citeauthoryear{Qian, Li, Zhang, and Chen}{Qian
  et~al\mbox{.}}{2016}]%
        {qian2016anonymizing}
\bibfield{author}{\bibinfo{person}{Jianwei Qian}, \bibinfo{person}{Xiang-Yang
  Li}, \bibinfo{person}{Chunhong Zhang}, {and} \bibinfo{person}{Linlin Chen}.}
  \bibinfo{year}{2016}\natexlab{}.
\newblock \showarticletitle{De-anonymizing social networks and inferring
  private attributes using knowledge graphs}. In \bibinfo{booktitle}{\emph{IEEE
  INFOCOM}}.
\newblock


\bibitem[\protect\citeauthoryear{Ramakrishnan, Keller, Mirza, Grama, and
  Karypis}{Ramakrishnan et~al\mbox{.}}{2001}]%
        {ramakrishnan2001privacy}
\bibfield{author}{\bibinfo{person}{Naren Ramakrishnan},
  \bibinfo{person}{Benjamin~J Keller}, \bibinfo{person}{Batul~J Mirza},
  \bibinfo{person}{Ananth~Y Grama}, {and} \bibinfo{person}{George Karypis}.}
  \bibinfo{year}{2001}\natexlab{}.
\newblock \showarticletitle{Privacy risks in recommender systems}.
\newblock \bibinfo{journal}{\emph{IEEE Internet Computing}}
  \bibinfo{volume}{5}, \bibinfo{number}{6} (\bibinfo{year}{2001}),
  \bibinfo{pages}{54}.
\newblock


\bibitem[\protect\citeauthoryear{Rao, Rohatgi, et~al\mbox{.}}{Rao
  et~al\mbox{.}}{2000}]%
        {rao2000can}
\bibfield{author}{\bibinfo{person}{Josyula~R Rao}, \bibinfo{person}{Pankaj
  Rohatgi}, {et~al\mbox{.}}} \bibinfo{year}{2000}\natexlab{}.
\newblock \showarticletitle{Can pseudonymity really guarantee privacy?}. In
  \bibinfo{booktitle}{\emph{USENIX Security}}.
\newblock


\bibitem[\protect\citeauthoryear{Rebollo-Monedero, Parra-Arnau, and
  Forn{\'e}}{Rebollo-Monedero et~al\mbox{.}}{2011}]%
        {rebollo2011information}
\bibfield{author}{\bibinfo{person}{David Rebollo-Monedero},
  \bibinfo{person}{Javier Parra-Arnau}, {and} \bibinfo{person}{Jordi
  Forn{\'e}}.} \bibinfo{year}{2011}\natexlab{}.
\newblock \showarticletitle{An information-theoretic privacy criterion for
  query forgery in information retrieval}. In
  \bibinfo{booktitle}{\emph{International Conference on Security Technology}}.
  Springer, \bibinfo{pages}{146--154}.
\newblock


\bibitem[\protect\citeauthoryear{Rizvi and Haritsa}{Rizvi and Haritsa}{2002}]%
        {rizvi2002maintaining}
\bibfield{author}{\bibinfo{person}{Shariq~J Rizvi} {and}
  \bibinfo{person}{Jayant~R Haritsa}.} \bibinfo{year}{2002}\natexlab{}.
\newblock \showarticletitle{Maintaining data privacy in association rule
  mining}. In \bibinfo{booktitle}{\emph{VLDB'02: Proceedings of the 28th
  International Conference on Very Large Databases}}. Elsevier,
  \bibinfo{pages}{682--693}.
\newblock


\bibitem[\protect\citeauthoryear{Rout, Bontcheva, Preo{\c{t}}iuc-Pietro, and
  Cohn}{Rout et~al\mbox{.}}{2013}]%
        {rout2013s}
\bibfield{author}{\bibinfo{person}{Dominic Rout}, \bibinfo{person}{Kalina
  Bontcheva}, \bibinfo{person}{Daniel Preo{\c{t}}iuc-Pietro}, {and}
  \bibinfo{person}{Trevor Cohn}.} \bibinfo{year}{2013}\natexlab{}.
\newblock \showarticletitle{Where's@ wally?: a classification approach to
  geolocating users based on their social ties}. In
  \bibinfo{booktitle}{\emph{Proceedings of Hypertext and Social Media}}. ACM.
\newblock


\bibitem[\protect\citeauthoryear{Ryoo and Moon}{Ryoo and Moon}{2014}]%
        {ryoo2014inferring}
\bibfield{author}{\bibinfo{person}{KyoungMin Ryoo} {and} \bibinfo{person}{Sue
  Moon}.} \bibinfo{year}{2014}\natexlab{}.
\newblock \showarticletitle{Inferring twitter user locations with 10 km
  accuracy}. In \bibinfo{booktitle}{\emph{Proceedings of the 23rd International
  Conference on World Wide Web}}. ACM, \bibinfo{pages}{643--648}.
\newblock


\bibitem[\protect\citeauthoryear{Sala, Zhao, Wilson, Zheng, and Zhao}{Sala
  et~al\mbox{.}}{2011}]%
        {sala2011sharing}
\bibfield{author}{\bibinfo{person}{Alessandra Sala}, \bibinfo{person}{Xiaohan
  Zhao}, \bibinfo{person}{Christo Wilson}, \bibinfo{person}{Haitao Zheng},
  {and} \bibinfo{person}{Ben~Y Zhao}.} \bibinfo{year}{2011}\natexlab{}.
\newblock \showarticletitle{Sharing graphs using differentially private graph
  models}. In \bibinfo{booktitle}{\emph{Proceedings of ACM SIGCOMM on Internet
  measurement conference}}.
\newblock


\bibitem[\protect\citeauthoryear{Sharad}{Sharad}{2016}]%
        {sharad2016change}
\bibfield{author}{\bibinfo{person}{Kumar Sharad}.}
  \bibinfo{year}{2016}\natexlab{}.
\newblock \showarticletitle{Change of Guard: The Next Generation of Social
  Graph De-anonymization Attacks}. In \bibinfo{booktitle}{\emph{Proceedings of
  the 2016 ACM Workshop on Artificial Intelligence and Security}}. ACM,
  \bibinfo{pages}{105--116}.
\newblock


\bibitem[\protect\citeauthoryear{Sharad and Danezis}{Sharad and
  Danezis}{2014}]%
        {sharad2014automated}
\bibfield{author}{\bibinfo{person}{Kumar Sharad} {and} \bibinfo{person}{George
  Danezis}.} \bibinfo{year}{2014}\natexlab{}.
\newblock \showarticletitle{An automated social graph de-anonymization
  technique}. In \bibinfo{booktitle}{\emph{Proceedings of the 13th Workshop on
  Privacy in the Electronic Society}}. ACM, \bibinfo{pages}{47--58}.
\newblock


\bibitem[\protect\citeauthoryear{Sharma, Gupta, and Bhatnagar}{Sharma
  et~al\mbox{.}}{2012}]%
        {sharma2012anonymisation}
\bibfield{author}{\bibinfo{person}{Sanur Sharma}, \bibinfo{person}{Preeti
  Gupta}, {and} \bibinfo{person}{Vishal Bhatnagar}.}
  \bibinfo{year}{2012}\natexlab{}.
\newblock \showarticletitle{Anonymisation in social network: A literature
  survey and classification}.
\newblock \bibinfo{journal}{\emph{International Journal of Social Network
  Mining}} \bibinfo{volume}{1}, \bibinfo{number}{1} (\bibinfo{year}{2012}),
  \bibinfo{pages}{51--66}.
\newblock


\bibitem[\protect\citeauthoryear{Shen and Jin}{Shen and Jin}{2014}]%
        {shen2014privacy}
\bibfield{author}{\bibinfo{person}{Yilin Shen} {and} \bibinfo{person}{Hongxia
  Jin}.} \bibinfo{year}{2014}\natexlab{}.
\newblock \showarticletitle{Privacy-preserving personalized recommendation: An
  instance-based approach via differential privacy}. In
  \bibinfo{booktitle}{\emph{Data Mining (ICDM), 2014 IEEE International
  Conference on}}. IEEE, \bibinfo{pages}{540--549}.
\newblock


\bibitem[\protect\citeauthoryear{Shokri, Theodorakopoulos, Le~Boudec, and
  Hubaux}{Shokri et~al\mbox{.}}{2011}]%
        {shokri2011quantifying}
\bibfield{author}{\bibinfo{person}{Reza Shokri}, \bibinfo{person}{George
  Theodorakopoulos}, \bibinfo{person}{Jean-Yves Le~Boudec}, {and}
  \bibinfo{person}{Jean-Pierre Hubaux}.} \bibinfo{year}{2011}\natexlab{}.
\newblock \showarticletitle{Quantifying location privacy}. In
  \bibinfo{booktitle}{\emph{Security and privacy (sp), 2011 ieee symposium
  on}}. IEEE, \bibinfo{pages}{247--262}.
\newblock


\bibitem[\protect\citeauthoryear{Shu, Wang, Tang, Zafarani, and Liu}{Shu
  et~al\mbox{.}}{2017}]%
        {shu2017user}
\bibfield{author}{\bibinfo{person}{Kai Shu}, \bibinfo{person}{Suhang Wang},
  \bibinfo{person}{Jiliang Tang}, \bibinfo{person}{Reza Zafarani}, {and}
  \bibinfo{person}{Huan Liu}.} \bibinfo{year}{2017}\natexlab{}.
\newblock \showarticletitle{User identity linkage across online social
  networks: A review}.
\newblock \bibinfo{journal}{\emph{ACM SIGKDD Explorations Newsletter}}
  \bibinfo{volume}{18}, \bibinfo{number}{2} (\bibinfo{year}{2017}),
  \bibinfo{pages}{5--17}.
\newblock


\bibitem[\protect\citeauthoryear{Srivatsa and Hicks}{Srivatsa and
  Hicks}{2012}]%
        {srivatsa2012deanonymizing}
\bibfield{author}{\bibinfo{person}{Mudhakar Srivatsa} {and}
  \bibinfo{person}{Mike Hicks}.} \bibinfo{year}{2012}\natexlab{}.
\newblock \showarticletitle{Deanonymizing mobility traces: Using social network
  as a side-channel}. In \bibinfo{booktitle}{\emph{Proceedings of the 2012 ACM
  conference on Computer and communications security}}. ACM,
  \bibinfo{pages}{628--637}.
\newblock


\bibitem[\protect\citeauthoryear{Stamatatos}{Stamatatos}{2009}]%
        {stamatatos2009survey}
\bibfield{author}{\bibinfo{person}{Efstathios Stamatatos}.}
  \bibinfo{year}{2009}\natexlab{}.
\newblock \showarticletitle{A survey of modern authorship attribution methods}.
\newblock \bibinfo{journal}{\emph{Journal of the Association for Information
  Science and Technology}} \bibinfo{volume}{60}, \bibinfo{number}{3}
  (\bibinfo{year}{2009}), \bibinfo{pages}{538--556}.
\newblock


\bibitem[\protect\citeauthoryear{Stone, Zickler, and Darrell}{Stone
  et~al\mbox{.}}{2008}]%
        {stone2008autotagging}
\bibfield{author}{\bibinfo{person}{Zak Stone}, \bibinfo{person}{Todd Zickler},
  {and} \bibinfo{person}{Trevor Darrell}.} \bibinfo{year}{2008}\natexlab{}.
\newblock \showarticletitle{Autotagging facebook: Social network context
  improves photo annotation}. In \bibinfo{booktitle}{\emph{IEEE Computer
  Society Conference on Computer Vision and Pattern Recognition Workshops}}.
\newblock


\bibitem[\protect\citeauthoryear{Sweeney}{Sweeney}{2002}]%
        {sweeney2002k}
\bibfield{author}{\bibinfo{person}{Latanya Sweeney}.}
  \bibinfo{year}{2002}\natexlab{}.
\newblock \showarticletitle{k-anonymity: A model for protecting privacy}.
\newblock \bibinfo{journal}{\emph{International Journal of Uncertainty,
  Fuzziness and Knowledge-Based Systems}} \bibinfo{volume}{10},
  \bibinfo{number}{05} (\bibinfo{year}{2002}), \bibinfo{pages}{557--570}.
\newblock


\bibitem[\protect\citeauthoryear{Tang and Wang}{Tang and Wang}{2016}]%
        {tang2016privacy}
\bibfield{author}{\bibinfo{person}{Qiang Tang} {and} \bibinfo{person}{Jun
  Wang}.} \bibinfo{year}{2016}\natexlab{}.
\newblock \showarticletitle{Privacy-preserving friendship-based recommender
  systems}.
\newblock \bibinfo{journal}{\emph{IEEE Transactions on Dependable and Secure
  Computing}} (\bibinfo{year}{2016}).
\newblock


\bibitem[\protect\citeauthoryear{Thomas, Grier, and Nicol}{Thomas
  et~al\mbox{.}}{2010}]%
        {thomas2010unfriendly}
\bibfield{author}{\bibinfo{person}{Kurt Thomas}, \bibinfo{person}{Chris Grier},
  {and} \bibinfo{person}{David~M Nicol}.} \bibinfo{year}{2010}\natexlab{}.
\newblock \showarticletitle{unfriendly: Multi-party privacy risks in social
  networks}. In \bibinfo{booktitle}{\emph{International Symposium on Privacy
  Enhancing Technologies Symposium}}. Springer, \bibinfo{pages}{236--252}.
\newblock


\bibitem[\protect\citeauthoryear{Thompson and Yao}{Thompson and Yao}{2009}]%
        {thompson2009union}
\bibfield{author}{\bibinfo{person}{Brian Thompson} {and}
  \bibinfo{person}{Danfeng Yao}.} \bibinfo{year}{2009}\natexlab{}.
\newblock \showarticletitle{The union-split algorithm and cluster-based
  anonymization of social networks}. In \bibinfo{booktitle}{\emph{Proceedings
  of Symposium on Information, Computer, and Communications Security}}.
\newblock


\bibitem[\protect\citeauthoryear{Tong, Faloutsos, and Pan}{Tong
  et~al\mbox{.}}{2006}]%
        {tong2006fast}
\bibfield{author}{\bibinfo{person}{Hanghang Tong}, \bibinfo{person}{Christos
  Faloutsos}, {and} \bibinfo{person}{Jia-Yu Pan}.}
  \bibinfo{year}{2006}\natexlab{}.
\newblock \showarticletitle{Fast random walk with restart and its
  applications}.
\newblock  (\bibinfo{year}{2006}).
\newblock


\bibitem[\protect\citeauthoryear{Verykios, Bertino, Fovino, Provenza, Saygin,
  and Theodoridis}{Verykios et~al\mbox{.}}{2004}]%
        {verykios2004state}
\bibfield{author}{\bibinfo{person}{Vassilios~S Verykios},
  \bibinfo{person}{Elisa Bertino}, \bibinfo{person}{Igor~Nai Fovino},
  \bibinfo{person}{Loredana~Parasiliti Provenza}, \bibinfo{person}{Yucel
  Saygin}, {and} \bibinfo{person}{Yannis Theodoridis}.}
  \bibinfo{year}{2004}\natexlab{}.
\newblock \showarticletitle{State-of-the-art in privacy preserving data
  mining}.
\newblock \bibinfo{journal}{\emph{ACM Sigmod Record}} \bibinfo{volume}{33},
  \bibinfo{number}{1} (\bibinfo{year}{2004}), \bibinfo{pages}{50--57}.
\newblock


\bibitem[\protect\citeauthoryear{Wang, Li, Wang, and Jin}{Wang
  et~al\mbox{.}}{2018}]%
        {wang2018you}
\bibfield{author}{\bibinfo{person}{Huandong Wang}, \bibinfo{person}{Yong Li},
  \bibinfo{person}{Gang Wang}, {and} \bibinfo{person}{Depeng Jin}.}
  \bibinfo{year}{2018}\natexlab{}.
\newblock \showarticletitle{You Are How You Move: Linking Multiple User
  Identities From Massive Mobility Traces}. In
  \bibinfo{booktitle}{\emph{Proceedings of SIAM SDM}}. Society for Industrial
  and Applied Mathematics.
\newblock


\bibitem[\protect\citeauthoryear{Wang, Guo, Lan, Xu, and Cheng}{Wang
  et~al\mbox{.}}{2016}]%
        {wang2016your}
\bibfield{author}{\bibinfo{person}{Pengfei Wang}, \bibinfo{person}{Jiafeng
  Guo}, \bibinfo{person}{Yanyan Lan}, \bibinfo{person}{Jun Xu}, {and}
  \bibinfo{person}{Xueqi Cheng}.} \bibinfo{year}{2016}\natexlab{}.
\newblock \showarticletitle{Your cart tells you: Inferring demographic
  attributes from purchase data}. In \bibinfo{booktitle}{\emph{Proceedings of
  WSDM}}. ACM.
\newblock


\bibitem[\protect\citeauthoryear{Wang and Wu}{Wang and Wu}{2013}]%
        {wang2013preserving}
\bibfield{author}{\bibinfo{person}{Yue Wang} {and} \bibinfo{person}{Xintao
  Wu}.} \bibinfo{year}{2013}\natexlab{}.
\newblock \showarticletitle{Preserving differential privacy in
  degree-correlation based graph generation}.
\newblock \bibinfo{journal}{\emph{Transactions on data privacy}}
  \bibinfo{volume}{6}, \bibinfo{number}{2} (\bibinfo{year}{2013}),
  \bibinfo{pages}{127}.
\newblock


\bibitem[\protect\citeauthoryear{Weinsberg, Bhagat, Ioannidis, and
  Taft}{Weinsberg et~al\mbox{.}}{2012}]%
        {weinsberg2012blurme}
\bibfield{author}{\bibinfo{person}{Udi Weinsberg}, \bibinfo{person}{Smriti
  Bhagat}, \bibinfo{person}{Stratis Ioannidis}, {and} \bibinfo{person}{Nina
  Taft}.} \bibinfo{year}{2012}\natexlab{}.
\newblock \showarticletitle{BlurMe: Inferring and obfuscating user gender based
  on ratings}. In \bibinfo{booktitle}{\emph{Proceedings of the sixth ACM
  conference on Recommender systems}}. ACM, \bibinfo{pages}{195--202}.
\newblock


\bibitem[\protect\citeauthoryear{Wu, Hu, Fu, Fu, Wang, and Lu}{Wu
  et~al\mbox{.}}{2018}]%
        {wu2018social}
\bibfield{author}{\bibinfo{person}{Xinyu Wu}, \bibinfo{person}{Zhongzhao Hu},
  \bibinfo{person}{Xinzhe Fu}, \bibinfo{person}{Luoyi Fu},
  \bibinfo{person}{Xinbing Wang}, {and} \bibinfo{person}{Songwu Lu}.}
  \bibinfo{year}{2018}\natexlab{}.
\newblock \showarticletitle{Social network de-anonymization with overlapping
  communities: Analysis, algorithm and experiments}. \bibinfo{publisher}{In
  Proceeding of INFOCOM}.
\newblock


\bibitem[\protect\citeauthoryear{Xiao, Chen, and Tan}{Xiao
  et~al\mbox{.}}{2014}]%
        {xiao2014differentially}
\bibfield{author}{\bibinfo{person}{Qian Xiao}, \bibinfo{person}{Rui Chen},
  {and} \bibinfo{person}{Kian-Lee Tan}.} \bibinfo{year}{2014}\natexlab{}.
\newblock \showarticletitle{Differentially private network data release via
  structural inference}. In \bibinfo{booktitle}{\emph{Proceedings of the 20th
  ACM SIGKDD international conference on Knowledge discovery and data mining}}.
  ACM, \bibinfo{pages}{911--920}.
\newblock


\bibitem[\protect\citeauthoryear{Xin and Jaakkola}{Xin and Jaakkola}{2014}]%
        {xin2014controlling}
\bibfield{author}{\bibinfo{person}{Yu Xin} {and} \bibinfo{person}{Tommi
  Jaakkola}.} \bibinfo{year}{2014}\natexlab{}.
\newblock \showarticletitle{Controlling privacy in recommender systems}. In
  \bibinfo{booktitle}{\emph{Advances in Neural Information Processing
  Systems}}. \bibinfo{pages}{2618--2626}.
\newblock


\bibitem[\protect\citeauthoryear{Yang and Leskovec}{Yang and Leskovec}{2013}]%
        {yang2013overlapping}
\bibfield{author}{\bibinfo{person}{Jaewon Yang} {and} \bibinfo{person}{Jure
  Leskovec}.} \bibinfo{year}{2013}\natexlab{}.
\newblock \showarticletitle{Overlapping community detection at scale: a
  nonnegative matrix factorization approach}. In
  \bibinfo{booktitle}{\emph{Proceedings of the sixth ACM international
  conference on Web search and data mining}}. ACM, \bibinfo{pages}{587--596}.
\newblock


\bibitem[\protect\citeauthoryear{Yang, McAuley, and Leskovec}{Yang
  et~al\mbox{.}}{2013}]%
        {yang2013community}
\bibfield{author}{\bibinfo{person}{Jaewon Yang}, \bibinfo{person}{Julian
  McAuley}, {and} \bibinfo{person}{Jure Leskovec}.}
  \bibinfo{year}{2013}\natexlab{}.
\newblock \showarticletitle{Community detection in networks with node
  attributes}. In \bibinfo{booktitle}{\emph{Data Mining (ICDM), 2013 IEEE 13th
  international conference on}}. IEEE, \bibinfo{pages}{1151--1156}.
\newblock


\bibitem[\protect\citeauthoryear{Yartseva and Grossglauser}{Yartseva and
  Grossglauser}{2013}]%
        {yartseva2013performance}
\bibfield{author}{\bibinfo{person}{Lyudmila Yartseva} {and}
  \bibinfo{person}{Matthias Grossglauser}.} \bibinfo{year}{2013}\natexlab{}.
\newblock \showarticletitle{On the performance of percolation graph matching}.
  In \bibinfo{booktitle}{\emph{Proceedings of the first ACM conference on
  Online social networks}}. ACM, \bibinfo{pages}{119--130}.
\newblock


\bibitem[\protect\citeauthoryear{Yin, Gupta, Weninger, and Han}{Yin
  et~al\mbox{.}}{2010a}]%
        {yin2010linkrec}
\bibfield{author}{\bibinfo{person}{Zhijun Yin}, \bibinfo{person}{Manish Gupta},
  \bibinfo{person}{Tim Weninger}, {and} \bibinfo{person}{Jiawei Han}.}
  \bibinfo{year}{2010}\natexlab{a}.
\newblock \showarticletitle{Linkrec: a unified framework for link
  recommendation with user attributes and graph structure}. In
  \bibinfo{booktitle}{\emph{Proceedings of WWW}}. ACM,
  \bibinfo{pages}{1211--1212}.
\newblock


\bibitem[\protect\citeauthoryear{Yin, Gupta, Weninger, and Han}{Yin
  et~al\mbox{.}}{2010b}]%
        {yin2010unified}
\bibfield{author}{\bibinfo{person}{Zhijun Yin}, \bibinfo{person}{Manish Gupta},
  \bibinfo{person}{Tim Weninger}, {and} \bibinfo{person}{Jiawei Han}.}
  \bibinfo{year}{2010}\natexlab{b}.
\newblock \showarticletitle{A unified framework for link recommendation using
  random walks}. In \bibinfo{booktitle}{\emph{Proceedings of ASONAM}}. IEEE,
  \bibinfo{pages}{152--159}.
\newblock


\bibitem[\protect\citeauthoryear{Ying and Wu}{Ying and Wu}{2009}]%
        {ying2009graph}
\bibfield{author}{\bibinfo{person}{Xiaowei Ying} {and} \bibinfo{person}{Xintao
  Wu}.} \bibinfo{year}{2009}\natexlab{}.
\newblock \showarticletitle{Graph generation with prescribed feature
  constraints}. In \bibinfo{booktitle}{\emph{Proceedings of the 2009 SIAM
  International Conference on Data Mining}}. SIAM, \bibinfo{pages}{966--977}.
\newblock


\bibitem[\protect\citeauthoryear{Yuan, Chen, and Yu}{Yuan
  et~al\mbox{.}}{2010}]%
        {yuan2010personalized}
\bibfield{author}{\bibinfo{person}{Mingxuan Yuan}, \bibinfo{person}{Lei Chen},
  {and} \bibinfo{person}{Philip~S Yu}.} \bibinfo{year}{2010}\natexlab{}.
\newblock \showarticletitle{Personalized privacy protection in social
  networks}.
\newblock \bibinfo{journal}{\emph{Proceedings of the VLDB Endowment}}
  \bibinfo{volume}{4}, \bibinfo{number}{2} (\bibinfo{year}{2010}),
  \bibinfo{pages}{141--150}.
\newblock


\bibitem[\protect\citeauthoryear{Zafarani, Abbasi, and Liu}{Zafarani
  et~al\mbox{.}}{2014}]%
        {zafarani2014social}
\bibfield{author}{\bibinfo{person}{Reza Zafarani},
  \bibinfo{person}{Mohammad~Ali Abbasi}, {and} \bibinfo{person}{Huan Liu}.}
  \bibinfo{year}{2014}\natexlab{}.
\newblock \bibinfo{booktitle}{\emph{Social media mining: an introduction}}.
\newblock \bibinfo{publisher}{Cambridge University Press}.
\newblock


\bibitem[\protect\citeauthoryear{Zhang, Xie, Gunter, Han, and Wang}{Zhang
  et~al\mbox{.}}{2014}]%
        {zhang2014privacy}
\bibfield{author}{\bibinfo{person}{Aston Zhang}, \bibinfo{person}{Xing Xie},
  \bibinfo{person}{Carl~A Gunter}, \bibinfo{person}{Jiawei Han}, {and}
  \bibinfo{person}{XiaoFeng Wang}.} \bibinfo{year}{2014}\natexlab{}.
\newblock \showarticletitle{Privacy Risk in Anonymized Heterogeneous
  Information Networks.}
\newblock \bibinfo{journal}{\emph{EDBT}} (\bibinfo{year}{2014}).
\newblock


\bibitem[\protect\citeauthoryear{Zhang, Sun, Zhang, and Zhang}{Zhang
  et~al\mbox{.}}{2018}]%
        {TextAnonymization}
\bibfield{author}{\bibinfo{person}{Jinxue Zhang}, \bibinfo{person}{Jingchao
  Sun}, \bibinfo{person}{Rui Zhang}, {and} \bibinfo{person}{Yanchao Zhang}.}
  \bibinfo{year}{2018}\natexlab{}.
\newblock \showarticletitle{Privacy-Preserving Social Media Data Outsourcing}.
  In \bibinfo{booktitle}{\emph{Proceedings of IEEE International Conference on
  Computer Communications (INFOCOM)}}.
\newblock


\bibitem[\protect\citeauthoryear{Zheleva and Getoor}{Zheleva and
  Getoor}{2009}]%
        {zheleva2009join}
\bibfield{author}{\bibinfo{person}{Elena Zheleva} {and} \bibinfo{person}{Lise
  Getoor}.} \bibinfo{year}{2009}\natexlab{}.
\newblock \showarticletitle{To join or not to join: the illusion of privacy in
  social networks with mixed public and private user profiles}. In
  \bibinfo{booktitle}{\emph{Proceedings of the 18th international conference on
  World wide web}}. ACM, \bibinfo{pages}{531--540}.
\newblock


\bibitem[\protect\citeauthoryear{Zheleva, Terzi, and Getoor}{Zheleva
  et~al\mbox{.}}{2012}]%
        {zheleva2012privacy}
\bibfield{author}{\bibinfo{person}{Elena Zheleva}, \bibinfo{person}{Evimaria
  Terzi}, {and} \bibinfo{person}{Lise Getoor}.}
  \bibinfo{year}{2012}\natexlab{}.
\newblock \showarticletitle{Privacy in social networks}.
\newblock \bibinfo{journal}{\emph{Synthesis Lectures on Data Mining and
  Knowledge Discovery}} \bibinfo{volume}{3}, \bibinfo{number}{1}
  (\bibinfo{year}{2012}), \bibinfo{pages}{1--85}.
\newblock


\bibitem[\protect\citeauthoryear{Zheng, Han, and Sun}{Zheng
  et~al\mbox{.}}{2018}]%
        {8295255}
\bibfield{author}{\bibinfo{person}{X. Zheng}, \bibinfo{person}{J. Han}, {and}
  \bibinfo{person}{A. Sun}.} \bibinfo{year}{2018}\natexlab{}.
\newblock \showarticletitle{A Survey of Location Prediction on Twitter}.
\newblock \bibinfo{journal}{\emph{IEEE Transactions on Knowledge and Data
  Engineering}} (\bibinfo{year}{2018}).
\newblock


\bibitem[\protect\citeauthoryear{Zhong, Yuan, Zhong, Zhang, and Xie}{Zhong
  et~al\mbox{.}}{2015}]%
        {zhong2015you}
\bibfield{author}{\bibinfo{person}{Yuan Zhong}, \bibinfo{person}{Nicholas~Jing
  Yuan}, \bibinfo{person}{Wen Zhong}, \bibinfo{person}{Fuzheng Zhang}, {and}
  \bibinfo{person}{Xing Xie}.} \bibinfo{year}{2015}\natexlab{}.
\newblock \showarticletitle{You are where you go: Inferring demographic
  attributes from location check-ins}. In \bibinfo{booktitle}{\emph{Proceedings
  of WSDM}}. ACM, \bibinfo{pages}{295--304}.
\newblock


\bibitem[\protect\citeauthoryear{Zhou and Pei}{Zhou and Pei}{2008}]%
        {zhou2008preserving}
\bibfield{author}{\bibinfo{person}{Bin Zhou} {and} \bibinfo{person}{Jian Pei}.}
  \bibinfo{year}{2008}\natexlab{}.
\newblock \showarticletitle{Preserving privacy in social networks against
  neighborhood attacks}. In \bibinfo{booktitle}{\emph{Data Engineering, 2008.
  ICDE 2008. IEEE 24th International Conference on}}. IEEE,
  \bibinfo{pages}{506--515}.
\newblock


\bibitem[\protect\citeauthoryear{Zhu, Li, Ren, Zhou, and Xiong}{Zhu
  et~al\mbox{.}}{2013}]%
        {zhu2013differential}
\bibfield{author}{\bibinfo{person}{Tianqing Zhu}, \bibinfo{person}{Gang Li},
  \bibinfo{person}{Yongli Ren}, \bibinfo{person}{Wanlei Zhou}, {and}
  \bibinfo{person}{Ping Xiong}.} \bibinfo{year}{2013}\natexlab{}.
\newblock \showarticletitle{Differential privacy for neighborhood-based
  collaborative filtering}. In \bibinfo{booktitle}{\emph{Proceedings of
  ASONAM}}. ACM, \bibinfo{pages}{752--759}.
\newblock


\bibitem[\protect\citeauthoryear{Zhu and Sun}{Zhu and Sun}{2016}]%
        {zhu2016differential}
\bibfield{author}{\bibinfo{person}{Xue Zhu} {and} \bibinfo{person}{Yuqing
  Sun}.} \bibinfo{year}{2016}\natexlab{}.
\newblock \showarticletitle{Differential privacy for collaborative filtering
  recommender algorithm}. In \bibinfo{booktitle}{\emph{Proceedings of the 2016
  ACM on International Workshop on Security And Privacy Analytics}}. ACM,
  \bibinfo{pages}{9--16}.
\newblock


\bibitem[\protect\citeauthoryear{Zou, Chen, and {\"O}zsu}{Zou
  et~al\mbox{.}}{2009}]%
        {zou2009k}
\bibfield{author}{\bibinfo{person}{Lei Zou}, \bibinfo{person}{Lei Chen}, {and}
  \bibinfo{person}{M~Tamer {\"O}zsu}.} \bibinfo{year}{2009}\natexlab{}.
\newblock \showarticletitle{K-automorphism: A general framework for privacy
  preserving network publication}.
\newblock \bibinfo{journal}{\emph{Proceedings of the VLDB Endowment}}
  \bibinfo{volume}{2}, \bibinfo{number}{1} (\bibinfo{year}{2009}),
  \bibinfo{pages}{946--957}.
\newblock


\end{thebibliography}

\end{document}